\documentclass[review]{elsarticle}
\PassOptionsToPackage{hyphens}{url}\usepackage{hyperref}
\usepackage[T1]{fontenc}%

\usepackage[per-mode=symbol]{siunitx}
\usepackage{subfig}
\usepackage{textcomp}%

\journal{Computer Physics Communications}

\usepackage{booktabs}
\usepackage{graphicx}
\usepackage{amssymb}
\usepackage[fleqn]{amsmath}
\usepackage{moresize}%

\newcommand{\vecki}{\ensuremath{\vec k_i}}
\newcommand{\veckf}{\ensuremath{\vec k_f}}
\newcommand{\ki}{\ensuremath{\vec k_i}}
\newcommand{\kf}{\ensuremath{\vec k_f}}
\newcommand{\Ei}{\ensuremath{E_i}}
\newcommand{\Ef}{\ensuremath{E_f}}

\newcommand{\targeti}{\ensuremath{\eta_i}}
\newcommand{\targetf}{\ensuremath{\eta_f}}
\newcommand{\targetEi}{\ensuremath{E_{\eta_i}}}
\newcommand{\targetEf}{\ensuremath{E_{\eta_f}}}
\newcommand{\operator}[1]{\ensuremath{\boldsymbol{#1}}}
\newcommand{\operatorvec}[1]{\ensuremath{\vec{\operator{#1}}}}
\newcommand{\kboltz}{\ensuremath{k_\text{B}}}
\newcommand{\order}[1]{\ensuremath{\mathcal{O}(#1)}}
\newcommand{\thalf}{\tfrac{1}{2}}

\newcommand{\thetabragg}{\ensuremath{\theta_\text{B}}}

\def\LRA{\Leftrightarrow}
\def\RA{\Rightarrow}
\newcommand\halfopen[2]{\ensuremath{[#1,#2)}}

\newcommand{\labsec}[1]{\label{sec:#1}}
\newcommand{\refsecnumonly}[1]{\ref{sec:#1}}
\newcommand{\refsec}[1]{Section~\refsecnumonly{#1}}
\newcommand{\Refsec}[1]{Section~\refsecnumonly{#1}}%
\newcommand{\reftwosections}[2]{Sections~\refsecnumonly{#1} and \refsecnumonly{#2}}
\newcommand{\labfig}[1]{\label{fig:#1}}
\newcommand{\reffignumonly}[1]{\ref{fig:#1}}
\newcommand{\reffig}[1]{Figure~\reffignumonly{#1}}
\newcommand{\Reffig}[1]{Figure~\reffignumonly{#1}}%
\newcommand{\refthreefigs}[3]{Figures~\reffignumonly{#1}, \reffignumonly{#2}, and \reffignumonly{#3}}

\newcommand{\labtab}[1]{\label{tab:#1}}
\newcommand{\reftabnumonly}[1]{\ref{tab:#1}}
\newcommand{\reftab}[1]{Table~\reftabnumonly{#1}}
\newcommand{\Reftab}[1]{Table~\reftabnumonly{#1}}%

\newcommand{\reflistingnumonly}[1]{\ref{lst:#1}}
\newcommand{\reflisting}[1]{Listing~\reflistingnumonly{#1}}
\newcommand{\Reflisting}[1]{Listing~\reflistingnumonly{#1}}%
\newcommand{\labeqn}[1]{\label{eqn:#1}}
\newcommand{\refeqnnumonly}[1]{\ref{eqn:#1}}
\newcommand{\refeqn}[1]{Eq.~\refeqnnumonly{#1}}
\newcommand{\Refeqn}[1]{Eq.~\refeqnnumonly{#1}}%
\newcommand{\reftwoeqns}[2]{Eqs.~\refeqnnumonly{#1} and \refeqnnumonly{#2}}

\usepackage{color}
\usepackage{listings}
\definecolor{lstcolstr}{rgb}{0,0.8,0}
\definecolor{lstcolkw}{rgb}{0,0,0.8}
\definecolor{lstcolcmt}{rgb}{0.8,0,0}
\lstdefinelanguage[ncrystal]{C}[ANSI]{C}{
  morekeywords=[1]{
    ncrystal_info_t,ncrystal_scatter_t,ncrystal_absorption_t,
    int32_t, uint32_t, int64_t, uint64_t, size_t
}}
\lstdefinelanguage[ncrystal]{C++}[]{C++}{
  morekeywords=[1]{
    NC,NCrystal,
    int32_t, uint32_t, int64_t, uint64_t, size_t,
    G4String, G4int, G4double, G4ThreeVector
}}
\lstdefinelanguage[ncrystal]{Python}[]{Python}{
  morekeywords=[1]{
    NC,NCrystal,
}}
\lstdefinelanguage[mccode]{C}[ANSI]{C}{
  morekeywords=[1]{
    DECLARE, DEFINE, END, FINALLY, INITIALIZE, MCDISPLAY, SAVE, SHARE,
    TRACE, DEFINITION, PARAMETERS, POLARISATION, SETTING,
    OUTPUT, INSTRUMENT, include,
    ABSOLUTE,AT,COMPONENT,EXTEND,GROUP,PREVIOUS,NEXT,MYSELF,RELATIVE,ROTATED,WHEN,JUMP,ITERATE,SPLIT,COPY
}}

\lstdefinelanguage[mccode]{C}[ANSI]{C}{
  morekeywords=[1]{
    DECLARE, DEFINE, END, FINALLY, INITIALIZE, MCDISPLAY, SAVE, SHARE,
    TRACE, DEFINITION, PARAMETERS, POLARISATION, SETTING,
    OUTPUT, INSTRUMENT, include,
    ABSOLUTE,AT,COMPONENT,EXTEND,GROUP,PREVIOUS,NEXT,MYSELF,RELATIVE,ROTATED,WHEN,JUMP,ITERATE,SPLIT,COPY
}}

\newcommand{\lstinlinecpp}[1]{\lstinline[columns=fixed,language={[ncrystal]C++}]{#1}}

\begin{document}

\begin{frontmatter}

\title{NCrystal : a library for thermal neutron transport}
\author[addressess,addressdtunutech]{X.-X.~Cai}
\author[addressess]{T.~Kittelmann\corref{mycorrespondingauthor}}
\cortext[mycorrespondingauthor]{Corresponding author. {\em Email address:} \texttt{thomas.kittelmann@esss.se}}

\address[addressess]{European Spallation Source ERIC, Sweden}
\address[addressdtunutech]{DTU Nutech, Technical University of Denmark, Denmark}

\begin{abstract}

  An open source software package for modelling thermal neutron transport is
  presented. The code facilitates Monte Carlo-based transport simulations and
  focuses in the initial release on interactions in both
  mosaic single crystals as well as polycrystalline materials and powders.
  Both coherent elastic (Bragg diffraction) and incoherent or
  inelastic (phonon) scattering are modelled, using basic parameters of the crystal unit
  cell as input.

  Included is a data library of validated crystal definitions, standalone
  tools and interfaces for \texttt{C++}, \texttt{C} and \texttt{Python}
  programming languages. Interfaces for two popular simulation packages,
  \texttt{Geant4} and \texttt{McStas}, are provided, enabling highly realistic
  simulations of typical components at neutron scattering instruments, including
  beam filters, monochromators, analysers, samples, and detectors. All interfaces
  are presented in detail, along with the end-user configuration procedure which
  is deliberately kept user-friendly and consistent across all interfaces.

  An overview of the relevant neutron scattering theory is provided, and the
  physics modelling capabilities of the software are discussed. Particular
  attention is given here to the ability to load crystal structures and form
  factors from various sources of input, and the results are benchmarked and
  validated against experimental data and existing crystallographic
  software. Good agreements are observed.

\end{abstract}

\end{frontmatter}

{\bf PROGRAM SUMMARY}

\begin{small}
\noindent
{\em Manuscript Title:} NCrystal : a library for thermal neutron transport                                 \\
{\em Authors:} X.-X.~Cai and T.~Kittelmann \\
{\em Program Title:} NCrystal                                          \\
{\em Journal Reference:}                                      \\
{\em Catalogue identifier:}                                   \\
{\em Licensing provisions:} Apache License, Version 2.0 (for core \texttt{NCrystal}).\\
{\em Programming language:} \texttt{C++}, \texttt{C} and \texttt{Python}       \\
{\em Operating system:} Linux, OSX, Windows                        \\
{\em Keywords:}  \\
Thermal neutron scattering, Simulations, Monte Carlo, Crystals, Bragg diffraction\\
{\em Classification:} 4, 7.6, 8, 17 \\
{\em External routines/libraries:} \texttt{Geant4}, \texttt{McStas}           \\
{\em Nature of problem:}\\ Thermal neutron transport in structured materials is
inadequately supported in popular Monte Carlo transport applications, preventing
simulations of a range of otherwise interesting setups.
   \\
{\em Solution method:}\\
Provide models for thermal neutron transport in flexible open source library, to be
used standalone or as backend in existing Monte Carlo packages. Facilitate
validation and work sharing by making it possible to share material
configurations between supported applications.
   \\
\end{small}

\section{Introduction}\labsec{intro}
The modelling of neutron transport through matter dates back to the efforts
aimed at tackling neutron diffusion problems in the middle of the twentieth
century, closely tied to the introduction of general purpose computers and the
inception of the method of Monte Carlo simulations~\cite{Metropolis:1987:BMC}.
Since then, the needs for transport simulations involving neutrons have
expanded, with a wide range of applications in radiation protection, reactor
physics, radiotherapy, and scattering instruments at spallation or
reactor sources. To facilitate such simulations a plethora of Monte Carlo
simulation applications exists today, which for the purposes here can be divided
into two categories: \emph{general purpose}
applications~\cite{mcnpx2006,mcnp5man,mcnp6,geant4a,geant4b,geant4c,fluka2014,phits2018,serpentmc,openmc},
capable of modelling a variety of particle types in flexible geometrical
layouts, and \emph{specialised}
applications~\cite{mcstas1,mcstas2,vitess1,vitess2,restrax,nispmc} dealing with
the design of neutron scattering instruments. Applications in the latter group
is focused on aspects of individual scatterings of thermal neutrons, but
generally lack capabilities for dealing with arbitrarily complex geometries and
the inclusion of physics of particles other than thermalised neutrons.

On the
other hand, the general purpose Monte Carlo applications are typically oriented
towards use-cases in fields such as reactor physics, high energy physics, and
radiation protection, and as a rule neglect all material structure and
effects of inter-atomic bindings -- resorting instead to the approximation of
treating all materials as a free gas of unbound atoms with no mutual
interactions.  This approximation is suitable at higher incident energies, but
for neutrons it breaks down at the thermal scale where their wavelengths and
kinetic energies are comparable to the typical distances and excitational energies
resulting from inter-atomic bindings.  The only presently available
option for improved modelling of thermal neutron interactions involves the
utilisation of specially prepared data files of differential cross section
data. These ``scattering kernels'' are only readily available in neutron evaluation
libraries, e.g.~\cite{endf8}, for the dozen or so materials which have traditionally
played an important role for shielding or moderation purposes at nuclear
facilities. For other materials, scattering kernels must be carefully
crafted using non-trivial procedures (like the application of delicate and computationally expensive molecular dynamics
models~\cite{marquezdamian2014_cabwater}) and in practice this is rarely
done. Furthermore, some aspects of thermal neutron scatterings are in practice
ignored when modelling is exclusively based on scattering kernels: since
nuclear reactor physics has historically mostly been focused on inelastic and
multiple-scattering phenomena, it is usually not possible to include a precise
model of the \emph{a priory} highly significant process of Bragg diffraction into the
setup for crystalline materials. At thermal energies, Bragg diffraction is the
dominant process for many relevant materials, and although it is an elastic
scattering process which does not change the neutron energy, it sends neutrons to characteristic solid angles
and can therefore critically influence the geometrical
reach of neutrons through a given geometry.

It is thus not currently
straight-forward to carry out simulations which at the same time incorporate
consistent and realistic physics models for neutrons at both high and low energy
scales in general materials, while simultaneously supporting flexible
geometrical layouts and the treatment of particles other than neutrons. Such
simulations would nonetheless be remarkably useful, in particular when
considering aspects of neutron scattering instruments~\cite{simtoolskelly2018,Cherkashyna_2014}. Here, precise
 modelling of individual neutron scatterings is crucial in all
beam-line components, while detailed understanding of background levels, the
effects of shielding, or the
workings of neutron detectors, require incorporation of detailed geometrical
layouts and additional physics such as modelling of gammas, fast neutrons, and
energy depositions resulting from secondary particles released upon
neutron capture. Other examples include the design of advanced moderators
or reflectors for novel reactors or neutron spallation sources, and proton or
neutron-based radiotherapy, in which adverse irradiation from neutrons are
abundant and an increasing concern~\cite{xu2008,Newhauser2011}.

The \texttt{NCrystal} toolkit presented here is aimed at remedying the
situation. Rather than introducing an entirely new application, it is designed
as a flexible backend, able to plug the gaps where existing Monte Carlo
applications are lacking in capabilities for treating thermal neutrons in
structured materials. It was originally developed to facilitate the design and
optimisation of
neutron detectors for the European Spallation Source, presently under
construction in Lund,
Sweden~\cite{esssourcedesign2018,esscdr,esstdr,dgcodechep2013,simtoolskelly2018},
but is intended to be as widely useful as possible -- not only as a service to the
community, but also to ultimately ensure a higher level of quality for the
toolkit itself due to feedback and validation from a larger potential
userbase. Although dedicated utilities for calculating various properties of
thermal neutrons exists~\cite{cripo,njoy2012,nxslib1,nxsg4}, the scope of
\texttt{NCrystal} is different, delivering at the same time an ambitious set of
physics models, a flexible object oriented design, multiple interfaces and
bindings, an open approach to development, and not the least a user-friendly
method of configuration. The core functionality is implemented as an open source
and highly portable \texttt{C++} library with no third-party dependencies, and
comes with language bindings for all versions of \texttt{C++}, \texttt{C} or
\texttt{Python} in widespread usage today. The current code is released with
version number 1.0.0 and under a highly liberal open source
license~\cite{apache2}.\footnote{Optional components adding support for file
  formats discussed in \reftwosections{data::nxs}{data::lazlau} rely on
  third-party code, available under different open source licenses.} Included in
the release is additionally tools, examples, data files, and a configuration
file for \texttt{CMake}~\cite{cmakebook2015} with which everything can be built
and installed. A
website~\cite{ncrystalwww} has been set up for the \texttt{NCrystal} project, on
which users will be able to interact with developers, locate future updates, and
access relevant documentation.

Concerning \texttt{NCrystal}'s capabilities for physics simulations, the work
has so far focused on facilitating realistic simulations of both inelastic and
elastic physics of thermal neutrons as they scatter in certain common bulk
crystalline materials. Namely those which can be described in terms of a
statistical arrangements of microscopic perfect crystals, and for which the
neutron's interaction probabilities are not so strong as to break down the Born
approximation. Additionally, material configuration can be carried out based
simply on a brief description of the crystal structures. All together, the
framework thus provides realistic thermal neutron physics in a large range of
powdered, polycrystalline or mosaic single crystals readily used at neutron
scattering facilities in beam filters, monochromators, analysers, samples,
detectors, and shielding. What is currently not supported, given the mentioned
constraints, is surface effects, materials that are either non-crystalline or
textured, and effects related to a break-down of the Born approximation
including the strong reflections found in neutron guide systems, or the strong
diffraction effects encountered in certain macroscopic crystal systems such as
synthetically grown Silicon. Future developments might address some of those
deficiencies, depending on community interest and resources.

The present paper will present the \texttt{NCrystal} framework itself, including
interfaces, configuration, and data input options.  Although a brief
overview of physics capabilities will be provided, detailed discussions of the
implementation of actual thermal neutron scattering models are beyond the scope
of the present paper, and will be the subject of dedicated publications at a later
date. After a review of the relevant theoretical concepts in \refsec{theory},
the design and implementation of the core \texttt{NCrystal} code is presented in
\refsec{coreframework}.  \Refsec{dataload} presents the capabilities for
initialising relevant parameters of modelled crystals from various data sources,
introduces the library of data files included with \texttt{NCrystal} and
discusses how the loaded results have been validated.
\Refsec{factoriesandunifiedcfg} presents the uniform method for material
configuration intended for most end-users, and an overview of the existing
language bindings and application interfaces is provided in
\refsec{bindingsandinterfaces}. Finally, a discussion of future directions and
planned improvements is carried out in \refsec{outlook}.

\section{Theoretical background}\labsec{theory}

Rather than intending to be an exhaustive treatment, this section will provide a
brief overview of crystals and neutron scattering, introducing relevant
concepts, models and terminology needed for the purposes of the present
discussions of \texttt{NCrystal}. Naturally, relevant references to more
extensive background literature are provided for readers seeking more
information.

\subsection{Crystals}\labsec{theory::crystals}

Although support for other types of materials is likely to be added eventually,
the present scope of \texttt{NCrystal} is restricted to crystals. Consequently,
a few basic concepts will be introduced in the following. Further
details can be found
in~\cite{sands1993crystallographyintro,charleskittel2004,waseda2011x,theo2002international}.

Informally, a crystal can be described as an arrangement of bound
atoms which is built up by the periodic repetition of a basic element in all
three spatial directions.  This repeated element is known as the
unit cell and is typically chosen such that it is the smallest such cell
that reflects the symmetry of the structure.  In an idealised setting, the
crystal structure would be infinite, but in practice it is sufficient that the
unit cell is much smaller than the entire structure. The choice of unit cell for a
given crystal is not always unique, but it is always possible to select one
which is a parallelepiped spanned by three linearly independent basis vectors
$\vec{a}$, $\vec{b}$ and $\vec{c}$.\footnote{Vectors are here and throughout the
  text donated with arrows ($\vec{a}$), while the absence of arrows indicate the
  corresponding scalar magnitudes ($a\equiv|\vec{a}|$). Additionally, unit vectors are donated with
  hats ($\hat{a}\equiv\vec{a}/a$) and quantum mechanical operators are shown in
  bold ($\operator{V}\!$, $\operatorvec{R}$).} By convention they are defined by their
respective lengths, $a$, $b$, and $c$, and the angles between them:
$\alpha\equiv\angle(\vec{b},\vec{c})$, $\beta\equiv\angle(\vec{a},\vec{c})$, and
$\gamma\equiv\angle(\vec{a},\vec{b})$. The crystal structure is completed by the
additional specification of the contents of the unit cell, often provided in
coordinates $(x,y,z)$ relative to the crystal axes. The position of the
$i$th constituent in the unit cell is thus given by:
\begin{align}
  \vec{p}_{i} = x_i\vec{a}+y_i\vec{b}+z_i\vec{c},\qquad x_i,y_i,z_i\in[0,1)
  \labeqn{crystals::ucpos}
\end{align}
Crystals exhibit a discrete symmetry under spatial translation with the vectors:
\begin{align}
  \vec{R}_{mno} = m\vec{a}+n\vec{b}+o\vec{c},\qquad m,n,o\in\mathbb{Z}
  \labeqn{Rmno}
\end{align}
Apart from a trivial global offset, coordinates of all constituents in the entire
crystal are given by $\vec{R}_{mno}+\vec{p}_i$ for all combinations of $mno$ and
$i$. The translational symmetry means that any function representing features of
the crystal system (such as particle densities) will obey:
\begin{align}
  f(\vec{r}) = f(\vec{r}+\vec{R}_{mno})
  \labeqn{crystals::translationalinvariance}
\end{align}
Additional symmetries under rotations, reflections or inversions are possible,
depending on the detailed shape and contents of the unit cell. The symmetry
properties of a crystal are described by the concept of space groups, and it
is possible to describe any spatial crystal symmetry by one of only 230 such
groups. These groups can be divided into 7 general classes of crystal
systems. Loosely ordered from lowest to high{}est symmetry they are: triclinic,
monoclinic, orthorhombic, tetragonal, trigonal, hexagonal, and cubic. The space
groups are defined and described exhaustively in~\cite{theo2002international}.

The space group of a given crystal structure naturally depends on the
unit cell positions of constituents, $\vec{p}_1,\ldots,\vec{p}_N$. Conversely, it
is possible to construct the full list of these positions by applying the
symmetry operators of a given space group to a smaller number of so-called
Wyckoff positions~\cite{wyckoff}. It is indeed very common to find crystal
structures defined solely in terms of their space group, unit cell shape and
dimensions, and a list of elements and associated Wyckoff positions.

The set of points at positions $\vec{R}_{mno}$ for all integers $m$, $n$, and
$o$, constitutes the crystal lattice. In each such lattice, there exist an
infinite number of families of evenly spaced parallel planes passing through the
lattice points. Each such family of lattice planes is described by a common
plane normal $\hat{n}$ and an interplanar distance $d$ (also known as the
$d$-spacing), with planes passing through the points $jd\hat{n}$ for all
$j\in\mathbb{Z}$. Particle diffraction in a crystal occurs in the form of
reflections from such families of lattice planes. The task of finding and
classifying these families is often done in the so-called \emph{reciprocal
  lattice} in momentum-space, which is the Fourier transform of the direct
lattice. The basis vectors of the reciprocal lattice are given by:
\begin{align}
\vec\tau_a=\frac{2\pi}{V_\text{uc}}\left(\vec{b}\times\vec{c}\right),\quad
\vec\tau_b=\frac{2\pi}{V_\text{uc}}\left(\vec{c}\times\vec{a}\right),\quad
\vec\tau_c=\frac{2\pi}{V_\text{uc}}\left(\vec{a}\times\vec{b}\right)
\labeqn{crystals::reclatbasisvectors}
\end{align}
Where $V_\text{uc}=\vec{a}\cdot(\vec{b}\times\vec{c})$ is the unit cell
volume. The points in the reciprocal lattice are thus given by:
\begin{align}
  \vec{\tau}_{hkl} = h\vec\tau_a+k\vec\tau_b+l\vec\tau_c,\qquad h,k,l\in\mathbb{Z}
\end{align}
As will be motivated in \refsec{theory::elastic}, it can be shown that each point in
the reciprocal lattice corresponds to a family of equidistant lattice planes in
the direct lattice. The family of planes corresponding to $\vec{\tau}_{hkl}$
will be indexed by the $hkl$ value (known as its Miller index), and has interplanar
spacing $d_{hkl}=2\pi/\tau_{hkl}$ and normal
$\hat{n}=\hat\tau_{hkl}=\vec\tau_{hkl}/\tau_{hkl}$.

In most real systems, regions in which the crystal lattice is continuous and
the defining translational invariance of
\refeqn{crystals::translationalinvariance} is unbroken, exists only at the
microscopic scale.\footnote{Exceptions to this exist, like in some synthetic
  silicon crystals. However, the scattering theory discussed in
  \refsec{theory::neutronscattering} is in any case not strictly applicable to
  such systems.}  Macroscopic crystals are then built up from these
independently oriented grains of perfect crystals (also known as
``crystallites''), and the actual distribution of orientations will not only be
related to various macroscopic material properties, but will influence
interaction probabilities when the system is probed with impinging particles. As long
as the grain sizes are small compared to the regions of material being probed,
these effects can be accounted for in a statistical manner -- for instance by
integrating microscopic interaction cross sections over the distribution of
crystallite orientations in order to arrive at effective macroscopic cross
sections. Such integrations are particularly trivial to perform in the extreme case where
the orientation of individual crystallites are completely independent and
uniformly distributed over all solid angles. This model, known as the powder
approximation, is not only suitable for modelling actual powdered crystalline
samples, but can also be used to approximate interactions in polycrystalline
materials like metals or ceramics -- especially when the level of correlation in
crystallite orientation (``texture'') is low or when the setup involves
sufficiently large geometries or spread in incoming particles that
effects due to local correlations are washed out. An example of this would be
the case where polycrystalline metal support structures are placed in various
places throughout a neutron instrument. On the other hand, interactions in a
highly textured polycrystalline sample placed in a tightly focused beam of probe
particles might not be very well described by the powder approximation.

In another extreme, all crystallites are almost co-aligned in so-called mosaic
crystals, with the exact distribution around the common axis of alignment given
by the mosaicity distribution. In the simple isotropic case, this distribution
is a Gaussian function of the angular displacement, with the corresponding width
(referred to as the mosaicity of the crystal) having typical values anywhere from
a few arc seconds to a few degrees. In \texttt{NCrystal}, such Gaussian mosaic
crystals are referred to as ``single crystals'', due to the large degree of
crystallite alignment throughout the material.

Some materials are better described by other mosaicity distributions, including anisotropic ones
where the distribution is not symmetric with respect to all axes. A particular
anisotropic distribution supported by \texttt{NCrystal} is one in which
crystallites are aligned around one particular axis, but appearing with random
rotations around the same axis. Such distributions occur in stratified or
layered crystals, like pyrolytic graphite, in which certain planes of atoms
tend to have strongly aligned normals in all crystallites, but no strong
alignment for the rotation of the same planes around their normals.

\subsection{Scattering theory}\labsec{theory::neutronscattering}

The theory of thermal neutron scatterings in condensed matter is thoroughly
treated in the literature, see for
instance~\cite{squires_2012,schober2014,marshalllovesery1971}. The discussion in
the present and following sections is to a large degree inspired
by~\cite{squires_2012,schober2014}.

Scattering in materials by thermal neutrons, taken here to loosely mean energies below
or at the \SI{}{\electronvolt} scale or wavelengths above or at the \SI{}{\nano\meter}
scale, can be described via point-like neutron-nuclei interactions or magnetic
dipole interactions between the neutron and any unpaired electrons in the target
material. For unpolarised neutrons and samples, the latter can be largely
ignored~\cite{Sears1986} and the current discussion will not concern itself with
such magnetic interactions, nor are they currently modelled in \texttt{NCrystal}
-- although there is nothing fundamental preventing their inclusion in the
future.

Thermal neutrons carry such low energy that in practice no emission of secondary
particles occurs during pure scattering interactions.  As a consequence,
scattering of unpolarised neutrons can be completely described in terms of the
differential cross section for scattering the incident neutron with wavevector
$\vecki$ into the wavevector $\veckf$. Based on the Born approximation or Fermi's
golden rule, this differential cross section is related to quantum mechanical
transition amplitudes by the so-called master equation for thermal neutron
scattering:
\begin{align}
  \frac{d^2\sigma_{\vecki\RA\veckf}}{d\Omega_f d\Ef} = \frac{k_f}{k_i}\frac{(2\pi)^4m^2}{\hbar^4}\sum_{\targeti,\targetf}{p(\targeti)|\langle\targetf,\veckf|\operator{V}|\targeti,\vecki\rangle|^2\delta(\hbar\omega+{\Delta}E_\eta)}
  \labeqn{mastereqn}
\end{align}
Here $m$ is the neutron mass, indices $i$ and $f$ denotes initial and final
states respectively, ${\eta_j}$ represents a complete set of eigenstates of the
target system, $p(\targeti)$ represents the initial occupation levels of target
states, and $\operator{V}$ is the interaction potential. Furthermore,
$\hbar\omega=\Ef-\Ei$ and ${\Delta}E_\eta\equiv\targetEf-\targetEi$ represents
the total energy change of the neutron and target system respectively. At this
point it might be useful to recall that in addition to wavenumber,
$k=|\vec{k}|$, the energy state of a non-relativistic neutron might equally be
characterised in terms of energy $E$, momentum $p$, velocity $v$, or wavelength
$\lambda$, related by $p={\hbar}k$, $2E=mv^2=p^2/m$, and $\lambda=2\pi/k$.

Before continuing with the evaluation of \refeqn{mastereqn}, it is important to
understand its validity and applicability. Based on the Born approximation, the
perturbation experienced by the initial neutron wave function during the
scattering is assumed to be small, so it is possible to neglect secondary
scattering of waves created as a result of the primary scattering. The sum in
\refeqn{mastereqn} implies that the resulting cross sections will grow with the
size of the target system, and therefore the validity of the equation will
eventually break down as the size of the target system grows and the probability
for secondary scattering becomes non-negligible. In reality this issue is
circumvented in the context of Monte Carlo transport simulations in which
neutron transport is modelled in a series of steps between independent
scatterings, with each simulated step length depending on the actual mean free
path length predicted by the cross section in
\refeqn{mastereqn}.\footnote{Specifically, if $L$ is the mean free path length
  (depending on cross sections and material density) and $R$ is a pseudo random
  number in the unit interval, then the step length until next interaction can
  be modelled as $-L\log(R)$.} High cross sections will automatically lead to
small step lengths, and therefore only a small part of the sample will be
traversed for each application of \refeqn{mastereqn}. Simulations employing such
stepping will therefore to some degree account for multiple scattering phenomena
like extinction effects in crystal diffraction. This effective subdivision of a
target system into smaller decoherent systems is, however, only possible if the
linear scale over which any symmetries in the material structure exists is small
compared to the mean free path lengths involved, so wavefunctions originating
from scattering in separate subsystems will be mutually incoherent.  As
described in \refsec{theory::crystals} most real macroscopic crystals consists of
independently aligned microscopic crystal grains, and in practice this ensures the necessary
decoherence between different parts of the macroscopic system. The exception is
the case of perfect macroscopic crystals (or mosaic crystals with vanishing
mosaic spread) like some synthetic silicon crystals. For an excellent and detailed
discussion about these issues, refer to~\cite[Sec.~11.5]{schober2014}.

In order to proceed with the evaluation of \refeqn{mastereqn}, one must provide
suitable representations of both interaction potential $\operator{V}$ and
distributions of target states, $p(\targeti)$. When dealing with thermal
neutrons and their relatively long wavelength, the short-range neutron-nuclei
interactions can effectively be described as being point-like. An effective
potential which describes them as such was introduced by
Fermi~\cite{Fermi:1936:SMN}. For a collection of $N$ nuclei located at positions
$\vec{R}_j$ it looks like:
\begin{align}
  V(\vec{r}) = \frac{2\pi\hbar^2}{m}\sum^N_{j=1} b_j\delta(\vec{r}-\vec{R}_j)
  \labeqn{fermipseudopotential}
\end{align}
Here $b_j$ is the scattering length of the $j$th nucleus, an effective parameter
capturing the details of the neutron-nucleus interaction which must be
determined through measurements. The term scattering
length stems from the fact that the corresponding cross section for scattering
from a single fixed nucleus is $4{\pi}b^2$, equal to the classical cross section
of a hard sphere of radius $2|b|$. In the presence of nuclear resonances, the
scattering length becomes strongly dependent on the neutron energy. However, at the
energy scale of thermal neutrons, resonances only exist for a few rare isotopes
-- and even then mostly at the high end of the thermal scale~\cite{sluijs2015}. For the present
purposes, the scattering length thus attain constant values, specific to each
type of nuclei. It is additionally possible to use non-zero imaginary components
of scattering lengths to represent absorption physics, however in many contexts
-- including the present -- the two types of interactions are dealt with
separately and the scattering lengths will therefore be treated as real and
constant numbers. Such separation between scattering and absorption processes is valid at the \order{\num{e-4}}
level for thermal neutrons~\cite{sears1984}, which is considered acceptable. In fact, as the current scope of \texttt{NCrystal} does not
cover physics of nuclear resonances, absorption processes will be dealt with
using the simple -- but in most cases accurate~\cite{sluijs2015} -- model of absorption cross sections being inversely
proportional to the neutron velocity. This $1/v$ scaling can be intuitively
interpreted as having the probability of absorption by a given nucleus
proportional to the time spent by the neutron near the nucleus, but also follows
from more careful reasoning~\cite{sears1984,breitwigner1936,bethe1935}.

It is possible to bring \refeqn{mastereqn} into a form in which the sum over
final states of the target system is removed and the potential is replaced with
its Fourier transform. This is particularly convenient for potentials involving
$\delta$-functions like \refeqn{fermipseudopotential}, as these transform into
constants. The resulting form is:
\begin{align}
 \frac{d^2\sigma_{\vecki\RA\veckf}}{d\Omega_f d\Ef}=\frac{k_f}{k_i}S(\vec Q,\omega)
\end{align}
where $\vec{Q}$ is the momentum transfer, $\vec{Q}\equiv\kf-\ki$, and $S(\vec Q,\omega)$ is the scattering function defined by:
\begin{align}
  S(\vec Q,\omega)\equiv
\frac{1}{2\pi\hbar}
\sum_{j,j'=1}^N{b_jb_{j'}}
\int_{-\infty}^{\infty}
dt\langle
e^{-i\vec{Q}\cdot\operatorvec{R}_{j'}(0)}
e^{i\vec{Q}\cdot\operatorvec{R}_{j}(t)}
\rangle
e^{-i\omega t}
\labeqn{sqwfirstform}
\end{align}
The $\langle\dots\rangle$ notation is here used to donate operator expectation values
in the target system,
$\langle{f}(\operator{A})\rangle=\sum_{\eta_i}p(\eta_i)\langle\eta_i|f(\operator{A})|\eta_i\rangle$,
with the target state weights, $p(\eta_i)$, usually defined by a requirement for
the target system to be in thermal equilibrium. The expectation value under the
integral in \refeqn{sqwfirstform} correlates the position of the nucleus $j$
at time $t$ with the position of the nucleus $j'$ at time $0$, and will here
be abbreviated as:
\begin{align}
  \langle{j}',j\rangle\equiv\langle{e}^{-i\vec{Q}\cdot\operatorvec{R}_{j'}(0)}e^{i\vec{Q}\cdot\operatorvec{R}_{j}(t)}\rangle
  \labeqn{jjprimedef}
\end{align}
Leading to the shorter expression for the scattering function:
\begin{align}
  S(\vec Q,\omega)\equiv
\frac{1}{2\pi\hbar}
\sum_{j,j'=1}^N{b_jb_{j'}}
\int_{-\infty}^{\infty}
dt\langle j',j
\rangle
e^{-i\omega t}
\end{align}
This definition of the scattering function contains a sum over the target system
under consideration (which as discussed must be of a linear size compatible with
the applied Born approximation). Most such systems of interests can be divided
into a number of statistically equivalent subsystems. That this is possible for
crystals is obvious, since each unit cell forms such a subsystem, but
subdivisions are generally possible in almost all systems, be they liquid,
polymeric or gaseous in nature. The scattering function for a subsystem can be
expressed as:
\begin{align}
  S(\vec Q,\omega)\equiv
\frac{1}{2\pi\hbar}
\sum_{j,j'=1}^N{\overline{b_jb_{j'}}}
\int_{-\infty}^{\infty}
dt\langle j',j
\rangle
e^{-i\omega t}
\labeqn{sqw_compact}
\end{align}
where now $N$ refers to the subsystem size and might for instance represent the
number of nuclei in a crystal unit cell, and $\overline{b_jb_{j'}}$ represents
an average performed over an ensemble of equivalent subsystems. Now,
neutron-nuclei scattering has the peculiarity that the scattering lengths are
isotope and spin-state dependent, whereas the positions of scatterers, the
nuclei, are determined by chemical properties, depending on the nuclear charge
but otherwise independent of isotope or spin state.\footnote{Of course, this
  statement becomes invalid for material temperatures near absolute zero where
  the energy difference between different nuclear spin states becomes comparable
  to thermal energies.} This independence means that the ensemble average of
products of scattering lengths found at positions $j$ and $j'$ obeys:
\begin{align}
  \overline{b_j b_{j'}} = \begin{cases}
      \overline{b_j}\cdot\overline{b_{j'}}, & \text{for}\ j\ne j' \\
      \overline{b^2_j}, & \text{for}\ j=j'
    \end{cases}
\labeqn{bjjensembleavprop}
\end{align}
With this, \refeqn{sqw_compact} can be written as the sum of two distinct contributions:
\begin{align}
  S(\vec Q,\omega) =\,& S_\text{coh}(\vec Q,\omega)+S_\text{inc}(\vec Q,\omega)\labeqn{sqw_split_incoh_coh}\\
  S_\text{coh}(\vec Q,\omega) \equiv\,& \frac{1}{2\pi\hbar}
\sum_{j,j'=1}^N{\overline{b_j}\cdot\overline{b_{j'}}}
\int_{-\infty}^{\infty}
dt\langle j',j
\rangle
e^{-i\omega t}\labeqn{sqw_coh}\\
  S_\text{inc}(\vec Q,\omega) \equiv\,& \frac{1}{2\pi\hbar}
\sum_{j=1}^N{\left(\overline{b_j^2}-\left(\overline{b_j}\right)^2\right)}
\int_{-\infty}^{\infty}
dt\langle j,j
\rangle
e^{-i\omega t}\labeqn{sqw_inc}
\end{align}
The terms in the coherent scattering function $S_\text{coh}(\vec Q,\omega)$
directly depend on material structure as they involve pair correlations between
nuclei at all positions in the material, whereas the incoherent scattering
function $S_\text{coh}(\vec Q,\omega)$ solely contains self-correlations terms.
The effective scattering length used for the nuclei at
position $j$ in $S_\text{coh}(\vec Q,\omega)$ is equal to the statistical
average of the scattering lengths at the position $j$ in the entire ensemble,
and is referred to as the coherent scattering length, $b_\text{coh}$. The
corresponding effective scattering length in $S_\text{inc}(\vec Q,\omega)$ is
instead given as the statistical spread of the scattering lengths at the
position $j$ in the entire ensemble, and is referred to as the incoherent
scattering length, $b_{\text{inc}}$. Using $\sigma=4{\pi}b^2$, one might also
talk about the related coherent and incoherent cross sections. Coherent and
incoherent scattering lengths or cross sections are defined for each natural
element (or any other well defined composition of isotopes), and can be directly
applied to target systems in which a given position
$j$ is always occupied by a nuclei of the same element in all subsystem
replicas. It can be shown that for thermal neutrons, the sum of the incoherent and coherent scattering
cross sections is to a good approximation equal to the unbound scattering cross section,
$\sigma_\text{free}$, for which the total scattering cross section will converge
at shorter wavelengths~\cite{sears1984}.\footnote{At \emph{very} short wavelengths this
  conversion is ultimately disrupted by nuclear resonance physics or P-wave contributions.}

The main difficulty in the evaluation of the scattering functions in
\reftwoeqns{sqw_coh}{sqw_inc} is the integral over states in
$\langle{j}',j\rangle$ implicit in \refeqn{jjprimedef}, whose evaluation in
principle relates to the potentially complicated time-dependent distribution of
nuclei in the target system. In a crystal, the nuclear positions can be decomposed
in terms of displacements $\operatorvec{u}$ from equilibrium positions
$\vec{d}$:
\begin{align}
  \operatorvec{R}_j(t)=\vec{d}_j+\operatorvec{u}_j(t)
\end{align}
And thus:
\begin{align}
  \langle{j}',j\rangle=\,& \langle{e}^{-i\vec{Q}\cdot\left(\vec{d}_{j'}+\operatorvec{u}_{j'}(0)\right)}e^{i\vec{Q}\cdot\left(\vec{d}_{j}+\operatorvec{u}_{j}(t)\right)}\rangle\nonumber\\
  =\,&{e}^{-i\vec{Q}\cdot\left(\vec{d}_{j'}-\vec{d}_{j}\right)}\langle{e}^{-i\vec{Q}\cdot\operatorvec{u}_{j'}(0)}e^{i\vec{Q}\cdot\operatorvec{u}_{j}(t)}\rangle
  \labeqn{jjprimefromdisplacements}
\end{align}
Under the assumption that displacements are small compared to inter-atomic
distances, motion of nuclei can be described with potentials which are quadratic
functions of the displacements. In this so-called \emph{harmonic approximation},
nuclei are essentially described as being pulled towards their equilibrium
positions by linear spring-like forces, with resulting harmonic
vibrations. Working in this approximation, it is possible to show that:
\begin{align}
  \langle j,j'\rangle =
  {e}^{-i\vec{Q}\cdot\left(\vec{d}_{j'}- \vec{d}_{j}\right)}
  {e}^{-W_{j'}(\vec{Q})}
  {e}^{-W_{j}(\vec{Q})}
  {e}^{\langle(\vec{Q}\cdot\operatorvec{u}_{j'}(0))(\vec{Q}\cdot\operatorvec{u}_{j}(t))\rangle}
\labeqn{jjharmonicdebyewaller}
\end{align}
where the Debye-Waller function gives a measure of the time-independent average
squared displacement of nuclei $j$ along $\vec{Q}$:
\begin{align}
  W_j(\vec{Q})\equiv\thalf\langle(\vec{Q}\cdot\operatorvec{u}_j(0))^2\rangle
  \labeqn{debyewallerfunction}
\end{align}
The time dependence in \refeqn{jjharmonicdebyewaller} enters exclusively in the
last exponential factor, which thus involves nuclear motion and can be expanded
in a Taylor series as:
\begin{align}
  {e}^{\langle(\vec{Q}\cdot\operatorvec{u}_{j'}(0))(\vec{Q}\cdot\operatorvec{u}_{j}(t))\rangle}
  =\sum_{n=0}^\infty{\frac{1}{n!}\left(\langle(\vec{Q}\cdot\operatorvec{u}_{j'}(0))(\vec{Q}\cdot\operatorvec{u}_{j}(t))\rangle\right)^n}
  \labeqn{jjexpansion}
\end{align}
The expectation value
$\langle(\vec{Q}\cdot\operatorvec{u}_{j'}(0))(\vec{Q}\cdot\operatorvec{u}_{j}(t))\rangle$
correlates linear displacements along $\vec{Q}$ of two nuclei at different
times. It is possible to show that this is directly related to an interaction in
which the neutron exchanges energy and momentum with a phonon
state, and the expansion in \refeqn{jjexpansion} thus lends itself to physics
interpretation in the phonon picture, with the $n$th term corresponding to
$n$-phonon interactions. At lower displacements or $Q$ values the expansion
converges more rapidly, and accordingly  multi-phonon
physics will be less important at lower neutron energies or material
temperatures.

\subsection{Elastic scattering}\labsec{theory::elastic}
The first term in the expansion of \refeqn{jjexpansion} gives rise to elastic
($\ki=\kf$) scattering when inserted into \refeqn{sqw_coh} or \refeqn{sqw_inc},
since:
\begin{align}
  \int_{-\infty}^{\infty}
dt e^{-i\omega t} = 2\pi\hbar\delta(\hbar\omega)=2\pi\hbar\delta(E_f-E_i)
\end{align}
With this, the partial differential cross section for incoherent elastic
scattering can be immediately found:
\begin{align}
 \frac{d^2\sigma^{\text{inc,el}}_{\vecki\RA\veckf}}{d\Omega_f d\Ef}=\frac{k_f}{k_i}S^\text{el}_\text{inc}(\vec Q,\omega) =
\sum_{j=1}^N{\left(\overline{b_j^2}-\left(\overline{b_j}\right)^2\right)}
{e}^{-2W_j(\vec{Q})}\delta(\hbar\omega)
\end{align}
The only directional dependency
enters here in the Debye-Waller factor, ${e}^{-2W_j(\vec{Q})}$, which approaches
unity when the neutron wavelength is much larger than the atomic displacements,
implying isotropic scattering. At higher neutron energies or
temperatures this approximation breaks down, and the scattering will be
increasingly focused in the forward direction at low values of the scatter angle
$\theta$, since for elastic scattering $Q=2k_i\sin(\theta/2)$.

The coherent scattering terms involve correlations between pairs of nuclei and
their evaluation will therefore be considerably more complex. In a crystal the
sum $\sum_j$ over all nuclei can be rewritten as a sum over unit cells first and
unit cell contents second: $\sum_{\vec{l}}\sum_i$. Here the index $\vec{l}$
assumes all $\vec{R}_{mno}$ values from \refeqn{Rmno} and the index $i$ runs
over all nuclei in the unit cell. Thus, the equilibrium positions formerly
denoted $d_j$ now becomes $\vec{l}+\vec{p}_i$, with $\vec{p}_i$ the
positions defined in \refeqn{crystals::ucpos}. Reordering sums appropriately and proceeding as for the incoherent
case, the expression for coherent elastic scattering in crystals
becomes:
\begin{align}
  S^\text{el}_\text{coh}(\vec Q,\omega) =\,& \delta(\hbar\omega)
\sum_{\vec{l},\vec{l}'}
  {e}^{-i\vec{Q}\cdot\left(\vec{l}'- \vec{l}\right)}
\sum_{i,i'}
  \overline{b_i}\cdot\overline{b_{i'}}
  {e}^{-W_{i'}(\vec{Q})}
  {e}^{-W_{i}(\vec{Q})}
  {e}^{-i\vec{Q}\cdot\left(\vec{p}_{i'}- \vec{p}_i\right)}
\nonumber\\
=\,&\delta(\hbar\omega)
\left(\sum_{\vec{l},\vec{l}'}{e}^{-i\vec{Q}\cdot\left(\vec{l}'- \vec{l}\right)}\right)
  \left|F(\vec{Q})\right|^2\labeqn{scat::cohelcrystalsstep1}
\end{align}
Where the form factor of the unit cell have been introduced:
\begin{align}
F(\vec{Q})\equiv\sum_{i}\overline{b_i}{e}^{-W_{i}(\vec{Q})}{e}^{i\vec{Q}\cdot\vec{p}_{i}}
\labeqn{unitcellformfactordef}
\end{align}
The parenthesised factor in \refeqn{scat::cohelcrystalsstep1} accounts for
interference between different unit cells. Given that all
$\vec{l}'=\vec{R}_{mno}$ for suitable choice of integers $m$, $n$ and $o$, the
translational invariance of \refeqn{crystals::translationalinvariance} implies:
\begin{align}
  \sum_{\vec{l},\vec{l}'}{e}^{-i\vec{Q}\cdot\left(\vec{l}'- \vec{l}\right)}
  = \sum_{\vec{l}'}\sum_{\vec{l}}{e}^{-i\vec{Q}\cdot\left(\vec{l}'- \vec{l}\right)}
  = N_\text{uc}\sum_{\vec{l}}{e}^{i\vec{Q}\cdot\vec{l}}
\end{align}
Where $N_\text{uc}$ is the number of unit cells in the crystal, which can be
removed by adopting the convention of providing cross sections normalised per
unit cell. The remaining factor can be investigated further by expressing
$\vec{Q}$ in terms of the basis vectors of the reciprocal lattice from
\refeqn{crystals::reclatbasisvectors}, $\vec{Q}=q_a\vec\tau_a+q_b\vec\tau_b+q_c\vec\tau_c$:
\begin{align}
  \sum_{\vec{l}}{e}^{i\vec{Q}\cdot\vec{l}} =\,&
  \sum_m\sum_n\sum_o{e}^{i(q_a\vec\tau_a+q_b\vec\tau_b+q_c\vec\tau_c)\cdot(m\vec{a}+n\vec{b}+o\vec{c})}\nonumber\\
  =\,&
  \left(\sum_m{e}^{i2{\pi}q_am}\right)
  \left(\sum_n{e}^{i2{\pi}q_bn}\right)
  \left(\sum_o{e}^{i2{\pi}q_co}\right)\labeqn{expqdotlintermidateresult}
\end{align}
Using:
\begin{align}
  \sum^{\infty}_{n=-\infty}e^{i2{\pi}xn}=\sum^{\infty}_{k=-\infty}\delta(x-k)
\end{align}
\Refeqn{expqdotlintermidateresult} becomes:
\begin{align}
  \left(\sum^{\infty}_{h=-\infty}\delta(q_a-h)\right)
  \left(\sum^{\infty}_{k=-\infty}\delta(q_b-k)\right)
  \left(\sum^{\infty}_{l=-\infty}\delta(q_c-l)\right)
\end{align}
Thus, coherent elastic scattering occurs only when
the momentum transfer, $\vec{Q}$, is exactly identical to one of the points in
the reciprocal lattice, $\vec\tau_{hkl}$, corresponding to interaction with the
associated family of lattice planes. The fact that $\vec{Q}=\vec\tau_{hkl}$
further supports the interpretation of reflection by lattice planes with normal
along $\vec\tau_{hkl}$, since elastic specular reflection by a mirror will
always have the normal proportional to the transferred momentum.
As $(\vec\tau_a,\vec\tau_b,\vec\tau_c)$ constitutes an orthogonal but not an
orthonormal base:
\begin{align}
  \delta(x\vec\tau_a&+y\vec\tau_b+z\vec\tau_c)
  =\delta(x|\vec\tau_a|\hat\tau_a+y|\vec\tau_b|\hat\tau_b+z|\vec\tau_c|\hat\tau_c)\nonumber\\
  &=\delta(x|\vec\tau_a|)\delta(y|\vec\tau_b|)\delta(z|\vec\tau_c|)
  =(|\vec\tau_a||\vec\tau_b||\vec\tau_c|)^{-1}\delta(x)\delta(y)\delta(z)\nonumber\\
  &=|\vec\tau_a\cdot(\vec\tau_b\times\vec\tau_c)|^{-1}\delta(x)\delta(y)\delta(z)
  =\frac{V_\text{uc}}{(2\pi)^3}\delta(x)\delta(y)\delta(z)
\end{align}
Where the last equality was found by inserting the definitions from
\refeqn{crystals::reclatbasisvectors} and evaluating the resulting cross product
of two cross products with the $BAC$--$CAB$ rule.  With this result, the coherent elastic
scattering function (normalised to the unit cell) can be written more compactly:
\begin{align}
  S^\text{el}_\text{coh}(\vec Q,\omega) =\,& \frac{(2\pi)^3\delta(\hbar\omega)}{V_\text{uc}}
  \sum_{hkl}\delta(\vec{Q}-\vec\tau_{hkl})
  \left|F(\vec\tau_{hkl})\right|^2\labeqn{scat::cohelcrystal}
\end{align}
The Debye-Waller factors will tend to suppress the form factors at high momenta,
equivalent to large absolute values of $h$, $k$, and $l$. The summation range of
these indices can therefore in practice be limited to a region around
0. This will be explored further in \refsec{data::ncmat} where the impact on
total coherent elastic cross sections due to a lower bound, $d_{\text{cut}}$, on $d_{hkl}$
(corresponding to upper bounds on $|\vec\tau_{hkl}|$) is investigated
systematically for a large number of crystals. For the $hkl$ values selected for
consideration it is then possible to pre-calculate the form factors, at the
corresponding $hkl$ indices. As will be discussed in
sections \reftwosections{coreframework}{dataload}, the initialisation of crystal
structures in \texttt{NCrystal} indeed involves the preparation of lists of
$hkl$ indices with corresponding $d$-spacings and pre-calculated form
factors. The Debye-Waller functions in the form factors are evaluated
using the Debye model presented in \refsec{theory::debye}.

The scattering described by \refeqn{scat::cohelcrystal} is usually referred to
as Bragg diffraction. Reflections from the family of lattice planes indexed by
$hkl$ requires $Q=|\vec{Q}|=|\vec\tau_{hkl}|=2\pi/d_{hkl}$. Since the scattering
is elastic $|\vec k_i|=|\vec k_f|=2\pi/\lambda$ and $Q=2k_i\sin(\theta/2)$ where
$\theta$ is the scattering angle. The maximal value of $Q$ is found when
$\theta=\pi$ and is $2k_i$. Thus, reflection is impossible unless (the Bragg condition):
\begin{align}
  2k_i\geq|\vec\tau_{hkl}|\quad\LRA\quad\lambda\leq2d_{hkl}
  \labeqn{braggcondition}
\end{align}
And the scattering angle $\theta$ will satisfy the Bragg equation:
\begin{align}
  \lambda=2d_{hkl}\sin(\theta/2)
  \labeqn{braggequation}
\end{align}
Often the above equation will be stated using the Bragg angle, defined as
$\thetabragg\equiv\theta/2$. It also often contains on the left hand side an
integer factor, $n\geq1$, denoting the so-called scattering
\emph{order}. However, this is merely a convenient manner in which to include
multiple co-oriented lattice plane families in a single equation, utilising the
fact that $\vec\tau_{nh,nk,nl}\propto\vec\tau_{hkl}$ and
$d_{nh,nk,nl}=d_{hkl}/n$.

For a crystal powder, in which crystal grains appear with uniformly randomised
orientations, the total coherent elastic cross section for scattering on a given
family of lattice planes satisfying \refeqn{braggcondition}, can be found as an
isotropic average over the orientation of $\vec\tau_{hkl}$ with respect to
$\vec{k}_i$. Considering just the $\delta$-functions, denoting the cosine of the angle between
$\vec\tau_{hkl}$ and $\vec{k}_i$ with $\mu$, and dropping the $hkl$ indices for brevity, this isotropic average gives:
\begin{align}
  \frac{1}{4\pi}\int d\Omega_\tau\int d\Omega_f\int dE_f\delta(\vec{Q}-\vec\tau)\delta(\hbar\omega)\nonumber
  =\,  \int d\Omega_\tau \frac{\delta(\tau^2+2k_i\tau\mu)}{2\pi\left|\vec{k}_i+\vec\tau\right|}\\
  =\,  2\pi\int^1_{-1} d\mu \frac{\delta(\tau^2+2k_i\tau\mu)}{2\pi\left|\vec{k}_i+\vec\tau\right|}
  =\,  \frac{1}{2k_i^2\tau} = \frac{\lambda^2d}{16\pi^3}\labeqn{isotropicaverageintegral}
\end{align}
Where it was used that
$\int{}d\Omega\delta(\vec{r}-\vec{a})=2a^{-1}\delta(r^2-a^2)$. Inserting the
omitted factors from \refeqn{scat::cohelcrystal} in \refeqn{isotropicaverageintegral}, it is
evident that the complete coherent elastic
cross section of a crystal powder can be written as the following sum over all lattice
plane families satisfying \refeqn{braggcondition}:
\begin{align}
  \sigma^\text{powder}_\text{el,coh}(\lambda) =\,& \frac{\lambda^2}{2V_\text{uc}}
  \sum^{\lambda\leq2d_{hkl}}_{hkl}d_{hkl}\left|F(\vec{\tau}_{hkl})\right|^2\labeqn{scat::powdertotcohelxs}
\end{align}
In case of a scattering event, the relative probability for it to
happen on a particular $hkl$ plane will depend on its contribution to the sum in
\refeqn{scat::powdertotcohelxs}, and the scattering angle $\theta$ will
subsequently be
determined by \refeqn{braggequation}. The azimuthal scattering angle around the
direction of $\ki$ is, however, not constrained and
all such angles contribute equally to the cross section. Thus, $\kf$ will be
distributed uniformly in a cone around $\ki$ with opening angle $\theta$. Such
cones are known as Debye-Scherrer cones, and are a prominent feature at any
powder diffractometer.

\subsection{Inelastic scattering}\labsec{theory::inelastic}

The $n\geq1$ terms in the expansion of \refeqn{jjexpansion} can be interpreted
as inelastic scatterings in which the incoming neutron exchanges energy and
momentum with $n$ phonon states. The evaluation of these terms can in principle be
very complicated depending on the material in question and the required
precision. For the purposes of the present publication, inelastic processes play
only a minor role, and the details of the implementations of such processes in
\texttt{NCrystal} is reserved for a future dedicated publication, with just a
brief overview provided here.

For non-oriented materials like liquids and crystal powders, rotational symmetry
implies that the scatter functions $S(\vec{Q},\omega)$ become independent of the
direction of $\vec{Q}$ around \ki, and thus become two-dimensional:
$S(\vec{Q},\omega)=S(Q,\omega)$. As inelastic scattering terms do not include
the factors of $\delta(\hbar\omega)$ found in elastic terms, $S(Q,\omega)$ will
generally be sufficiently smooth that it is possible to interpolate it from its
values on a discrete grid in $(Q,\omega)$-space.\footnote{While phonon terms
  contain $\delta$-functions ensuring momentum and energy conservation, the
  convolution with phonon state densities and integration over grain orientations
  in the powder approximation eliminates them from the final scatter functions.} Such
grids of tabulated values, usually referred to as \emph{scattering kernels} in the
context of Monte Carlo applications like \texttt{MCNP}, can in principle provide
a high degree of detail, but their construction and subsequent validation require significant
efforts. One appealing option is to directly measure $S(Q,\omega)$ in a neutron
scattering experiment, which has obvious advantages but also disadvantages
depending on experimental precision and coverage of $(Q,\omega)$ space, as well
as access to an appropriate neutron instrument in the first place. The other
option is theoretical calculations, which usually must employ some
approximations due to the complexity involved in an \emph{ab initio} approach. One such
powerful if computationally expensive numerical approach is molecular
dynamics simulations in which semi-classical modelling of atomic trajectories
are used to provide expectation values for atomic displacements and
correlations~\cite{marquezdamian2014_cabwater}.

On the other end of the accuracy spectrum is the usage of various empirical
closed form expressions~\cite{freund,cassels1950,binder} describing the total
inelastic cross section as a function of neutron energy or wavelength. Not only
do such formulas not provide details on scatter angles or energy transfers, they
usually require additional tuned parameters and tend to be valid either at very
long or very short neutron wavelengths. Although it is possible to patch two
such formulas together~\cite{nxsg4,adib} in order to cover both ends of the
wavelength spectrum, the resulting behaviour at intermediate wavelengths tends
in general to be highly inaccurate.

Returning to \refeqn{jjexpansion}, it is possible to employ various
approximations in order to extract results. This approach is in
particular useful at longer neutron wavelengths, where the single-phonon ($n=1$)
term dominates, and indeed experimental techniques like neutron spectroscopy
tend to focus on using the single-phonon term to access information about the
dynamics of investigated samples, with contributions from multi-phonon terms
($n\geq2$) usually seen as unwanted background to be estimated and subtracted.

It is planned for \texttt{NCrystal} to support the modelling of inelastic
scattering with enhanced data like scattering kernels or phonon density
histograms when available, and prototype code with novel capabilities for
precision sampling already exist~\cite{xxcaisampling2018}. It is
nonetheless desirable that reasonably accurate modelling of inelastic scattering
components should be provided by \texttt{NCrystal} even for materials where no
more data than that required for Bragg diffraction is provided. Thus, an
iterative procedure suggested in~\cite{sjolander1958} has been adopted for
\texttt{NCrystal}, in which the contribution to the scatter function involving
$n$ phonons are estimated from the contribution involving $n-1$
phonons, starting from single-phonon contributions determined by the Debye model
(cf.~\refsec{theory::debye}). Additionally working under the
so-called incoherent approximation~\cite{Placzek1952}, in which
off-diagonal elements of $\langle{j},j'\rangle$ are assumed to cancel each other
out in coherent inelastic scattering, allows the prescribed method to estimate
both coherent and incoherent components of inelastic scattering.

\subsection{The Debye model}\labsec{theory::debye}

It is clear from the discussion so far that any actual evaluation of scattering
functions or cross sections involves evaluation of the Debye-Waller function
defined in \refeqn{debyewallerfunction}, which is not surprising since this
function captures the dynamics of the system due to thermal fluctuations. In
principle the evaluation requires highly
non-trivial material-specific information, either based on theory, numerical
work, measurements, or a combination of those. In order to be able to provide
meaningful results for user defined materials which might lack such specialised
information, \texttt{NCrystal} code is currently evaluating the Debye-Waller
factors using a simplified model, in which expected displacements are isotropic
around the equilibrium positions and whose dynamics are otherwise governed by a
model introduced by Debye~\cite{debyeheatcapacity1912}. The Debye model assumes that phonons in crystals
always propagate at a fixed velocity (neglecting certain effects like anisotropy and
polarisation) and only exist below some frequency
threshold, $\omega_\text{D}$, with the associated Debye temperature,
$T_\text{D}=\hbar\omega_\text{D}/\kboltz$, corresponding to the
temperature of the most energetic phonon. Loosely speaking, a high Debye
temperature indicates a strongly bound material structure and vice versa. The
model is additionally based on derivations by Glauber~\cite{glauber1955}, who applied the Debye model in
order to estimate non-isotropic displacements.

With $\alpha$ denoting a Cartesian coordinate and averaging over phonon
polarisations, \cite[Eqs.~4-5]{glauber1955} in the notation of the present paper
implies:
\begin{align}
  \langle \operator{u}^2_\alpha \rangle
   =\,& \sum_i \frac{\hbar}{2NM\omega_i} \left(\frac{2}{\exp(\hbar\omega_i/{\kboltz}T)-1}+1\right)\nonumber\\
   =\,& \sum_i \frac{\hbar}{2NM\omega_i} \coth(\hbar\omega_i/2{\kboltz}T)
\end{align}
Where the index $i$ runs over all phonon states,  $N$ is the number of phonon
states, $M$ is the atomic mass and $T$ is the material temperature.
The sum over phonon states can be replaced by an integral with the state density,
$\rho(\hbar\omega)=N^{-1}\sum_i\delta(\hbar\omega-\hbar\omega_i)$:
\begin{align}
  \langle \operator{u}^2_\alpha \rangle
   =\,& \int^{\hbar\omega_D}_0 \frac{\hbar}{2M\omega}\coth(\hbar\omega/2{\kboltz}T)\rho(\hbar\omega) d(\hbar\omega)
   \labeqn{msdintegralwithrho}
\end{align}
Now, as phonons in the Debye model propagate at constant velocity, the
frequency of a phonon will be proportional to its momentum. Assuming that phonon
states carry momenta distributed uniformly in momentum space, this implies
$\rho(\hbar\omega)\propto\omega^2$. Fixing the normalisation
$\int^{\hbar\omega_\text{D}}_0\rho(\hbar\omega)d(\hbar\omega)=1$ yields
$\rho(\hbar\omega)=3\omega^2/\hbar\omega^3_\text{D}$. Inserting into
\refeqn{msdintegralwithrho} and changing the integration variable
from $\hbar\omega\rightarrow{u}$ to $u\equiv\hbar\omega/{\kboltz}T$, this can be
written as:
\begin{align}
  \langle \operator{u}^2_\alpha \rangle = \frac{3\hbar^2}{M{\kboltz}T_\text{D}}f(T/T_\text{D})
  \labeqn{msdintegralwithf}
\end{align}
Where:
\begin{align}
  f(x) =\,& x^2\int^{1/x}_0 \coth\left(\frac{u}{2}\right)\frac{u}{2}du\nonumber\\
  =\,& x^2\int^{1/x}_0 \left(1+\frac{2}{e^u-1}\right)\frac{u}{2}du\nonumber\\
  =\,& \frac{1}{4} + x^2\int^{1/x}_0 \frac{udu}{e^u-1}
  \labeqn{msdf}
\end{align}
A result which is identical to~\cite[Eq.~11]{glauber1955}. It is interesting to
note that the displacements predicted by \refeqn{msdintegralwithf} do not vanish
at \SI{0}{\kelvin}, which is a quantum mechanical effect related to
the wave functions in the ground states. For a given atomic mass, temperature and Debye
temperature, it is straight-forward to evaluate $f(T/T_\text{D})$ numerically
and thus approximate the Debye-Waller function as:
\begin{align}
  W_j(\vec{Q})\equiv\thalf\langle(\vec{Q}\cdot\operatorvec{u}_j(0))^2\rangle
  \approx \thalf Q^2\delta^2_j
  \labeqn{dwfuncapprox}
\end{align}
where $\delta^2_j$ is the (isotropic) mean-squared displacement provided by
\refeqn{msdintegralwithf}. In principle the Debye temperature could be allowed
to vary for each site $j$ in the unit cell, but in practice it is usual to allow
it to be specified for each type of element found in the crystal and
\texttt{NCrystal} accordingly supports the specification
of Debye temperatures either globally or per type of element. It is often the case that Debye temperatures for a
given crystal can be found in the literature
(e.g.\ \cite{debyewallerfactors_peng1996,uraniumdioxide_properties_wang_2013}). If
not, one might instead be able to obtain values for mean-squared displacements,
which can then be used with \refeqn{msdintegralwithf} to estimate the Debye
temperature, or one might alternatively return to the original purpose of the
Debye model and determine the global Debye temperature for the material from its
heat capacity.

\Reffig{rmsd} shows the displacements predicted by the Debye model as a function
of temperature for a number of crystals. Indeed, as required by the harmonic
approximation, it is generally true that the predicted displacements are much
smaller than typical inter-atomic distances, although the validity is strongest at
lower temperatures.

\begin{figure}
  \centering
  \includegraphics[width=1.0\textwidth]{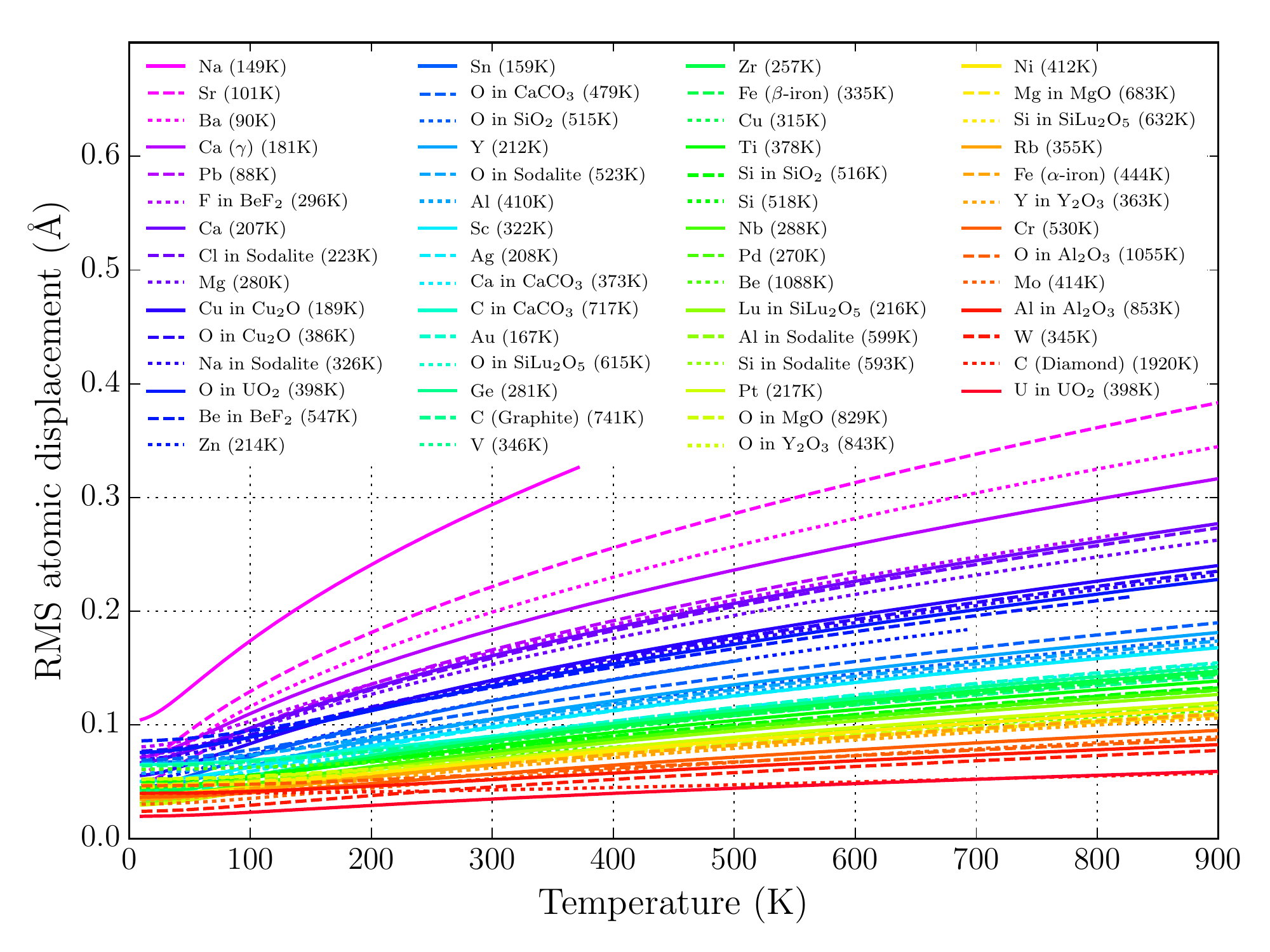}
  \caption{Root-mean-squared atomic displacements in a number of crystals
    predicted by the Debye model as discussed in the text. The applied Debye
    temperatures are indicated in parentheses, and curves are terminated at the
    melting points where relevant.}
  \labfig{rmsd}
\end{figure}

\section{Core framework overview}\labsec{coreframework}

At its core, all capabilities of the \texttt{NCrystal} toolkit are implemented
in an object oriented manner in a \texttt{C++} library, providing both clearly
defined interfaces for clients and internal separation between code
implementing physics models, code loading data, and code providing
infrastructure needed for integration into final applications. As will be
discussed in further detail in
\reftwosections{factoriesandunifiedcfg}{bindingsandinterfaces}, it should be
noted that while it is certainly possible to use the \texttt{C++} classes
discussed in the present section directly, typical users are expected to use
other more suitable interfaces for their work -- employing also a simpler and
generic approach to material configuration.

In \reffig{diagram_classes} is shown the most important classes available in the
release of \texttt{NCrystal} presented here, along with their internal
inheritance relationships. At the root of the tree sits a few infrastructure classes not directly
related to actual physics modelling, starting with the universal base class
\texttt{RCBase}, which provides reference counted memory management for all
derived classes.\footnote{Although \texttt{C++11} provides alternatives for such
  reference counting in the form of modern smart pointers, it is for the time
  being the aim of \texttt{NCrystal} to support also \texttt{C++98}, in which
  support for such is incomplete.}  One level below this sits the
\texttt{CalcBase} class, from which all classes actually implementing physics
modelling code must inherit. The main feature of \texttt{CalcBase} is currently
that it provides derived classes with access to pseudo-random numbers, in a
manner which can be configured either globally or separately for each
\texttt{CalcBase} instance. In order to do so, each \texttt{CalcBase} instance
keeps a reference to an instance of a class deriving from \texttt{RandomBase} --
a class which provides a generic interface for pseudo-random number
generation. The primary purpose of this setup is to let \texttt{NCrystal} code
use external sources of random numbers when embedded into existing frameworks
managing their own random number generators, like those discussed in
\reftwosections{interfaces::geant4}{interfaces::mcstas}. If no particular random
generator is otherwise enabled, the system will fall back to using a
\texttt{xoroshiro128+}~\cite{xoroshiro128plus_2018preprint} generator implemented in the
\texttt{RandXRSR} class.
\begin{figure}
  \centering
  \includegraphics[width=1.0\textwidth]{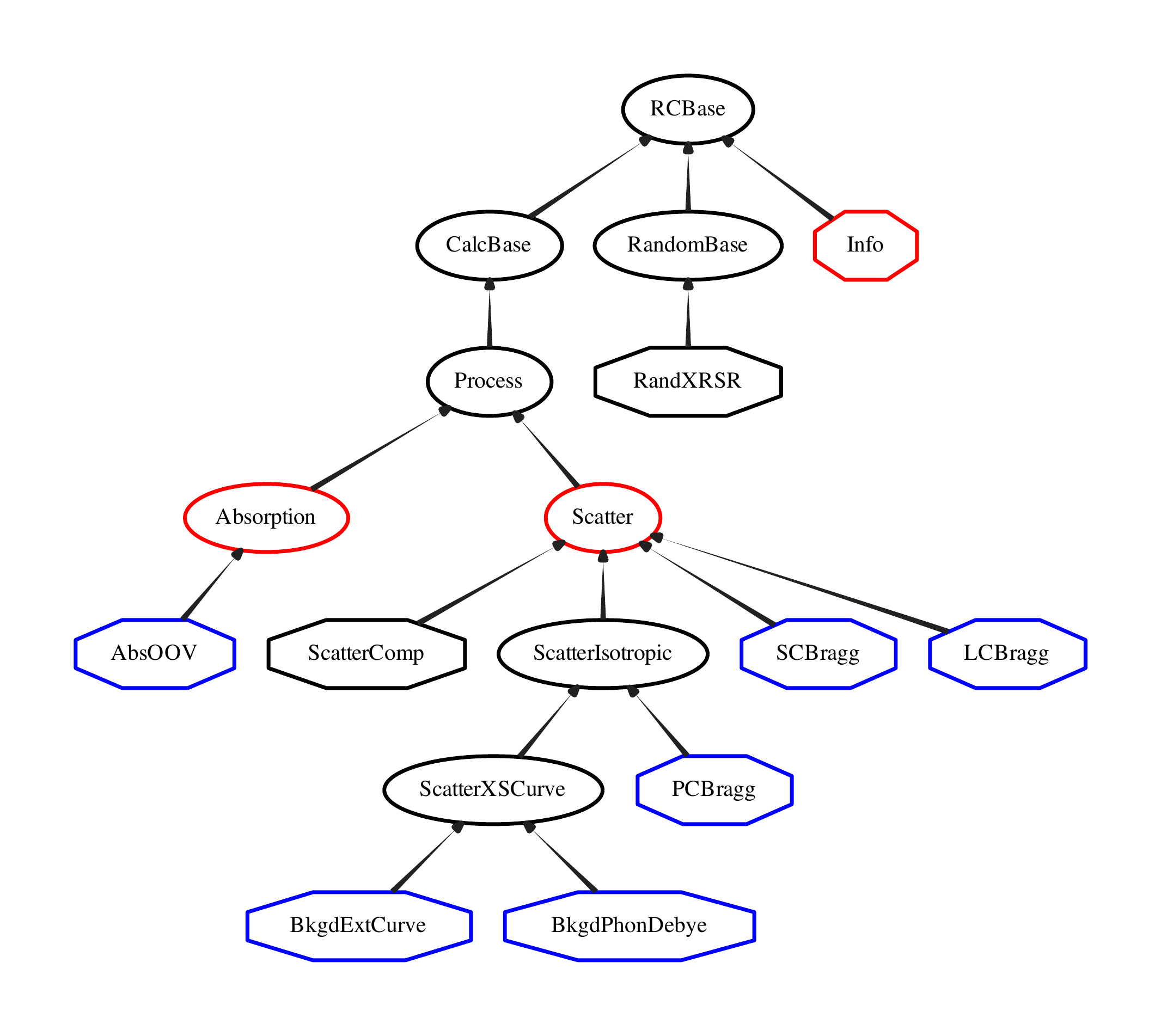}
  \caption{\texttt{NCrystal} class hierarchy.  Abstract classes are indicated
    with rounded edges while concrete classes are shown in octagons. The most
    important interfaces providing end-user results are indicated in red and
    final classes providing physics modelling are shown in blue. Internal
    utility and factory classes are not shown.}
  \labfig{diagram_classes}
\end{figure}

Currently, the only class deriving directly from \texttt{CalcBase} is
\texttt{Process}. This is an abstract class representing a physics process,
specifying a general interface for calculation of cross sections as a function
of incident neutron state. The returned cross section values are normalised to
the number of atoms in the sample, and the neutron states are specified in terms
of kinetic energy ($E_i$) and direction ($\hat{k}_i$), which is equivalent to
providing $\ki$ but using parameters typically available in general purpose
Monte Carlo applications without conversions. Additionally, for reasons of
convenience and computational efficiency, it is possible for processes to
indicate if they are only relevant for a given energy range (domain), and
processes are divided into those that are \emph{oriented} in the sense that they
actually depend on $\hat{k}_i$ and those (isotropic) ones that do not. For the
latter, one can access the cross section without specifying a direction. For
reference, the precise methods available via the \texttt{Process} interface can
be seen in \reftab{cppinterfacemethods}.

\begin{table}[tp]
\centering
\resizebox{\textwidth}{!}{%
\begin{tabular}{@{}l@{}}
\toprule
\textbf{Methods available via the \texttt{Process} interface:} \\ \midrule
\lstinlinecpp{bool isOriented() const;}\\[-0.5ex]
\lstinlinecpp{double crossSection(double ekin, const double (&indir)[3] ) const;}\\[-0.5ex]
\lstinlinecpp{double crossSectionNonOriented( double ekin ) const;}\\[-0.5ex]
\lstinlinecpp{void domain(double& ekin_low, double& ekin_high) const;}\\ \midrule
\textbf{Additional methods available via the \texttt{Scatter} interface:} \\ \midrule
\lstinlinecpp{void generateScattering( double ekin, const double (&indir)[3],}\\[-1.5ex]
\lstinlinecpp{\ \ \ \ \ \ \ \ \ \ \ \ \ \ \ \ \ \ \ \ \ \ \ \ \ double (&outdir)[3], double& delta_ekin ) const;}\\[-0.5ex]
\lstinlinecpp{void generateScatteringNonOriented( double ekin, double& angle,}\\[-1.5ex]
\lstinlinecpp{\ \ \ \ \ \ \ \ \ \ \ \ \ \ \ \ \ \ \ \ \ \ \ \ \ \ \ \ \ \ \ \ \ \ \ \ double& delta_ekin ) const;}\\
 \bottomrule
\end{tabular}}
\caption{\texttt{C++} methods provided on the \texttt{Process} and
  \texttt{Scatter} interface classes.}
\labtab{cppinterfacemethods}
\end{table}

Two abstract classes further specialise the \texttt{Process} interface, with
roles indicated by their names: \texttt{Absorption} and \texttt{Scatter}. The
\texttt{Absorption} class does not currently provide any functionality over the
\texttt{Process} class, since the current scope of \texttt{NCrystal} does not
include any particular description of absorption reactions apart from their
cross sections. In principle it would of course be possibly to extend this class
in the future, should the need ever arise for \texttt{NCrystal} to be able to model the actual outcome
of such events in terms of secondary particles produced.

The \texttt{Scatter} class does on the other hand extend the \texttt{Process}
interface, adding methods for random sampling of final states, as can also be
seen in \reftab{cppinterfacemethods}.  Specifically, this sampling generates
both energy transfers, $\Ef-\Ei$, and final direction of the neutron, $\hat
k_f$. In the case of isotropic processes, only the energy transfer and the polar
scattering angle, $\theta$, from $\hat{k}_i$ to $\hat{k}_f$ will depend on the
actual physics implemented, and will itself be independent of $\hat{k}_i$. The
azimuthal scattering angle will be independent and uniformly distributed in
$\halfopen{0}{2\pi}$. For isotropic processes, it is thus again possible if
desired to avoid the methods with full directional vectors and instead sample
energy transfer and polar scatter angle given just the kinetic energy.  The
specialised base class \texttt{ScatterIsotropic} is provided for developers in
order to simplify implementation of isotropic scatter processes. This somewhat
complicated design was chosen in order to make it straight-forward to
accommodate both oriented and isotropic scatter processes in the same manner,
avoiding unnecessary duplication of code and interfaces while still making it
possible to retain full computational efficiency allowed by the symmetries in
the isotropic cases. The next class in \reffig{diagram_classes},
\texttt{ScatterXSCurve}, exists in order to facilitate scatter models which only
provide cross sections, adding simple fall-back models for sampling of final
states. Rounding up the scattering infrastructure, \texttt{ScatterComp} is a
wrapper class playing a special role, representing the composition of multiple
\texttt{Scatter} objects into a single one. It is used whenever multiple
independent components representing partial cross sections must be combined in
order to fully describe the scattering process of interest -- such as when a
Bragg diffraction component is combined with a component providing inelastic or
incoherent scattering. At the bottom of the class hierarchy in
\reffig{diagram_classes} sits classes actually implementing physics models of
absorption or scattering, shown in blue. They will be discussed in further
detail in \refsec{physmodels}.

Finally, the \texttt{Info} class is a data structure containing information
about crystal structure at the microscopic scale of crystallites. As illustrated
in \reffig{ncinfo}, the \texttt{Info} class serves the role of separating
sources of crystal definitions from the physics models using the data as input.
The data fields available on the \texttt{Info} class are provided in
\reftab{ncinfo}. Not all input sources will be able to provide all the data
fields shown, nor will a particular physics modelling algorithm require all
fields to be available. Thus, to ensure maximal flexibility for both providers
and consumers of \texttt{Info} objects, all data fields are considered optional. If a
given physics algorithm does not find the information it needs, it should simply
indicate a configuration error, by throwing an appropriate \texttt{C++}
exception. When configuring materials via the recommended method for end-users
(cf.\ \refsec{factoriesandunifiedcfg}), the factory code will generally try to
avoid triggering undesired errors when the user intent seems obvious: if for
instance the input data does not provide any information out of which it would
be possible to provide cross sections for inelastic scattering processes, it is
most likely that the user is simply only interested in elastic physics and no
instantiation of inelastic processes will be attempted.

\begin{figure}
  \centering
  \includegraphics[width=0.8\textwidth]{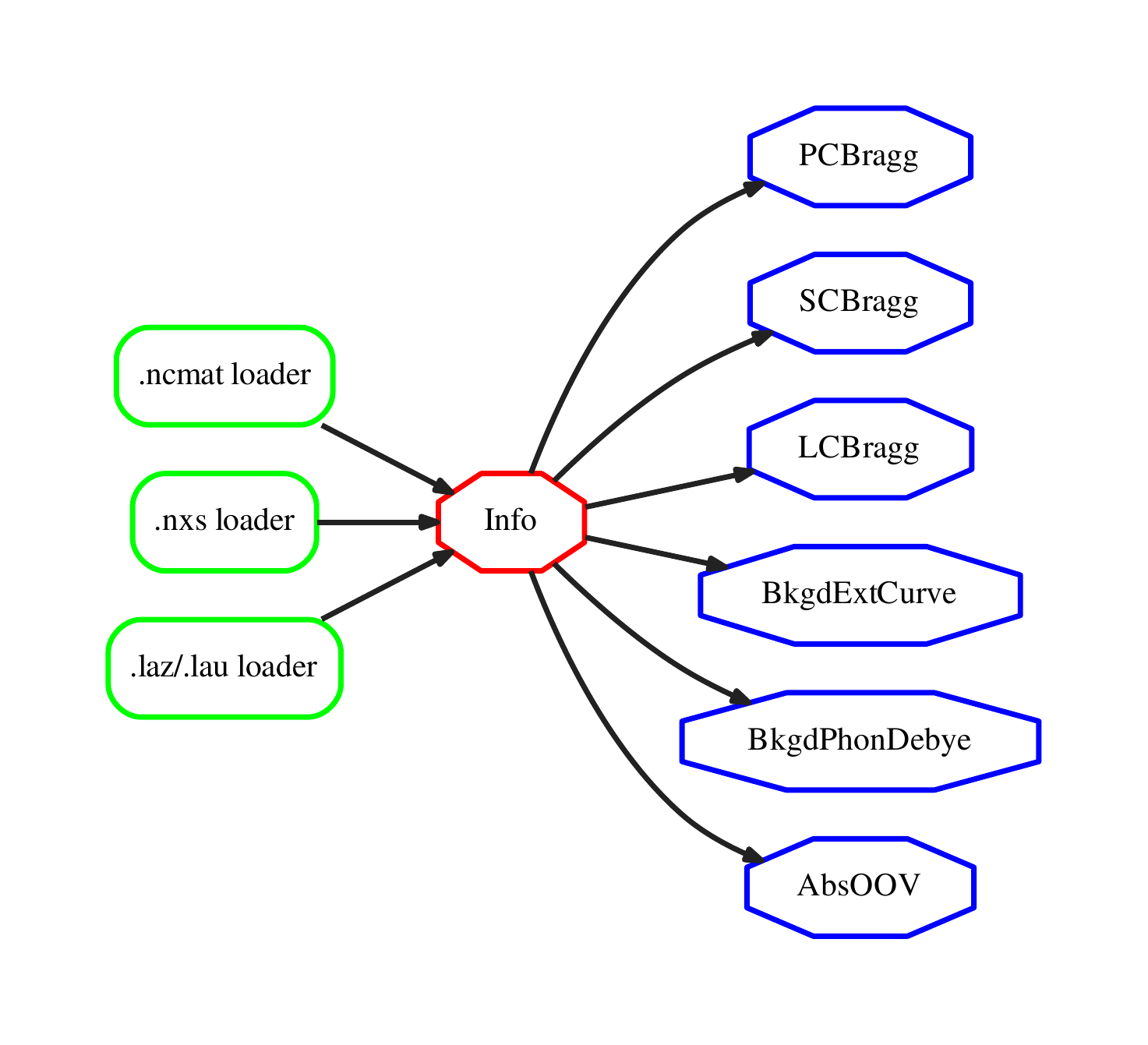}
  \caption{Flow of crystal data in \texttt{NCrystal}. Different factories are
    responsible for turning crystal definitions from a variety of sources into
    full-fledged instantiations of the Info class, which is then consumed by
    various physics models.}
  \labfig{ncinfo}
\end{figure}

\begin{table}[tp]
\centering
\begin{tabular}{@{}lll@{}}
\toprule
\textbf{Field} & \textbf{Contents} & \textbf{Units} \\ \midrule
\texttt{StructureInfo} & Basic info about unit cell:&\\[-0.5ex]
& \hspace*{1em}Lattice parameters ($a$, $b$, $c$, $\alpha$, $\beta$,
$\gamma$)&\SI{}{\angstrom}, \SI{}{\degree} \\[-1ex]
& \hspace*{1em}Volume & \SI{}{\cubic\angstrom}\\[-1ex]
& \hspace*{1em}Number of atoms&\\[-1ex]
& \hspace*{1em}Space group number$^\dagger$ &\\[0.5ex]
\texttt{AtomInfo} (list) & List of atoms in unit cell, each with:&\\[-0.5ex]
& \hspace*{1em}Atomic number (Z)&\\[-1ex]
& \hspace*{1em}Number per unit cell&\\[-1ex]
& \hspace*{1em}Per-element Debye temperature$^\dagger$ &\SI{}{\kelvin}\\[-1ex]
& \hspace*{1em}Mean squared displacement$^\dagger$ &\SI{}{\angstrom}\\[-1ex]
& \hspace*{1em}List of element positions$^\dagger$ &\\[0.5ex]
\texttt{HKLInfo} (list) & List of $hkl$ families, each with:&\\[-0.5ex]
& \hspace*{1em}$hkl$ value (representative)&\\[-1ex]
& \hspace*{1em}$d$-spacing&\SI{}{\angstrom}\\[-1ex]
& \hspace*{1em}Form factor, $|F(\vec{\tau}_{hkl})|^2$&\SI{}{\barn}\\[-1ex]
& \hspace*{1em}Multiplicity (family size)&\\[-1ex]
& \hspace*{1em}List of all $hkl$ values in family$^\dagger$&\\[-1ex]
& \hspace*{1em}List of all normals in family$^\dagger$&\\[0.5ex]
$d_{\text{cut}}$ & Threshold $d$-spacing value for $hkl$ list&\SI{}{\angstrom}\\[0.5ex]
$\sigma_\text{abs}$ & Absorption cross section at $\SI{2200}{\meter/\second}$&\SI{}{\barn}\\[0.5ex]
$\sigma_\text{free}$ & Unbound scattering cross section &\SI{}{\barn}\\[0.5ex]
$\sigma_\text{bkgd}(\lambda)$ & Cross section curve for
inelastic/incoh. &$\SI{}{\angstrom}\to\SI{}{\barn}$\\[-1ex]
& scattering components (function object) &\\[0.5ex]
Temperature & Material temperature &\SI{}{\kelvin}\\[0.5ex]
Debye temperature & Global Debye temperature of material &\SI{}{\kelvin}\\[0.5ex]
Density & Material density &\SI{}{\gram\per\cubic\cm}\\[0.5ex]
 \bottomrule
\end{tabular}
\caption{Data fields of \texttt{Info} objects. A field can represent either a
  scalar value or a block of related data, and the availability of all
  fields is optional. Additionally, some parameters inside block fields might themselves be optional, as indicated with $\dagger$'s. 
  Cross sections are given as per-atom values, averaged
  over the atoms of the unit cell. An ``$hkl$ family'' is here a group of $hkl$ indices for which form factors and $d$-spacings have
  identical values.}
\labtab{ncinfo}
\end{table}

This scheme of decoupling data loading code from physics models allows \texttt{NCrystal} to
easily support data read from a variety of input files, and makes it simple to
add support for additional input sources in the future. If desired, it would even
be possible to specify the information directly in program code or load it from
a database. Currently supported data sources are discussed in \refsec{dataload}.

In \reflisting{ncrystalrawcpp} is shown an example of \texttt{C++} code loading
a crystal definition and constructing objects able to model absorption and
scattering in a polycrystalline or powdered material. The example illustrates in
practice how some of the classes discussed in this section are used, but also
indicates the complexity arising out of the need to provide model-specific
parameters. In \refsec{factoriesandunifiedcfg}, a more convenient method for
initialisation and configuration will be introduced.

\begin{lstlisting}[float,language={[ncrystal]C++},
  label={lst:ncrystalrawcpp},
  caption={\texttt{C++} code loading a crystal definition and
    creating related \texttt{NCrystal} objects.}
]
#include "NCrystal/NCrystal.hh"
namespace NC = NCrystal;//alias for convenience

int main() {

  //Create crystal Info based on a datafile and other
  //parameters (here temperature and d-spacing cutoff):

  std::string datafile = "somefile.ncmat";
  double temp_kelvin = 400;
  double dcutoff_angstrom = 0.5;
  const NC::Info * info = NC::loadNCMAT( datafile,
                                         temp_kelvin,
                                         dcutoff_angstrom );

  //Compose Scatter and Absorption objects, passing the Info
  //object and other parameters as needed:

  bool bkgd_thermalise = true;
  NC::ScatterComp * scat = new NC::ScatterComp();
  scat->addComponent(new NC::PCBragg(info));
  scat->addComponent(new NC::BkgdPhonDebye( info,
                                            bkgd_thermalise ));
  NC::Absorption * absn = new NC::AbsOOV(info);

  // ...
  // code here using info, scat and absn pointers
  // ...

  return 0;
}
\end{lstlisting}

\subsection{Physics models}\labsec{physmodels}

At the heart of \texttt{NCrystal} is the actual modelling of specific physics
processes, provided via the classes shown with blue outlines in
\reffig{diagram_classes}. The only modelling of absorption cross sections
presently available is provided in the \texttt{AbsOOV} class, which implements
the $1/v$ scaling model discussed in \refsec{theory::neutronscattering}, by
scaling the value of $\sigma_\text{abs}$ given at the reference velocity of
$\SI{2200}{\meter/\second}$ (cf.\ \reftab{ncinfo}).  Despite being implemented
with a simple model, absorption cross sections are nonetheless provided in
\texttt{NCrystal} for completeness, in order to carry out validation against data from
measurements of total cross sections like those presented in
\refsec{data::validation} or to facilitate the creation of plugins for Monte
Carlo simulations like the one presented in \refsec{interfaces::mcstas}. Additionally, it is
important to reiterate that for most nuclear isotopes, the $1/v$ modelling is
actually remarkably accurate at thermal neutron energies (cf.~\refsec{theory::neutronscattering}).

The remaining physics models all concern scattering processes. Three processes
implement the coherent elastic physics of Bragg diffraction discussed in
\refsec{theory::elastic}, differing by which distribution of crystallite grains
they assume as discussed in \refsec{theory::crystals}. \texttt{PCBragg}
implements an ideal powder model, with cross sections given by
\refeqn{scat::powdertotcohelxs} and scattering into Debye-Scherrer cones. It can
be used to model powders and polycrystalline materials. \texttt{SCBragg}
implements a model for single crystals with isotropic Gaussian mosaicity, and
\texttt{LCBragg} provides a special anisotropic model of layered crystals
similar to pyrolytic graphite. For reasons of scope, the three models of Bragg
diffraction will be presented in detail in a future dedicated publication,
covering details of both theory, implementation and validation.

Two models currently provide incoherent and inelastic physics:
\texttt{BkgdExtCurve} and \texttt{BkgdPhonDebye}.\footnote{The term ``Bkgd'' is
  here used as a short-hand for inelastic and incoherent processes. It is merely
  used to lighten the notation, and of course is not meant to imply that
  e.g.\ inelastic neutron scattering is always to be considered ``background'' to
  some other signal (which would be incorrect).}  The former is a simple wrapper
of externally provided $\sigma_\text{bkgd}(\lambda)$ cross section curves
(cf.\ \reftab{ncinfo}), and exists mostly for reference, allowing
\texttt{nxslib}-provided cross section curves to be exposed via
\texttt{NCrystal} when using \texttt{.nxs} files
(cf.\ \refsec{data::nxs}). Unless selected explicitly, the factories discussed
in \refsec{factoriesandunifiedcfg} will always prefer the more accurate model
provided by \texttt{BkgdPhonDebye}. This improved model relies on the incoherent
approximation and the iterative procedure discussed in
\refsec{theory::inelastic}, and will be presented in detail in a future
dedicated publication. In the current release of \texttt{NCrystal} it does not
provide any detailed sampling of scatter angles or energy transfers, and hence
like \texttt{BkgdExtCurve} it derives from the \texttt{ScatterXSCurve} class.
When asked to generate a scattering, \texttt{ScatterXSCurve} will scatter
isotropically and either model energy transfers as absent (i.e.\ elastic), or as
``fully thermalising''. In the latter case, the energy of the outgoing neutron
will be sampled from a thermal (Maxwell) energy distribution specified solely by
the temperature of the material simulated. Both of these choices use well
defined distributions, but are clearly lacking in realism. It is the plan
to improve this situation in the  future, in order to provide a complete and
consistent treatment of all components involved in thermal neutron
scattering. For most materials, this will be achieved by adopting proper sampling models in
\texttt{BkgdPhonDebye}, still based on the currently used iterative procedure
for cross section estimation. As already discussed in
\refsec{theory::inelastic}, it is additionally planned to allow even higher
accuracy and reliability when modelling inelastic scattering in select
materials, by introducing data-driven models which are able to utilise
pre-tabulated scatter-kernels or phonon state densities, if available.

\section{Crystal data initialisation}\labsec{dataload}

Currently, \texttt{NCrystal} supports the loading of crystal information from
four different file formats: the native and recommended \texttt{.ncmat} format
which will be introduced in \refsec{data::ncmat}, and the existing
\texttt{.nxs}, \texttt{.laz}, and \texttt{.lau} formats which will be discussed
in \reftwosections{data::nxs}{data::lazlau}. The latter formats are mostly
supported as a service to the neutron scattering community already using these,
for instance in the context of configuring instrument simulations in
\texttt{McStas}~\cite{mcstas1,mcstas2} or
\texttt{VitESS}~\cite{vitess1,vitess2}. Support for direct loading of
\texttt{CIF} (``Crystallographic Information File'') files~\cite{Hall:es0164}
was considered as they are readily available in online databases~\cite{grazulis2009,amcsd}, but ultimately abandoned due to the flexible nature of such
files, many of which were found in practice to not contain suitable information
for the purposes of \texttt{NCrystal}. Manually selected \texttt{CIF} files were,
however, read with \cite{dtuase2017} and used to assemble a library of
\texttt{.ncmat} files which is shipped with \texttt{NCrystal}. These files,
providing the crystal structures listed in \reftab{datalib}, first and foremost
describe many materials of interest to potential \texttt{NCrystal} users,
serving both as a convenient starting point and point of reference. Moreover,
the library encompasses six of the seven crystal systems, includes mono- and
poly-atomic crystals, and includes materials with large variations in quantities
like Debye temperatures, neutron scattering lengths and number of atoms per unit
cell. Thus, it provides a convenient ensemble for benchmarking and validating
\texttt{NCrystal} code. Additionally, where not prevented for technical reasons
like usage of per-element Debye temperatures in poly-atomic systems,
\texttt{.nxs} files were automatically generated from the provided
\texttt{.ncmat} files. \Refsec{data::validation} will describe the validation
carried out as concerns both the data files themselves, as well as the code
responsible for loading crystal information from them. Naturally, it is fully
expected that the list of files in \reftab{datalib} will be expanded in the
future, in response to requests from the user community.

\begin{table}[tp]
\ssmall
\centering
\begin{tabular}{@{}llllc@{}}
\toprule
\textbf{Crystal} & \textbf{Space group} & \textbf{References} & \textbf{Formats} & \textbf{Validations}  \\ \midrule
Ag & 225 (Cubic) & \cite[11135]{amcsd}\cite{wyckoff1963crystal}\cite{debyewallerfactors_peng1996} & \texttt{.ncmat}, \texttt{.nxs} & N$\overline{\text{T}}$ \\
Al$_{2}$O$_{3}$ (Corundum) & 167 (Trigonal) & \cite[09327]{amcsd}\cite{kirfel1990} & \texttt{.ncmat} & $\overline{\text{N}}$$\overline{\text{T}}$GF  \\
Al & 225 (Cubic) & \cite[11136]{amcsd}\cite{wyckoff1963crystal}\cite{debyewallerfactors_peng1996} & \texttt{.ncmat}, \texttt{.nxs} & NT \\
Au & 225 (Cubic) & \cite[11140]{amcsd}\cite{wyckoff1963crystal}\cite{debyewallerfactors_peng1996} & \texttt{.ncmat}, \texttt{.nxs} & N$\overline{\text{T}}$ \\
Ba & 229 (Cubic) & \cite[11207]{amcsd}\cite{wyckoff1963crystal}\cite{debyewallerfactors_peng1996} & \texttt{.ncmat}, \texttt{.nxs} & NF \\
BeF$_{2}$ (Be. fluoride) & 152 (Trigonal) & \cite{pallavi2011} & \texttt{.ncmat} & $\overline{\text{N}}$F \\
Be & 194 (Hexagonal) & \cite[11165]{amcsd}\cite{wyckoff1963crystal}\cite{debyewallerfactors_peng1996} & \texttt{.ncmat}, \texttt{.nxs} & NT \\
C (Pyr. graphite) & 194 (Hexagonal) & \cite[14675]{amcsd}\cite{trucano1975} & \texttt{.ncmat} & NTF \\
C (Diamond) & 227 (Cubic) & \cite[11242]{amcsd}\cite{wyckoff1963crystal}\cite{debyewallerfactors_peng1996} & \texttt{.ncmat}, \texttt{.nxs} & NF \\
CaCO$_{3}$ (Aragonite) & 62 (Orthorhombic) & \cite[06300]{amcsd}\cite{antao2009} & \texttt{.ncmat} & $\overline{\text{N}}$GF \\
Ca & 225 (Cubic) & \cite[11141]{amcsd}\cite{wyckoff1963crystal}\cite{debyewallerfactors_peng1996} & \texttt{.ncmat}, \texttt{.nxs} & N$\overline{\text{T}}$ \\
Ca ($\gamma$-calcium) & 229 (Cubic) & \cite[11208]{amcsd}\cite{wyckoff1963crystal}\cite{debyewallerfactors_peng1996} & \texttt{.ncmat}, \texttt{.nxs} & NF \\
Cr & 229 (Cubic) & \cite[11209]{amcsd}\cite{wyckoff1963crystal}\cite{debyewallerfactors_peng1996} & \texttt{.ncmat}, \texttt{.nxs} & NT \\
Cu$_{2}$O (Cuprite) & 224 (Cubic) & \cite[09326]{amcsd}\cite{kirfel1990} & \texttt{.ncmat} & $\overline{\text{N}}$F \\
Cu & 225 (Cubic) & \cite[11145]{amcsd}\cite{wyckoff1963crystal}\cite{debyewallerfactors_peng1996} & \texttt{.ncmat}, \texttt{.nxs} & NT \\
Fe ($\alpha$-iron) & 229 (Cubic) & \cite[11214]{amcsd}\cite{wyckoff1963crystal}\cite{debyewallerfactors_peng1996} & \texttt{.ncmat}, \texttt{.nxs} & N$\overline{\text{T}}$ \\
Fe ($\beta$-iron) & 229 (Cubic) & \cite[11215]{amcsd}\cite{wyckoff1963crystal}\cite{debyewallerfactors_peng1996} & \texttt{.ncmat}, \texttt{.nxs} & NF \\
Ge & 227 (Cubic) & \cite[11245]{amcsd}\cite{wyckoff1963crystal}\cite{debyewallerfactors_peng1996} & \texttt{.ncmat}, \texttt{.nxs} & NT \\
MgO (Periclase) & 225 (Cubic) & \cite[00501]{amcsd}\cite{hazen1976} & \texttt{.ncmat} & $\overline{\text{N}}$F \\
Mg & 194 (Hexagonal) & \cite[11183]{amcsd}\cite{wyckoff1963crystal}\cite{debyewallerfactors_peng1996} & \texttt{.ncmat}, \texttt{.nxs} & N$\overline{\text{T}}$ \\
Mo & 229 (Cubic) & \cite[11221]{amcsd}\cite{wyckoff1963crystal}\cite{debyewallerfactors_peng1996} & \texttt{.ncmat}, \texttt{.nxs} & NT \\
Na & 229 (Cubic) & \cite[11223]{amcsd}\cite{wyckoff1963crystal}\cite{debyewallerfactors_peng1996} & \texttt{.ncmat}, \texttt{.nxs} & NF \\
Na$_{4}$Si$_{3}$Al$_{3}$O$_{12}$Cl (Sodalite)  & 218 (Cubic) & \cite[06211]{amcsd}\cite{antao2008} & \texttt{.ncmat} & $\overline{\text{N}}$F \\
Nb & 229 (Cubic) & \cite[11224]{amcsd}\cite{wyckoff1963crystal}\cite{debyewallerfactors_peng1996} & \texttt{.ncmat}, \texttt{.nxs} & NT \\
Ni & 225 (Cubic) & \cite[11153]{amcsd}\cite{wyckoff1963crystal}\cite{debyewallerfactors_peng1996} & \texttt{.ncmat}, \texttt{.nxs} & NT \\
Pb & 225 (Cubic) & \cite[11154]{amcsd}\cite{wyckoff1963crystal}\cite{debyewallerfactors_peng1996} & \texttt{.ncmat}, \texttt{.nxs} & N$\overline{\text{T}}$ \\
Pd & 225 (Cubic) & \cite[11155]{amcsd}\cite{wyckoff1963crystal}\cite{debyewallerfactors_peng1996} & \texttt{.ncmat}, \texttt{.nxs} & N$\overline{\text{T}}$ \\
Pt & 225 (Cubic) & \cite[11157]{amcsd}\cite{wyckoff1963crystal}\cite{debyewallerfactors_peng1996} & \texttt{.ncmat}, \texttt{.nxs} & N$\overline{\text{T}}$ \\
Rb & 229 (Cubic) & \cite[11228]{amcsd}\cite{wyckoff1963crystal}\cite{debyewallerfactors_peng1996} & \texttt{.ncmat}, \texttt{.nxs} & NF \\
Sc & 194 (Hexagonal) & \cite[11192]{amcsd}\cite{wyckoff1963crystal}\cite{debyewallerfactors_peng1996} & \texttt{.ncmat}, \texttt{.nxs} & N$\overline{\text{T}}$ \\
SiLu$_{2}$O$_{5}$ & 15 (Monoclinic) & \cite{gustafsson2001} & \texttt{.ncmat} & $\overline{\text{N}}$RF \\
SiO$_{2}$ (Quartz) & 154 (Trigonal) & \cite[06212]{amcsd}\cite{antao2008} & \texttt{.ncmat} & $\overline{\text{N}}$GF \\
Si & 227 (Cubic) & \cite[11243]{amcsd}\cite{wyckoff1963crystal}\cite{debyewallerfactors_peng1996} & \texttt{.ncmat}, \texttt{.nxs} & NT \\
Sn & 141 (Tetragonal) & \cite[11248]{amcsd}\cite{wyckoff1963crystal}\cite{debyewallerfactors_peng1996} & \texttt{.ncmat}, \texttt{.nxs} & NT \\
Sr & 225 (Cubic) & \cite[11161]{amcsd}\cite{wyckoff1963crystal}\cite{debyewallerfactors_peng1996} & \texttt{.ncmat}, \texttt{.nxs} & NF \\
Ti & 194 (Hexagonal) & \cite[11195]{amcsd}\cite{wyckoff1963crystal}\cite{debyewallerfactors_peng1996} & \texttt{.ncmat}, \texttt{.nxs} & N$\overline{\text{T}}$ \\
UO$_{2}$ (Uraninite) & 225 (Cubic) & \cite[11728]{amcsd}\cite{wyckoff1963crystal}\cite{uraniumdioxide_properties_wang_2013} & \texttt{.ncmat}, \texttt{.nxs} & NT \\
V & 229 (Cubic) & \cite[11235]{amcsd}\cite{wyckoff1963crystal}\cite{debyewallerfactors_peng1996} & \texttt{.ncmat}, \texttt{.nxs} & N$\overline{\text{T}}$ \\
W & 229 (Cubic) & \cite[11236]{amcsd}\cite{wyckoff1963crystal}\cite{debyewallerfactors_peng1996} & \texttt{.ncmat}, \texttt{.nxs} & N$\overline{\text{T}}$ \\
Y$_{2}$O$_{3}$ (Ytr. oxide) & 206 (Cubic) & \cite{park2011} & \texttt{.ncmat} & $\overline{\text{N}}$GF \\
Y & 194 (Hexagonal) & \cite[11199]{amcsd}\cite{wyckoff1963crystal}\cite{debyewallerfactors_peng1996} & \texttt{.ncmat}, \texttt{.nxs} &  NT \\
Zn & 194 (Hexagonal) & \cite[11200]{amcsd}\cite{wyckoff1963crystal}\cite{debyewallerfactors_peng1996} & \texttt{.ncmat}, \texttt{.nxs} & NT \\
Zr & 194 (Hexagonal) & \cite[11201]{amcsd}\cite{wyckoff1963crystal}\cite{debyewallerfactors_peng1996} & \texttt{.ncmat}, \texttt{.nxs} & N$\overline{\text{T}}$ \\
 \bottomrule
\end{tabular}
\caption{Crystal definitions currently shipped with \texttt{NCrystal}. Structures and Debye temperatures were
  obtained from the listed references, with specific database IDs
  for~\cite{amcsd}. The last column indicates
  validations, as discussed in \refsec{data::validation}:
  checks with \texttt{nxslib} of symmetries and form factors (N,$\overline{\text{N}}$); with experimental
  total cross sections (T,$\overline{\text{T}}$);
  with standard refinement software on simulated powder patterns
  (G,F); with measured form factors (R).}
\labtab{datalib}
\end{table}

\subsection{\texttt{NCrystal} material files (\texttt{.ncmat})}\labsec{data::ncmat}

Intended to be straight-forward to parse algorithmically, while
at the same time being directly legible to scientists with crystallographic
knowledge, the layout of \texttt{NCrystal} material files is deliberately kept
both simple and intuitive, as
illustrated with the sample file in \reflisting{ncmatexample}.  These simple
text files always begin with the keyword \texttt{NCMAT} and the version of the
format. Next comes optionally a number of \texttt{\#}-prefixed lines with
free-form comments, intended primarily as a place to document the origin,
purpose or suitability of the file. The data itself follows after this
introduction, and is placed in four clearly denoted sections as shown in the
listing. The \texttt{@CELL} and \texttt{@ATOMPOSITIONS} sections define the unit
cell layout directly, and the definition must be compatible with the space group
number which can optionally be provided in the \texttt{@SPACEGROUP}
section. Finally, the \texttt{@DEBYETEMPERATURE} section contains the effective
Debye temperature of the crystal (cf.\ \refsec{theory::debye}) -- either a
single global value, or as per-element values, indicated with the name of each
element (for mono-atomic crystals, there is of course no difference between
specifying a per-element or a global value). Where relevant, values are specified
in units of \si{\angstrom}, \si{\kelvin} and degrees for lengths,
temperature and angles respectively.

\begin{lstlisting}[float,language={},
  label={lst:ncmatexample},
  caption={Sample \texttt{.ncmat} file defining the face-centred cubic
    structure of a pure aluminium crystal, with lattice lengths of \SI{4.04958}{\angstrom} and a Debye
    temperature of \SI{410.35}{\kelvin}.}
]
NCMAT v1
#Converted from the CIF file of the entry 0011136 in the AMCSD
#reference: Wyckoff R W G, Crystal Structures, vol. 1, p. 7-83,
#1963. The Debye temperature is derived from the Debye-Waller
#factor at 293K compiled in the supplement of Acta Cryst., A52,
#p. 456-470, 1996.
@CELL
    lengths 4.04958 4.04958 4.04958
    angles 90. 90. 90.
@SPACEGROUP
    225
@ATOMPOSITIONS
    Al 0. 0.5 0.5
    Al 0. 0. 0.
    Al 0.5 0.5 0.
    Al 0.5 0. 0.5
@DEBYETEMPERATURE
    Al 410.35
\end{lstlisting}

The atomic positions are specified using relative lattice coordinates
(cf.\ \refeqn{crystals::ucpos}), and the name of an element is also required at
each such position. It is not necessary, nor currently possible, to specify
element-specific data such as masses, cross sections or scattering
lengths. Instead, \texttt{NCrystal} currently includes an internal database of
such numbers for all natural elements with $Z\le92$, based
on~\cite{nscatlentable_2001,Rauch2000,KOESTER199165}. Although somewhat
inflexible, this scheme provides a high level of convenience, consistency and
robustness by significantly lowering the amount of parameters required in each
\texttt{.ncmat} file. It is envisioned that a future version of \texttt{NCrystal}
would support more specialised use-cases by supporting not only the optional
specification of element- and isotope-specific data in \texttt{.ncmat} files,
but also to allow for specification of features like chemical disorder,
impurities, doping or enrichment.

With the information parsed directly from the \texttt{.ncmat} file and the
specification of a material temperature, the remaining information shown in
\reftab{ncinfo} will be derived numerically at initialisation time (with the exception of
$\sigma_\text{bkgd}(\lambda)$ which is not relevant for \texttt{.ncmat} files).
The involved calculations are mostly trivial, with one exception being the
atomic mean-squared displacements which are calculated based on the isotropic
Debye model discussed in \refsec{theory::debye}. The other exception is the
creation of the $hkl$ lists (\texttt{HKLInfo}) and the associated $d$-spacing
threshold, $d_{\text{cut}}$. For a given value of $d_{\text{cut}}$, all $hkl$ points in
the reciprocal lattice with $d_{hkl}\geq{d}_{\text{cut}}$ are considered. The
number of such points can be considerable, as shown in \reffig{datalib_counts},
and considerable care is taken to control the initialisation time. At each given
$hkl$ point the squared form factors, $|F(\vec{\tau}_{hkl})|^2$, are calculated
directly via a numerical evaluation of \refeqn{unitcellformfactordef}, requiring
relatively expensive sine and cosine function evaluations with the phase specific to
each atomic position, $\vec\tau_{hkl}\cdot\vec{p}_{i}$. Naturally, the
$\overline{b_i}{e}^{-W_{i}(\vec{Q})}$ factors are computed just once and reused,
and half of the $hkl$ points are dealt with by using the symmetry
$|F(\vec{\tau}_{hkl})|^2=|F(-\vec{\tau}_{hkl})|^2$. Additionally, entries
with squared form factors less than a fixed threshold value of
$f_{\text{cut}}=\SI{e-5}{\barn}$ are discarded. This keeps forbidden
$hkl$ entries out of the final lists, i.e.\ those which for reasons of symmetry
should have vanishing form factors in the strict mathematical sense, but which
nevertheless acquire tiny non-zero but negligible values during the numerical
evaluations.  More importantly, the non-zero value of $f_{\text{cut}}$ enables
significant improvements in initialisation time through an early-abort
technique. Specifically, an upper bound on the squared form factor value can
be calculated without any looping over atom positions or expensive trigonometric
function evaluations,
by replacing all involved sines and cosines calls with $1$. This can make it
possible to predict without expensive calculations that a value $f_{\text{cut}}$
is unreachable. In fact, only $hkl$ points near the origin of the reciprocal
lattice need detailed consideration, because the Debye-Waller factors suppress
entries with smaller $d$-spacings (cf.\ \refeqn{dwfuncapprox}):
$\exp(-W_{i}(\vec\tau_{hkl}))=\exp(-2\pi^2\delta^2_i/d^2_{hkl})$. In particular, it means that the time required for
$hkl$ list initialisation practically tends towards a constant as
$d_{\text{cut}}$ is decreased to ever smaller values. This behaviour is much preferable to the
$\order{1/d^3_{\text{cut}}}$ behaviour of an implementation without such an
early-abort.

In addition to the calculation of squared form-factors at all
considered $hkl$ points, points must also be sorted into families of points
sharing $d$-spacing and squared form-factor values. This is done with a
map-based $\order{\log(d_\text{cut})}$ algorithm. All together, the final result
is an $hkl$ list initialisation algorithm which is fast enough that the
trade-off between accuracy and initialisation time inherent in the choice of
$d_\text{cut}$ is not very severe.

\begin{figure}
  \centering
  \includegraphics[width=1.0\textwidth]{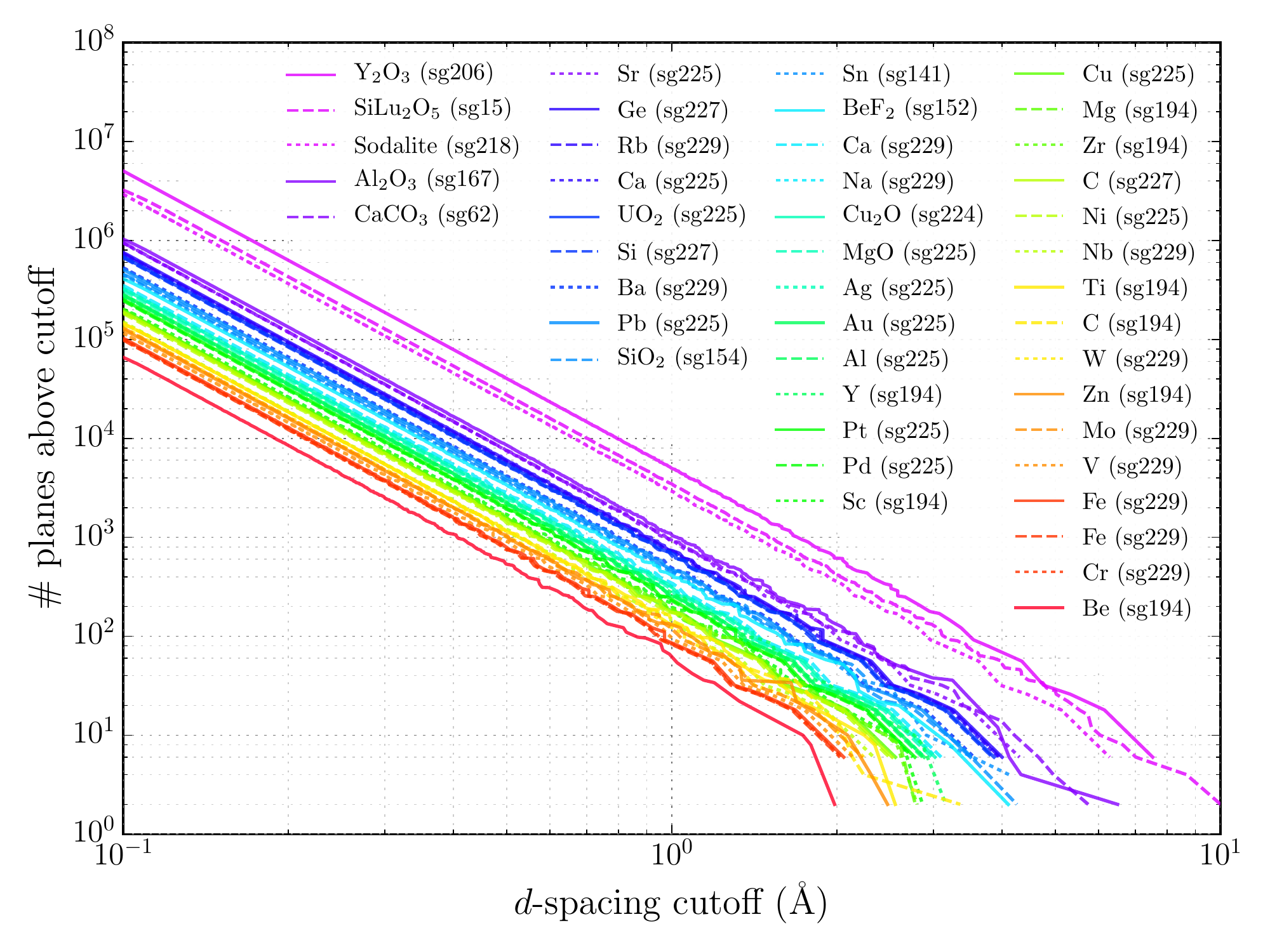}
  \caption{Number of $hkl$ planes above a given $d$-spacing threshold for
    crystal structures in the \texttt{NCrystal} data library.}
  \labfig{datalib_counts}
\end{figure}

The value of $d_\text{cut}$ can always be set directly by the users, but in
order to determine appropriate default values for the majority of
users who are not expected to do so, the impact on Bragg diffraction cross sections in the
powder approximation (cf.\ \refeqn{scat::powdertotcohelxs}) was
investigated. Although most powder diffraction experiments would concentrate on
wavelengths longer than $2d_\text{cut}$, and therefore not be directly affected
by it, the total cross section for diffraction in a powder is still a meaningful
benchmark. After all, if a too high cut-off value is chosen, too many planes
would be left out and the total cross section would be underestimated at shorter
wavelengths -- with corresponding degradation of realism in a simulation of for
instance beam filters or shielding based on powders or polycrystalline
materials. Thus, \reffig{datalib_contribs} shows the relative impact of
$d_\text{cut}$ on the cross section in the limiting case $\lambda\rightarrow0$. The
impact levels gauged in this limit are clearly very conservative, since at this
wavelength all omitted planes satisfy \refeqn{braggcondition} resulting in a
large relative impact, while the factor of $\lambda^2$ in
\refeqn{scat::powdertotcohelxs} removes any absolute impact. Still, to be
conservative, a fairly aggressive default value of
$d_\text{cut}=\SI{0.1}{\angstrom}$ was chosen. This is a sensible choice, since
the typical initialisation time of most materials with this value range from a
few milliseconds to approximately one second, depending on unit cell
complexity.\footnote{Timings were carried out on a mid-range laptop from 2014
  with a \SI{2}{\giga\hertz} CPU.  More details or accuracy in stated numbers
  are on purpose not provided, as such timings are notorious for their
  dependency on both platform and system state, and only rough magnitudes of
  timings are important for the present discussion.}  Two of the tested
materials in the data library stood out, however, with initialisation times of
\SI{6}{\second} (SiLu$_{2}$O$_{5}$) and \SI{31}{\second} (Y$_{2}$O$_{3}$)
respectively. These crystal structures are distinguished by their large unit
cell volumes and number of atoms per unit cell, 64 and 80 respectively. In order to keep default load
times at approximately \SI{1}{\second} or less, a number which is unlikely to
cause concern for casual users, the default value is raised to
$d_\text{cut}=\SI{0.25}{\angstrom}$ for materials with more than 40 atoms in the
unit cell.

\begin{figure}
  \centering
  \includegraphics[width=1.0\textwidth]{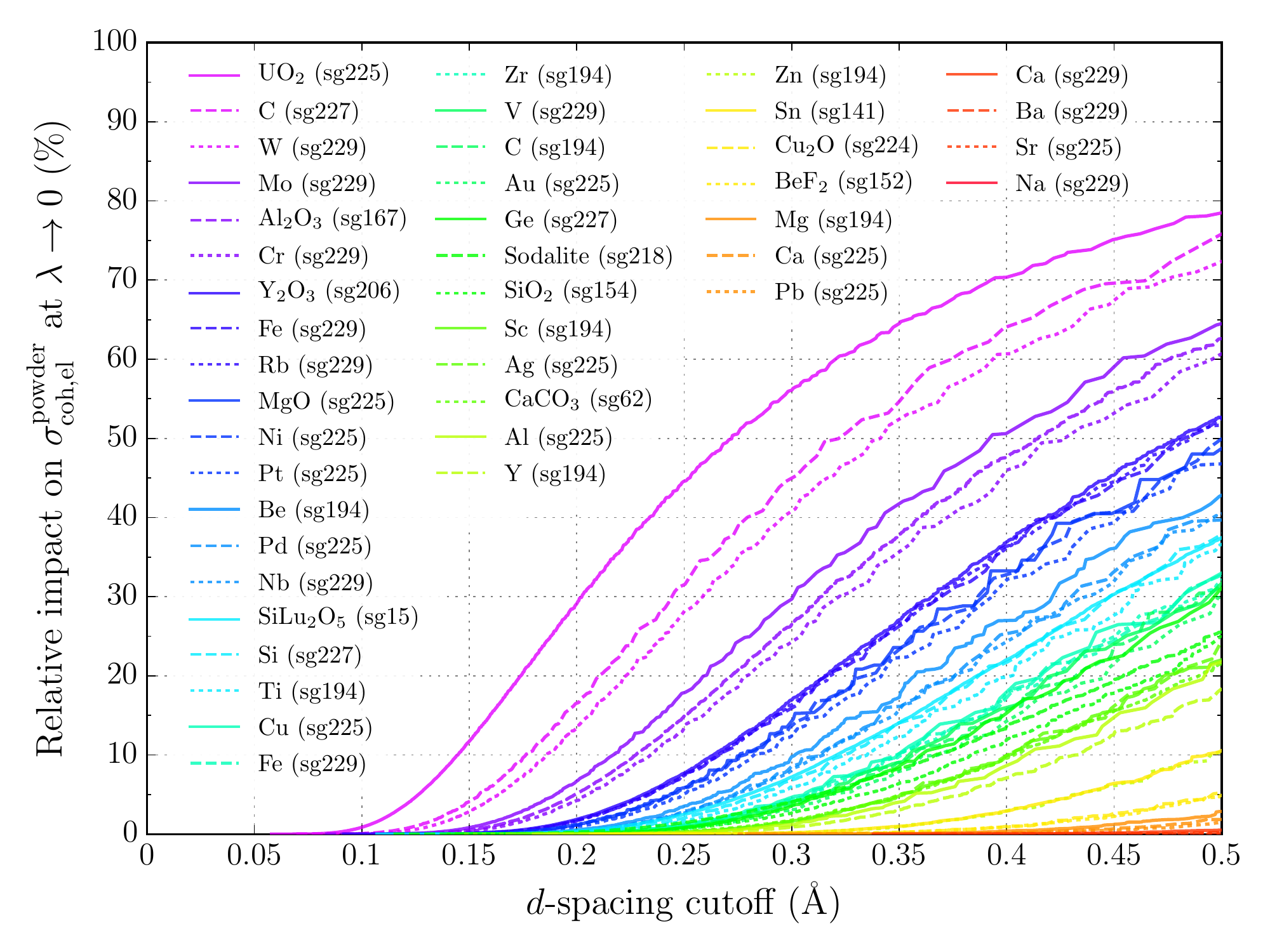}
  \caption{Relative impact in the short-wavelength limit of $d$-spacing
    threshold on the cross section for Bragg diffraction in a powder,
    for crystal structures in the \texttt{NCrystal} data library.}
  \labfig{datalib_contribs}
\end{figure}

The resulting impact on the cross section curve of both thresholds applied by
default, $f_{\text{cut}}=\SI{e-5}{\barn}$ and $d_\text{cut}=\SI{0.1}{\angstrom}$
or $\SI{0.25}{\angstrom}$, is shown in
\reffig{datalib_cutoffimpactonpowderxs}. For most materials the effect is
completely negligible, and even in the worst case of yttrium-oxide, the effect
exists only at very short wavelengths and is hardly noticeable.

\begin{figure}
  \centering
  \includegraphics[width=1.0\textwidth]{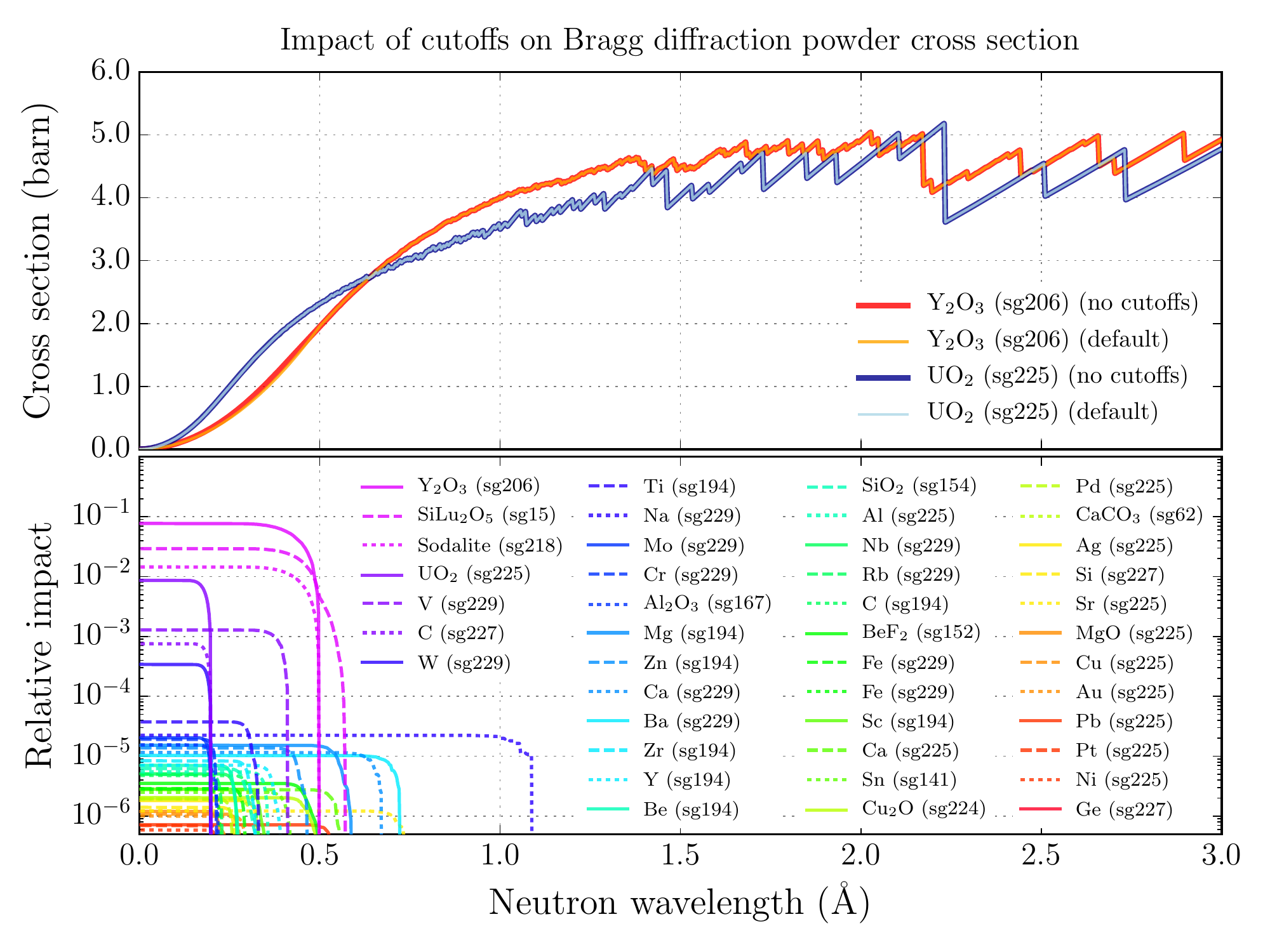}
  \caption{Impact of the default thresholds ($f_{\text{cut}}=\SI{e-5}{\barn}$
    and $d_\text{cut}=\SI{0.1}{\angstrom}$ or $\SI{0.25}{\angstrom}$) on the
    cross section for Bragg diffraction in a powder, for crystal structures in
    the \texttt{NCrystal} data library. The plot on top compares cross section
    curves directly for two of the most affected materials, while the plot below
    shows the relative impact for all structures in the \texttt{NCrystal} data
    library.}
  \labfig{datalib_cutoffimpactonpowderxs}
\end{figure}

\subsection{\texttt{.nxs} file loader}\labsec{data::nxs}

Routines in the \texttt{nxs} library~\cite{nxslib1,nxslib2,nxslib3}, used in
\texttt{McStas}~\cite{mcstas1,mcstas2}, \texttt{VitESS}~\cite{vitess1,vitess2}
and \texttt{NXSG4}~\cite{nxsg4} to provide cross sections in crystal powders, comes with an associated text-based file format
for crystal structure definition, recognised by the extension \texttt{.nxs}. The
file format, described in~\cite{nxsg4}, contains information similar to that in
the \texttt{.ncmat} file of \reflisting{ncmatexample} with a few notable
differences. The first one is that element-specific cross sections and masses
must be specified directly in the files themselves, and the second is that the
files do not contain the full list of atomic positions in the unit cell, but
rather just the Wyckoff positions of the atoms. Finally, only a global Debye temperature
can be specified, meaning that per-element Debye Temperatures in poly-atomic
systems are unsupported. Upon loading the file, the \texttt{nxs} library
creates the full list of atomic positions in the unit cell by application of the
relevant symmetry operators of the specified crystal space group, which is
carried out through an internal dependency on the
\texttt{SgInfo}~\cite{sginfowebsite} library. Support of the \texttt{.nxs}
format in \texttt{NCrystal} thus introduce dependencies on these two external
libraries, each with their own unique open source license. In order to ensure
that only users actually interested in \texttt{.nxs} file support would be
exposed to these extra license requirements, the \texttt{NCrystal} code
implementing the support for \texttt{.nxs} file loading is kept clearly
separated from the rest, and it is straight-forward to build \texttt{NCrystal}
without it.\footnote{Note that users wanting to use the \texttt{.nxs} loading
  capabilities are not required to install additional libraries by themselves,
  as the optional code supporting \texttt{.nxs} files in \texttt{NCrystal} already embeds
  versions of \texttt{SgInfo-1.0.1} and \texttt{nxs-1.5} -- with a few custom
  patches correcting issues concerning monoclinic and triclinic crystals.}

Upon loading \texttt{.nxs} files, most information shown in \reftab{ncinfo} will
be loaded directly from the input file. The $hkl$ lists in the \texttt{HKLInfo}
section are, however, provided by calculations in the \texttt{nxs} library, and
are not calculated directly in \texttt{NCrystal} code.  On one hand this ensures
that users loading \texttt{.nxs} files will get the exact same form factors when
loading their files with \texttt{NCrystal} as when loading them elsewhere.  More
importantly, however, it makes it possible to perform meaningful cross checks of
the results of loading equivalent \texttt{.ncmat} and \texttt{.nxs} files by
comparing the resulting form factors with those resulting from loading an
equivalent \texttt{.nxs} file.  As the \texttt{nxs} library evaluates
Debye-Waller factors using a model~\cite{vogel_thesis} which is based on the
same underlying approximations of isotropic displacements and the Debye Model
(cf.\ \refsec{theory::debye}), the resulting form factors should be directly
comparable. However, as the $hkl$ list creation in the \texttt{nxs} library
relies on application of space group symmetries, employing selection rules and
starting from Wyckoff positions rather than the full list of atomic positions,
such comparisons are highly non-trivial, validating both the compared files and
the code loading them. The result of such validations are discussed in
\refsec{data::validation}.

An important detail in the construction of $hkl$ lists is that the \texttt{nxs}
library does not directly support a specification of a $d$-spacing threshold,
instead limiting the range of Miller indices probed by specification of a
parameter $N_\text{max}$, resulting in the consideration of all $hkl$ points for
which $|h|,|k|,|l|\le{N}_\text{max}$. Consequently, the $d$-spacing threshold is
implemented on the \texttt{NCrystal} side by first calculating the value
of $N_\text{max}$ needed to contain all points withing $d_\text{cut}$, and
subsequently ignoring all entries inevitably generated with a $d$-spacing beyond
this value. As the \texttt{nxs} library code for grouping $hkl$ entries into
families is implemented with a relatively slow linear algorithm (resulting in an
overall algorithmic complexity of $\order{N^4_\text{max}}$), the $d_\text{cut}$
value selected by default for \texttt{.nxs} files is somewhat higher than for
\texttt{.ncmat} files. It is chosen so as to correspond to $N_\text{max}=20$ but
with $d_\text{cut}$ at most \SI{0.5}{\angstrom} and at least
\SI{0.1}{\angstrom}. To prevent perceived programme lockups, it will result in
an error if a user requests a $d_\text{cut}$ value which requires
$N_\text{max}>50$.

Another important point is that the \texttt{nxs} code was not written to support
single crystal modelling, and therefore the loaded \texttt{HKLInfo} objects will
contain no lists of normals or $hkl$ indices.  However, when $hkl$ family
composition corresponds to symmetry equivalence groups (as is indeed the case for
\texttt{.nxs} files), the \texttt{NCrystal} single crystal code is able to
reconstruct such lists of normals on demand if absent, and it will therefore
still be possible to use \texttt{.nxs} files for single crystal
simulations in \texttt{NCrystal}. Finally, the \texttt{nxs} library provides
estimates of inelastic/incoherent scattering cross sections based on various
empirical formulas, and these will be provided in the
$\sigma_\text{bkgd}(\lambda)$ field of \texttt{Info} objects when an
\texttt{.nxs} file is loaded. By default, the cross section curve provided is
similar to the one discussed in \cite{nxsg4}, representing an ad hoc combination
of empirical formulas due to Freund~\cite{freund} and Cassels~\cite{cassels1950}.
As discussed in \refsec{physmodels}, these curves are provided for
reference: the native \texttt{NCrystal} algorithms provide more robust
predictions for inelastic/incoherent cross sections and is used by default also
when working with \texttt{.nxs} files.

\subsection{\texttt{.laz} and \texttt{.lau} file loader}\labsec{data::lazlau}

Due to the widespread usage in various \texttt{McStas} components for modelling
of Bragg diffraction, \texttt{NCrystal} also supports the loading of
\texttt{.laz} file from~\cite{lazy_Yvon:a15322} and \texttt{.lau} files
from~\cite{crystallographica_Siegrist:wt0001}. Both of these very similar
text-based file formats can be generated from a \texttt{CIF} file by the
\texttt{cif2hkl} application, part of \texttt{iFit}~\cite{iFit_code1}, and their most notable feature is that they directly contain $hkl$
lists with $d$-spacings and form factors. Typically, \texttt{.laz} files are used
in the context of powder diffraction, while \texttt{.lau} files can be used to deal
with single crystal diffraction as well. This distinction implies that the
latter files are larger, since they break down each $hkl$ family into
multiple lines of data, in order to provide enough information that all plane
normals associated with a given family can be directly inferred.

The \texttt{NCrystal} code loading such files is rather simple, since no special
calculations are needed to fill the \texttt{HKLInfo} section. By default, no thresholds
are applied upon loading the $hkl$ information from the file, but if desired it
is of course possible to specify a custom $d$-spacing threshold in order to
ignore some $hkl$ families in the file. Other information in \reftab{ncinfo}
loaded from information at the beginning of the files are \texttt{StructureInfo},
$\sigma_\text{abs}$, and density. Finally, it is also possible to specify the
temperature when loading the file, but it is important to note
that since form factors are hard-coded in the file itself, their values are
unaffected by the actual value provided. All in all, the loaded information is
sufficient for modelling of Bragg diffraction, but absent is information which
could be used to model inelastic or incoherent components. Thus, the factories
described in \refsec{factoriesandunifiedcfg} will create processes without such
components for \texttt{.laz} or \texttt{.lau} files.

\subsection{Validation}\labsec{data::validation}

In order to simultaneously validate not only the code responsible for
initialising crystal structure information from \texttt{.ncmat} and
\texttt{.nxs} files, but also the individual data files provided with
\texttt{NCrystal}, multiple approaches were pursued. One particular concern is
the code responsible for creating $hkl$ lists with $d$-spacings and squared form
factors from \texttt{.ncmat} files. Another is the fact that the crystal
symmetry in \texttt{.ncmat} files is contained \emph{implicitly} in the full
list of atomic unit cell positions, as opposed to \texttt{.nxs} or \texttt{CIF}
files in which the symmetry is expressed \emph{explicitly} in terms of space
group number and Wyckoff positions. The optional inclusion of the space group
number in \texttt{.ncmat} files is mostly cosmetic, and there is currently no
guarantee that the indicated space group actually corresponds to the symmetries
expressed by the unit cell shape and atomic position.

The first validation carried out for all files requires that the atomic
positions in a given \texttt{.ncmat} file are compatible with those generated by
\texttt{nxslib} from the Wyckoff positions in the corresponding \texttt{.nxs}
file -- which behind the scenes implements symmetry operations for the indicated
space group via the \texttt{SgInfo} library. Next, as discussed in
\refsec{data::nxs}, the $hkl$ lists created from the two files are also compared
systematically. This validates the Debye model and direct approach utilised
in the custom $hkl$ list creation code used for \texttt{.ncmat} files against
the corresponding but different implementation in \texttt{nxslib}, which relies
on symmetry information from the \texttt{SgInfo} library for selection rules and
multiplicities. All provided files were validated in this manner, as indicated
with N or $\overline{\text{N}}$ in the last column of \reftab{datalib}. Files
marked with $\overline{\text{N}}$ are poly-atomic crystals with per-element
Debye temperatures, which is not supported in \texttt{.nxs} files and which
accordingly had to be compared using files in which a global Debye temperature
was substituted.

Next, the $hkl$ lists created from \texttt{.ncmat} files are validated again by using
them to generate artificial powder diffraction spectrums and testing them with
software which is normally used to decode crystal structures from such spectrums
at real powder diffraction experiments. First, a simple neutron
diffraction instrument was simulated with \texttt{McStas}
(cf.~\refsec{interfaces::mcstas}), using \texttt{NCrystal} and the relevant
\texttt{.ncmat} file to model a crystal powder sample.\footnote{For reference,
  the instrument setup used was similar to the one shown in
  \reflisting{examplemcstasinstr}, but with a few modifications carried out in
  order to increase the quality of the produced diffraction patterns. Thus, the
  sample size was reduced slightly, the number of detector bins increased,
  linear collimators were added before and after the monochromator, and a radial
  collimator was placed in front of the detector. Finally, the monochromator was
  changed to Germanium-511, in order to select wavelengths around
  \SI{1.54}{\angstrom}.  Computing resources for the simulations were provided
  by the ESS DMSC Computing Centre.}  As a result, a powder spectrum was
produced for each of
the four tested materials -- marked with G in the last column of
\reftab{datalib}. These spectra were then used as input to
\texttt{GSAS-II}~\cite{gsasiitoby2013}, which in all cases managed to recover
the crystal structure which was present in the original \texttt{CIF} file from
which the relevant \texttt{.ncmat} file was produced. As an example,
\reffig{gsas:al2o3pattern} shows a refinement for a powder spectrum with
\num{9.1e7} simulated neutrons collected in the detector array after interaction with a
sapphire sample, and
\reftab{gsas:al2o3refinement} shows the corresponding parameters extracted by  \texttt{GSAS-II}, which in addition
to space group and atomic (Wyckoff) positions also include atomic mean-squared
displacements. The goodness-of-fit is provided by \texttt{GSAS-II} in terms of
$R$-factor, which in all cases was less than 1.9\%, indicating a very
high degree of compatibility. Due to the usage of a full-scale simulation with
actual \texttt{NCrystal} components enabled, these comparisons incidentally
validate features of \texttt{NCrystal} beyond just the initialisation of crystal
structure information -- but  dedicated future publications will document
additional validations performed for these more thoroughly.

\begin{figure}
  \centering
  \includegraphics[width=0.99\textwidth]{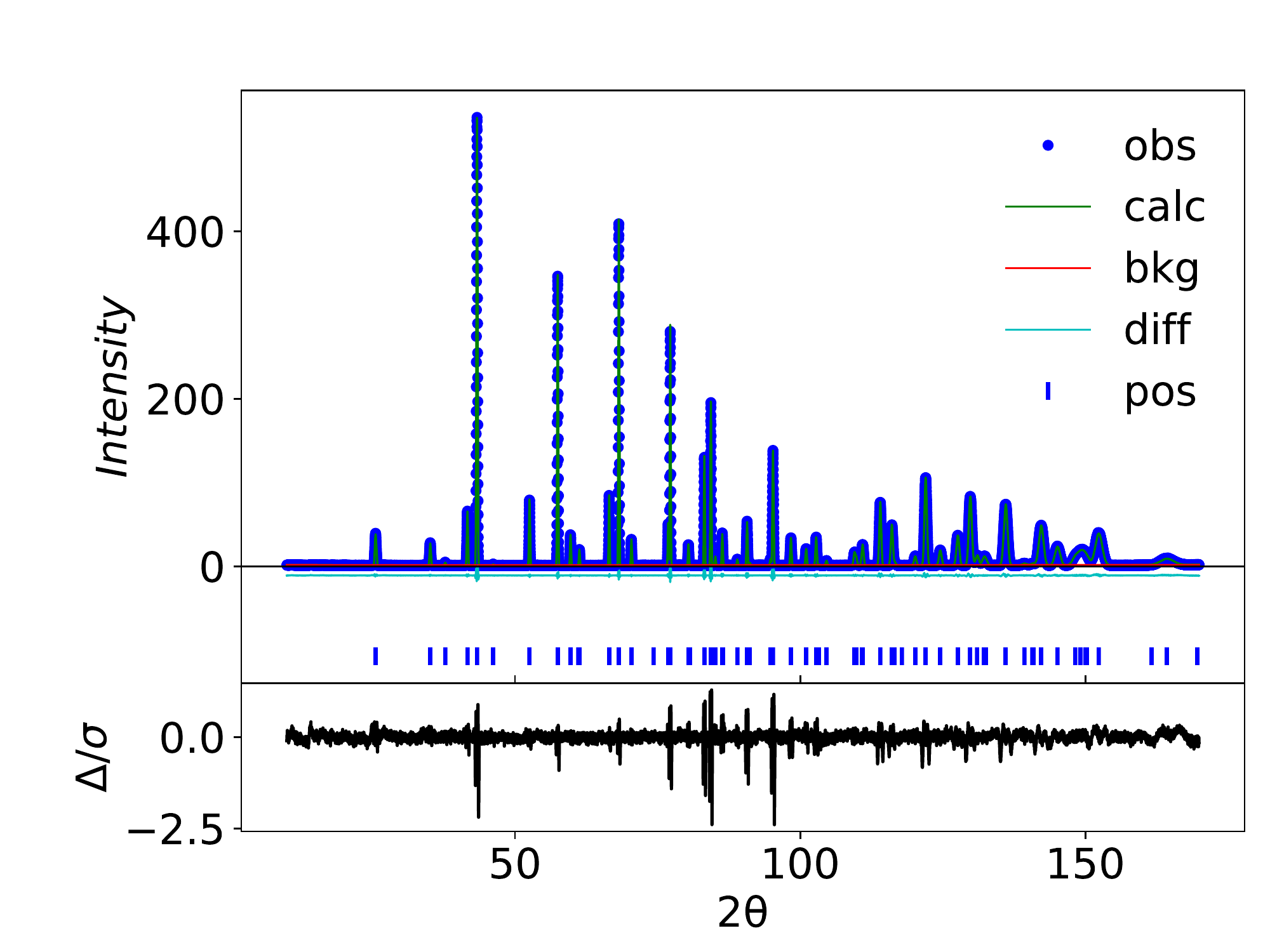}
  \caption{Comparison of sapphire (corundum) powder diffraction pattern
    simulated by \texttt{McStas} and \texttt{NCrystal} (``obs'') and refined
    curves from \texttt{GSAS-II}. Refined peak positions are shown in the middle
    section of the plot and differences (normalised to peak widths) are shown at
    the bottom.  Plot generated by \texttt{GSAS-II}.}
  \labfig{gsas:al2o3pattern}
\end{figure}

\begin{table}
\centering
\begin{tabular}{lll}
\toprule
 &  Original & Refined \\
\midrule
Lattice $a$ (\SI{}{\angstrom})& 4.757 & 4.75701 \\
Lattice $c$ (\SI{}{\angstrom})& 12.9877 & 12.98773 \\
Al position $x$ & 0.0 & 0.00000 \\
Al position $y$ & 0.0 & 0.00000 \\
Al position $z$ & 0.35218 & 0.35221\\
O position $x$ & 0.30625 & 0.30613 \\
O position $y$ & 0.0 & 0.00000 \\
O position $z$ & 0.25 & 0.25000 \\
Al MSD (\SI{}{\square\angstrom}) & 0.002644 & 0.00259 \\
O MSD (\SI{}{\square\angstrom})& 0.003167 & 0.00305 \\
$R$-factor & & 0.82\% \\
\bottomrule
\end{tabular}
\caption{Original parameters and refined results from \texttt{GSAS-II} for sapphire (corundum).}
\labtab{gsas:al2o3refinement}
\end{table}

In addition to performing a full-blown instrument simulation with \texttt{McStas},
an idealised diffraction spectrum for a given neutron wavelength
can be constructed directly from the $d$-spacings, multiplicities and squared form factors in a
given $hkl$ list by referring to
\reftwoeqns{braggequation}{scat::powdertotcohelxs}.  Furthermore, geometrical coverage of a typical array of detector
tubes at a powder diffractometer is accounted for by the introduction of an acceptance
factor of $1/\sin\theta$ -- and all other effects related to sample size, beam
spread and divergence, etc. are modelled in a simplistic manner by replacing
the ideal $\delta$-function line shapes with Gaussian distributions of  \SI{0.1}{\degree} fixed
width. Despite the simplicity, the resulting artificial powder spectra are
sufficiently realistic that it is possible to extract the crystal
structure from them using \texttt{FullProf}~\cite{fullprof}.  For the 17
materials tested in this manner, marked with F in the last column of
\reftab{datalib}, \texttt{FullProf} was able to precisely recover their crystal
structures -- returning $R$-factors which were in all cases better than
2.7\%. \Reffig{fullprof:aragonitepattern} shows an example of a powder spectrum
fitted with \texttt{FullProf} and \reftab{fullprof:aragoniterefinement} the
corresponding extracted parameters.

\begin{figure}
  \centering
  \includegraphics[width=0.99\textwidth]{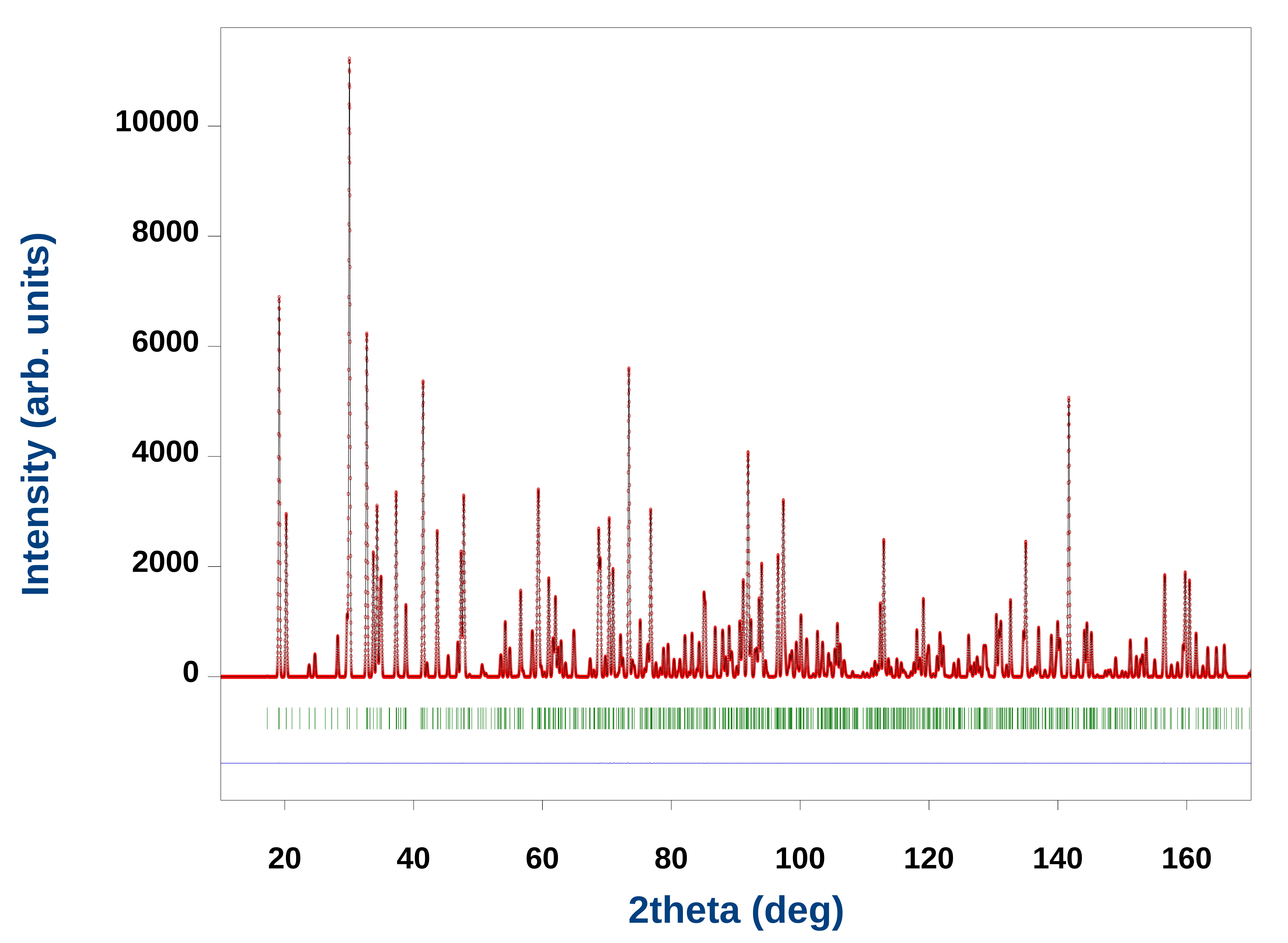}
  \caption{Comparison of artificial aragonite (CaCO$_{3}$) powder diffraction pattern constructed
    based on \texttt{NCrystal} $hkl$ lists (red) and refined
    curves from \texttt{FullProf} (black). Below are shown refined peak
    positions (green) and differences between input and refined patterns (blue).
    Plot generated by \texttt{FullProf}.}
  \labfig{fullprof:aragonitepattern}
\end{figure}

\begin{table}
\centering
\begin{tabular}{lll}
\toprule
 &  Original & Refined \\
\midrule
Lattice $a$ (\SI{}{\angstrom})& 4.96062 & 4.960621   \\
Lattice $b$ (\SI{}{\angstrom})& 7.97006 & 7.970061  \\
Lattice $c$ (\SI{}{\angstrom})& 5.74181 &  5.741809  \\
Ca position $x$ & 0.25000     & 0.25000 \\
Ca position $y$ & 0.41500   & 0.41500  \\
Ca position $z$ & 0.75960  & 0.75960\\
C position $x$ &  0.25000   &  0.25000 \\
C position $y$ & 0.76190   &  0.76190\\
C position $z$ &  -0.08490  & -0.08490 \\
O$^a$ position $x$ & 0.25000    &  0.25000 \\
O$^a$ position $y$ & 0.92240     &  0.92240  \\
O$^a$ position $z$ &  -0.09560    & -0.09560 \\
O$^b$ position $x$ &  0.47380   & 0.47380 \\
O$^b$ position $y$ &   0.68040    & 0.68040 \\
O$^b$ position $z$ &  -0.08710    & -0.08710\\
Ca MSD (\SI{}{\square\angstrom}) &0.0079889 &   0.00800  \\
C MSD (\SI{}{\square\angstrom})&  0.007991  & 0.00799 \\
O$^a$ MSD (\SI{}{\square\angstrom})& 0.0124836 & 0.012482 \\
O$^b$ MSD (\SI{}{\square\angstrom})& 0.0124836 & 0.012482 \\
$R$-factor & & 1.96\% \\
 \bottomrule
\end{tabular}
\caption{Original parameters and refined results from \texttt{FullProf} for aragonite.}
\labtab{fullprof:aragoniterefinement}
\end{table}

As another independent validation of squared form factor predictions, measured
and refined values for the only monoclinic crystal structure in the
considered files (dilutetium silicon pentaoxide, SiLu$_{2}$O$_{5}$) were taken
from~\cite{gustafsson2001} (entry 2012009 in~\cite{grazulis2009}), and directly
compared against the ones predicted by \texttt{NCrystal}. The result for
$d$-spacings larger than \SI{1.5}{\angstrom} is shown in \reffig{SiLu2O5},
indicating a good agreement. Accordingly, this material is marked with R in the
last column of \reftab{datalib}.

\begin{figure}
  \centering
  \includegraphics[width=0.99\textwidth]{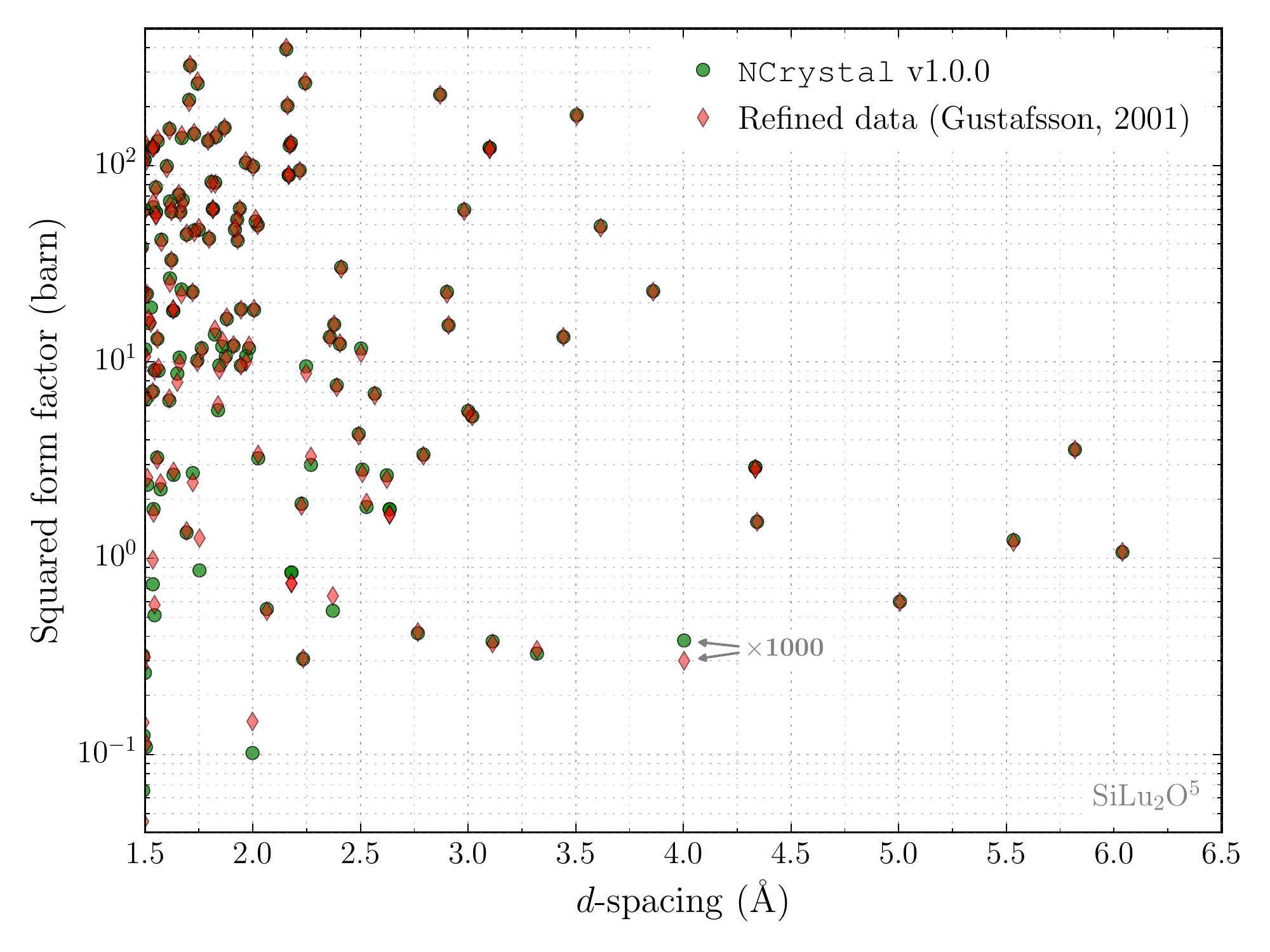}
  \caption{Refined measured squared form factors for monoclinic dilutetium
    silicon pentaoxide (SiLu$_{2}$O$_{5}$)~\cite{gustafsson2001} compared
    against those predicted by \texttt{NCrystal}.
    Note that for visualisation purposes, the weak reflection plane at
    approximately \SI{4.0}{\angstrom} is shown after multiplication with a factor of 1000.}
  \labfig{SiLu2O5}
\end{figure}

As a final validation, experimental measurements of energy-dependent total cross sections were obtained from
EXFOR~\cite{exfor2014} where available and tested against predictions from
\texttt{NCrystal}. This tests not only \texttt{.ncmat} files and the associated
crystal initialisation code, but also partly the implemented physics processes for
powders. Most materials were validated in this manner, as indicated with T or
$\overline{\text{T}}$ in the last column of \reftab{datalib}. Where files are
marked with $\overline{\text{T}}$, it indicates a rather weak validation of the
resulting Bragg edges -- either due to low quality data or because that material
has a relatively weak coherent elastic component, making the cross section
dominated by inelastic or absorption physics. For reasons of space, all
resulting validation plots are provided on the Data Library sub-section
of~\cite{ncrystalwww}, with just a few examples included here.

First, \refthreefigs{valtotxsMo}{valtotxsNi}{valtotxsZn} shows comparisons for
molybdenum, nickel, and zinc respectively. In all cases, some experimental data sets clearly
support the predicted cross sections, but occasionally some data sets provide
inconsistent results -- in mutual contradiction with not only \texttt{NCrystal}
but also other data sets. This underlines the inherent difficulty in performing
such validations with pre-existing measurements which might have been performed
under different or unclear conditions, with textured samples, etc. Nonetheless,
the general agreement with the predictions of \texttt{NCrystal} is clear -- and
it is interesting to note that it is particularly good for the magnetic
material, Nickel. This seems to support the validity of ignoring magnetic
interactions in the modelling, as long as both the sample and the incident neutrons themselves are
unpolarised. \Reffig{valtotxsUO2} shows another good agreement, this time for a
poly-atomic crystal:  uranium oxide. Finally, the plot for tin in
\reffig{valtotxsSn} is interesting in that it confirms a good agreement of
Bragg edges and single phonon scattering (despite some unclear result from one
data set from an unpublished measurement which could be due to texture). But
around \SI{1.3}{\electronvolt}, effects of a nuclear resonance can be seen, which
is not currently included in the modelling provided by \texttt{NCrystal}.

\begin{figure}
  \centering
  \includegraphics[width=0.85\textwidth]{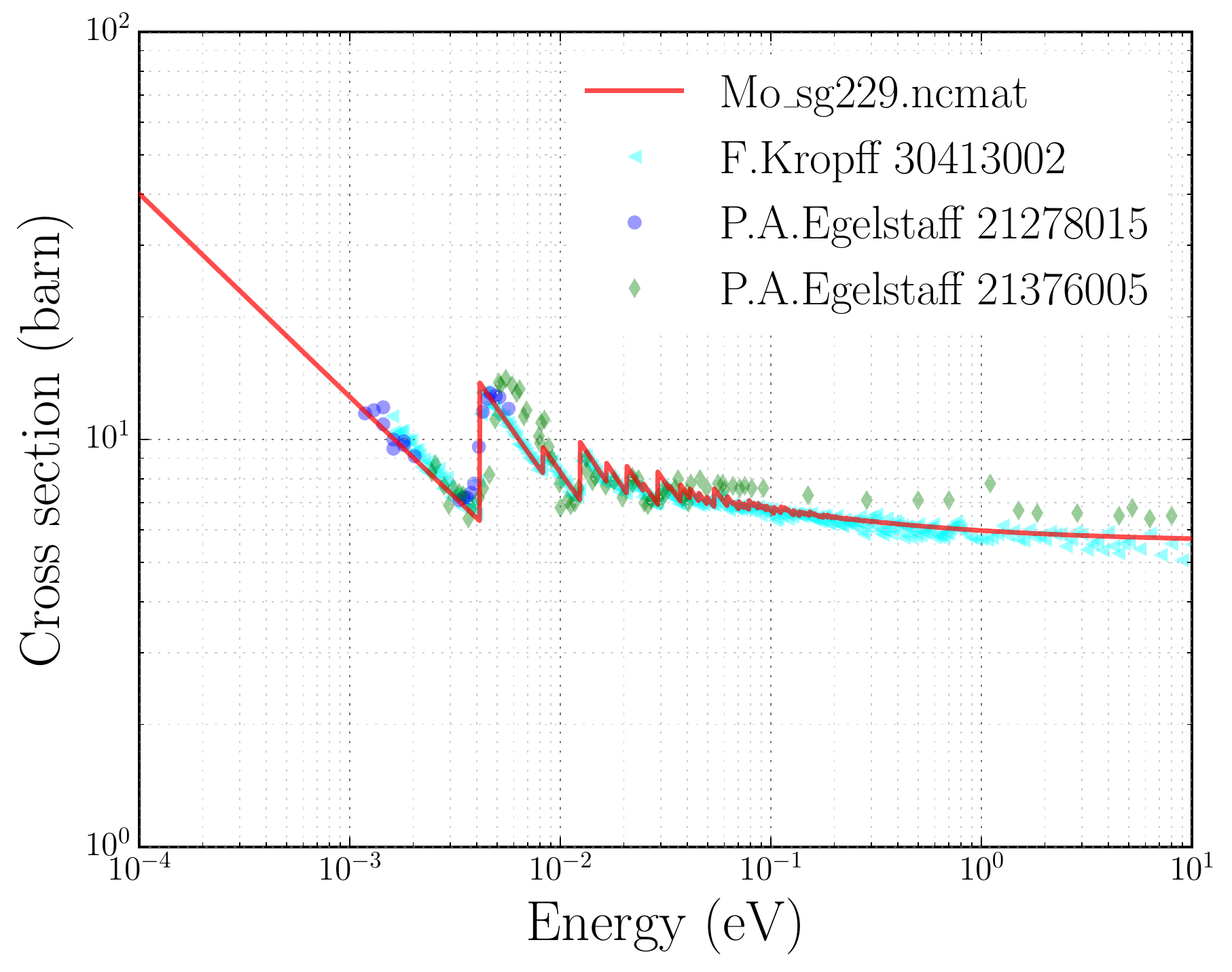}
  \caption{Validation of the total interaction cross section of molybdenum with
    experimental data from EXFOR~\cite{exfor2014}.}
  \labfig{valtotxsMo}
\end{figure}

\begin{figure}
  \centering
  \includegraphics[width=0.85\textwidth]{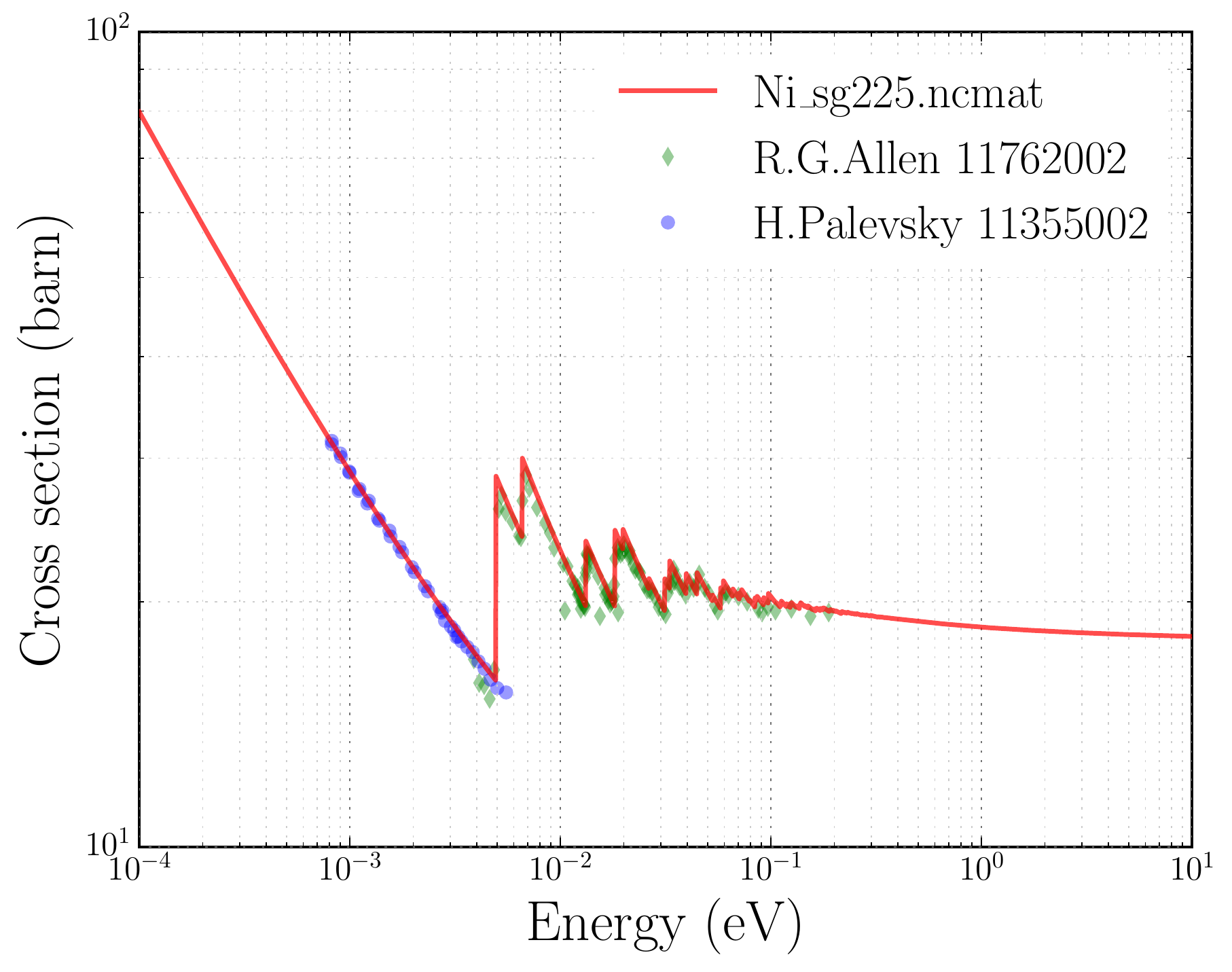}
  \caption{Validation of the total interaction cross section of nickel with
    experimental data from EXFOR~\cite{exfor2014}.}
  \labfig{valtotxsNi}
\end{figure}

\begin{figure}
  \centering
  \includegraphics[width=0.85\textwidth]{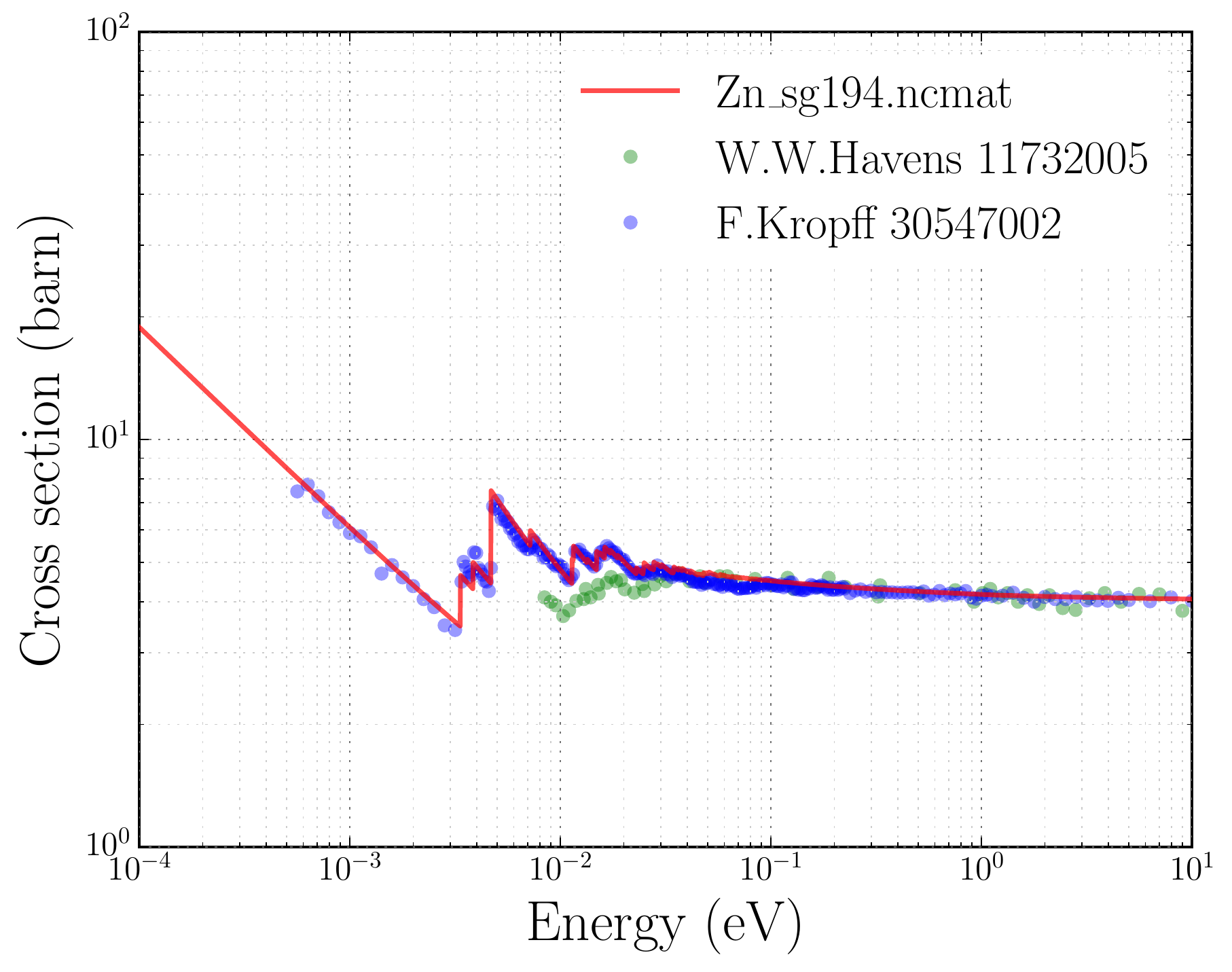}
  \caption{Validation of the total interaction cross section of zinc with
    experimental data from EXFOR~\cite{exfor2014}.}
  \labfig{valtotxsZn}
\end{figure}

\begin{figure}
  \centering
  \includegraphics[width=0.85\textwidth]{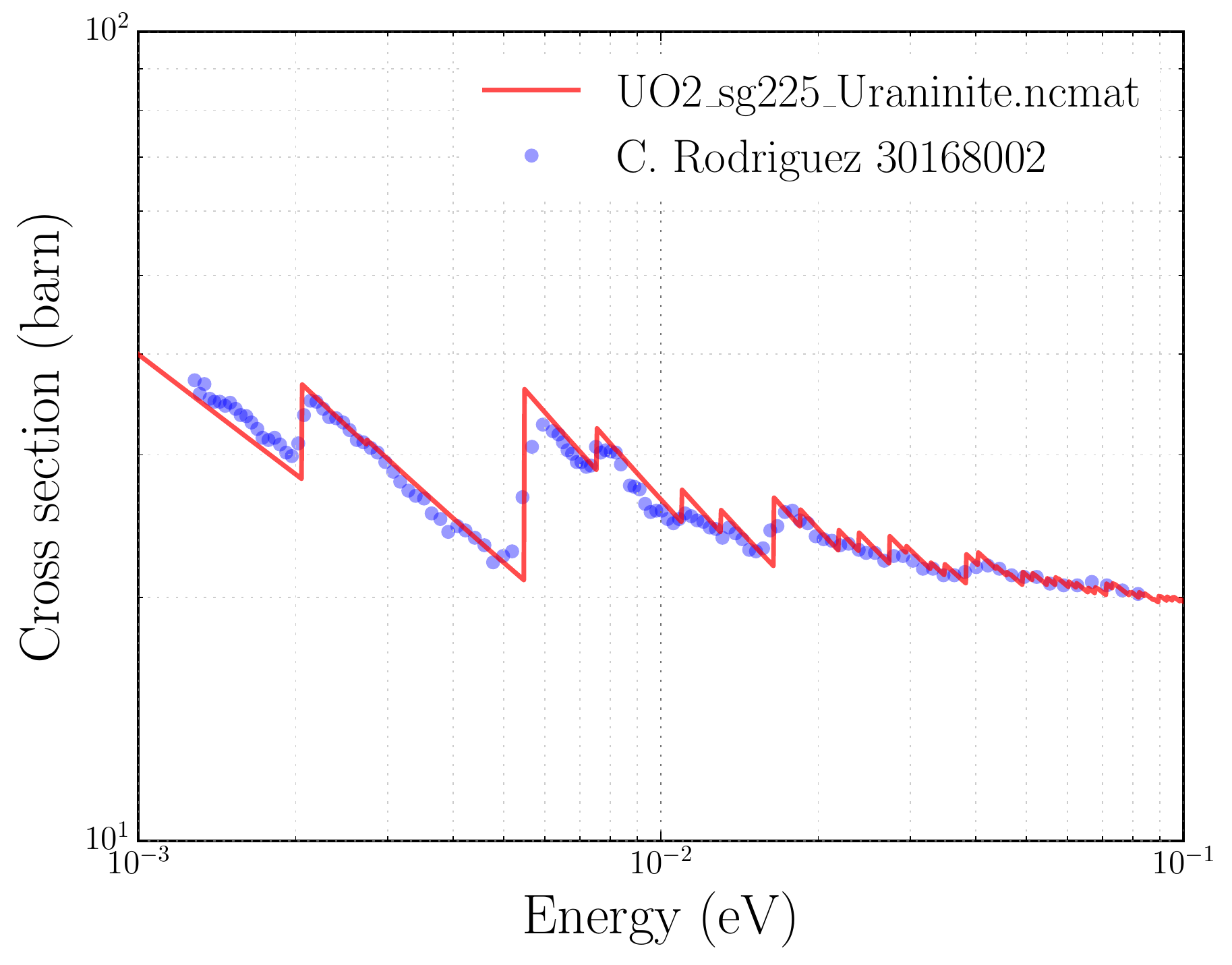}
  \caption{Validation of the total interaction cross section of uranium oxide
    with experimental data from EXFOR~\cite{exfor2014}.}
  \labfig{valtotxsUO2}
\end{figure}

\begin{figure}
  \centering
  \includegraphics[width=0.85\textwidth]{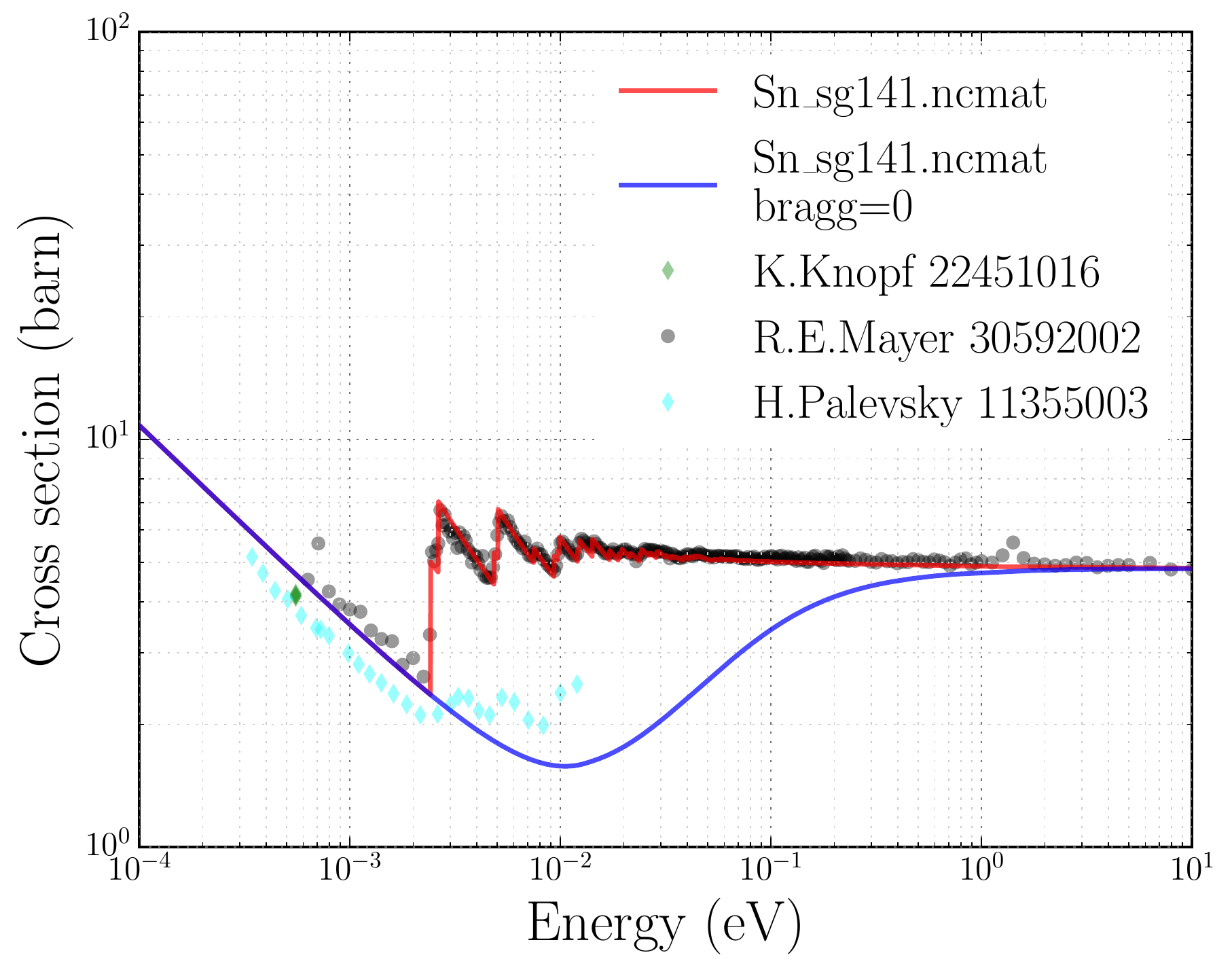}
  \caption{Validation of the total interaction cross section of tin with
    experimental data from EXFOR~\cite{exfor2014}.}
  \labfig{valtotxsSn}
\end{figure}

\section{Factories and unified configuration}\labsec{factoriesandunifiedcfg}

The object oriented implementation of \texttt{NCrystal} in a \texttt{C++} class
hierarchy described in \refsec{coreframework}, provide a high degree of
flexibility. This flexibility is an advantage for experts wishing to extend or
customise \texttt{NCrystal}, benefiting from low-level access to the various
classes and utilities exposed to developers as part of the \texttt{NCrystal}
API. However, as demonstrated by the example in
\reflisting{ncrystalrawcpp}, this flexibility comes at a cost of complexity
which it is desirable to contain. Not only in order to make the usage of
\texttt{NCrystal} simpler and less error prone for non-expert users, but also
because it is possible to use \texttt{NCrystal} in a variety of different
contexts, including several different programming languages and Monte Carlo
simulation frameworks as discussed in \refsec{bindingsandinterfaces}. The burden
of having to expose, support, document, validate and maintain all customisation
options of the core \texttt{C++} class library in all of these would be highly
non-trivial.

Thus, a simplified method for high-level material configuration was adapted for
\texttt{NCrystal} and employed in a consistent manner across all supported
interfaces and plugins. Specifically, it takes the form of a simple string in
which the name of a data file is combined with other parameters as needed. Not
only is it technically trivial to accept and pass around such a string in
essentially any programming language or user interface, but it is also by
definition easily persistifiable and shareable between different users and
frameworks. Furthermore, as \texttt{NCrystal} evolves and new material
parameters are introduced, existing interfaces and plugins will not require
corresponding updates, since they will simply continue to pass along any
specified strings to the \texttt{NCrystal} library just as before. The strings
are composed according to the following syntax, in which a file name is
optionally followed by a semicolon separated list of parameter
assignments:\footnote{To make it easier to embed such configuration strings into
  other text-based data sources, it is mandated that they are only valid if they
  consist of simple \texttt{ASCII} characters (excluding control-codes) and
  without any of the following characters:
  \texttt{"\textquotesingle|><()\{\}[]}. The only exception is in the filename,
  which might have to contain characters in other encodings if such are used in
  the filesystem.}
\begin{center}
  \texttt{<FILENAME>[;ignorefilecfg][;PARNAME1=VAL1][..][;PARNAMEN=VALN]}
\end{center}
Unless the filename is given as an absolute path, \texttt{NCrystal} will first
search the current working directory for the data file, then any directory
indicated with the environment variable \texttt{NCRYSTAL\_DATADIR}, before
finally looking in the directory where data files from the data library shipped
with \texttt{NCrystal} itself were copied as part of the \texttt{NCrystal}
installation procedure. The optional and rarely used \texttt{ignorefilecfg}
keyword indicates to \texttt{NCrystal} that it should not search the data files
themselves for additional parameter settings. This is needed, because
\texttt{NCrystal} allows users to embed default parameter values directly inside
the input files (usually in comments), by placing them inside square braces
following the keyword \texttt{NCRYSTALMATCFG}. Currently, the only data file in
the data library shipped with \texttt{NCrystal} employing this is the file
\texttt{C\_sg194\_pyrolytic\_graphite.ncmat}, which contains a comment embedding
the following contents:
\begin{center}
  \texttt{NCRYSTALMATCFG[lcaxis=0,0,1]}
\end{center}
Due to this, the configuration of a single crystal material from that file will
result in the anisotropic mosaicity distribution appropriate for pyrolytic
graphite to be used rather than the standard isotropic one.

\Reftab{ncmatcfg} documents the configuration parameters most likely to be of
interest to typical \texttt{NCrystal} users, but for clarity a few example
strings and their implications will be presented in the following.

\begin{table}[tp]
\centering
\resizebox{\textwidth}{!}{%
\begin{tabular}{@{}llll@{}}
\toprule
\textbf{Parameter} & \textbf{Default} & \textbf{Description}
& \textbf{Units ($^\dagger$=default)} \\ \midrule
\texttt{temp}    & \SI{293.15}{\kelvin} & Temperature of material. & \texttt{K}$^\dagger$, \texttt{C}, \texttt{F} \\
\texttt{dcutoff} & \num{0} &
  Value of $d$-spacing threshold, $d_\text{cut}$. A value of 0 im- & \texttt{Aa}$^\dagger$, \texttt{nm}, \texttt{mm}, \texttt{cm}, \texttt{m} \\
&& plies automatic threshold selection. &\\
\texttt{bragg} & \texttt{true} &  Disable to exclude Bragg diffraction. &\\
\texttt{bkgd} & \texttt{"best"} &  Influence incoherent/inelastic models chosen
by fac-
  &\\
&& tories. The default, \texttt{"best"}, implies most realistic   &\\
&& available model, while a value of \texttt{"none"} or \texttt{"0"} ex-  &\\
&& cludes such processes. Other values allow advanced&\\
&& users to select or tune models. &\\
[2ex]\multicolumn{4}{@{}l@{}}{\textbf{Powders only:}} \\\cline{1-2}\\[-2ex]
\texttt{packfact} & \num{1.0} & Powder packing factor. & \\
[2ex]\multicolumn{4}{@{}l@{}}{\textbf{Single crystals only:}}\\\cline{1-2}\\[-2ex]
\texttt{mos} & & Gaussian spread in mosaic crystals (FWHM). & \texttt{rad}$^\dagger$, \texttt{deg}, \texttt{arcmin}, \texttt{arcsec} \\
\texttt{dir1}
&& Primary orientation of crystal given by specifying &\\
&& directions in both crystal and lab frames with the &\\
&& syntax: \texttt{"@crys:cx,cy,cz@lab:lx,ly,lz"}.&\\
&& Alternatively, the crystal direction can be specified &\\
&& in reciprocal lattice space rather than direct space&\\
&& as \texttt{"@crys\_hkl:ch,ck,cl@lab:lx,ly,lz"}.  &\\
\texttt{dir2}
&& Secondary orientation of crystal, specified using the &\\
&& same syntax as for the \texttt{dir1} parameter. Only com- &\\
&& ponents of \texttt{dir2} orthogonal to \texttt{dir1} will actually  &\\
&& influence the orientation (but see \texttt{dirtol}). &\\
\texttt{dirtol} & \SI{e-4}{\radian}
 & Specification of \texttt{dir1} and \texttt{dir2} are accepted only if   & \texttt{rad}$^\dagger$, \texttt{deg}, \texttt{arcmin}, \texttt{arcsec} \\
&& the angle between \texttt{dir1} and \texttt{dir2} is similar in crys- &\\
&& tal and lab frames within this tolerance. &\\
\texttt{lcaxis}
&& Symmetry axis of anisotropic layered crystals simi-&\\
&& lar to pyrolytic graphite (as vector, e.g.\ \texttt{"0,0,1"}).  &\\
\bottomrule
\end{tabular}}
\caption{Some of the most important parameters available in
  \texttt{NCrystal} configuration strings.}
\labtab{ncmatcfg}
\end{table}

\begin{itemize}
  \small \item \textbf{\texttt{"Al\_sg225.ncmat"}}, \textbf{\texttt{"Al\_sg225.nxs"}}, or
  \textbf{\texttt{"Al.laz"}}\\ Polycrystalline or powdered aluminium at room
  temperature, using appropriate code for loading \texttt{.ncmat}, \texttt{.nxs}
  or \texttt{.laz} files respectively. Although in principle representing the
  same material, the realism will depend on the type of input file, as
  \texttt{NCrystal} automatically selects the most realistic modelling possible
  while also considering issues like load times. In this particular case, the
  scattering models based on the \texttt{.laz} file will not contain any
  incoherent or inelastic components, and the Bragg diffraction based on the
  \texttt{.nxs} file will contain fewer reflection planes at shorter
  $d$-spacings than the one based on the \texttt{.ncmat} file (cf.~\refsec{data::nxs}).
\item \textbf{\texttt{"Be\_sg194.ncmat;temp=100K"}}\\
  Polycrystalline or powdered beryllium at a low temperature, useful for modelling of beryllium filters.
\item \textbf{\texttt{"Be\_sg194.ncmat;temp=100K;bkgd=phonondebye:elastic"}}\\
  The same beryllium, but forcing the modelling of inelastic scattering with the
  \texttt{BkgdPhonDebye} class to neglect energy transfers (cf.\ \refsec{physmodels}).
\item \textbf{\texttt{"Be\_sg194.ncmat;bkgd=0;temp=100K"}}\\ The same
  beryllium again, but this time the only
  scattering being modelled is that of Bragg diffraction. While clearly
  decreasing the modelled realism, avoiding incoherent or inelastic scatterings
  is occasionally useful for reasons of clarity or validation.
\item \textbf{\texttt{"SiO2\_sg154\_Quartz.ncmat;packfact=0.7"}}\\
  Loosely packed quartz powder.
\item \textbf{\texttt{"C\_sg227\_Diamond.ncmat"}}\\
  Powdered diamond.
\item  \textbf{\texttt{"Cu\_sg225.ncmat;mos=0.5deg;}}\\
  \mbox{\textbf{\texttt{~dir1=@crys\_hkl:2,2,0@lab:0,0,1;dir2=@crys\_hkl:0,0,1@lab:0,1,0"}}}\\
  Setup for modelling of a Cu220  monochromator with mosaicity 0.5\SIUnitSymbolDegree.
\item \textbf{\texttt{"Ge\_sg227.ncmat;mos=40.0arcsec;dirtol=180deg;dcutoff=0.5Aa;bkgd=0;}}\\
  \mbox{\textbf{\texttt{~dir1=@crys\_hkl:5,1,1@lab:0,0,1;dir2=@crys\_hkl:1,0,0@lab:1,0,0"}}}\\
  Setup for modelling of a Ge511 monochromator with mosaicity
  40\SIUnitSymbolArcsecond. The user was not overly concerned with the
  secondary direction of the crystal in this case and just decided to align the
  unspecified parts of the first crystal axis with the laboratory $x$-axis,
  increasing the \texttt{dirtol} parameter in order to allow this. Furthermore,
  the modelling was made faster at the cost of realism by increasing the
  $d$-spacing cutoff and disabling incoherent and inelastic components with \texttt{bkgd=0}.
\item
  \textbf{\texttt{"C\_sg194\_pyrolytic\_graphite.ncmat;mos=3deg;}}\\
  \mbox{\textbf{\texttt{~dir1=@crys\_hkl:0,0,1@lab:0,0,1;dir2=@crys\_hkl:1,1,0@lab:0,1,0"}}}\\
  Setup for modelling of a highly ordered pyrolytic graphite (HOPG) single
  crystal. Note that as discussed above, the \texttt{lcaxis=0,0,1} parameter is
  embedded into the file itself,  enabling the proper anisotropic mosaicity distribution.
\item
  \textbf{\texttt{"C\_sg194\_pyrolytic\_graphite.ncmat;ignorefilecfg;mos=3deg;}}\\
  \mbox{\textbf{\texttt{~dir1=@crys\_hkl:0,0,1@lab:0,0,1;dir2=@crys\_hkl:1,1,0@lab:0,1,0"}}}\\
  The same as the previous material, but ignoring the embedded \texttt{lcaxis}
  parameter, resulting in an isotropic Gaussian mosaicity distribution.
\item \textbf{\texttt{"Al2O3\_sg167\_Corundum.ncmat;bragg=0;temp=-50C"}}\\
  Setup for modelling a cooled sapphire crystal without any Bragg
  diffraction. This approximation could be useful in order to efficiently model
  a single crystal sapphire filter in a beam, assumed to be positioned such that
  the Bragg condition is unsatisfied for all planes.
\end{itemize}

\section{Interfaces and bindings}\labsec{bindingsandinterfaces}

The method of material configuration presented in
\refsec{factoriesandunifiedcfg} facilitates the usage of \texttt{NCrystal} in a
variety of contexts, consistently employing the same configuration strings and
data files everywhere. This allows the sharing of material configurations
between applications and users, and implies a freedom of choice when tuning or
validating such configurations -- irrespective of which Monte Carlo application is
ultimately used to study a given problem. Accordingly, \refsec{interfaces::cmdline}
will present a tool for quick inspection and tuning of material configurations,
while \refsec{interfaces::bindings} will present language bindings available
to advanced users needing direct access to \texttt{NCrystal} functionality --
including the case of developers wishing to create new \texttt{NCrystal} plugins for Monte
Carlo applications. Finally, the \emph{raison d'etre} of \texttt{NCrystal},
\reftwosections{interfaces::geant4}{interfaces::mcstas} respectively presents
plugins for \texttt{Geant4} and \texttt{McStas}.  Although not discussed further here, it should
be noted that at present \texttt{NCrystal} can additionally be used in
\texttt{ANTS2}~\cite{antstwo2016} thanks to work by A. Morozov (University of Coimbra, Portugal), and work is
in progress by J.I. M\'{a}rquez Dami\'{a}n (Centro At\'{o}mico Bariloche, CNEA, Argentina) to enable its
usage with \texttt{OpenMC}~\cite{openmc} and \texttt{NJOY}~\cite{njoy2012} as well.

\subsection{Command-line inspection}\labsec{interfaces::cmdline}

A command-line utility, \texttt{ncrystal\_inspectfile}, which can be used to
inspect and tune material configurations, is included in the \texttt{NCrystal}
release. It accepts one or more \texttt{NCrystal} configuration strings as
arguments, and either plots or prints relevant information in response. Detailed
instructions are available on demand by specification of the \texttt{-{}-help}
flag, but a few usage examples will be provided here for illustration.

First of all, when providing just one configuration string, the corresponding
material will be created as a powder, and two plots will be produced: the
components of the resulting neutron cross sections and a distribution of randomly
sampled scatter angles. For instance, the following command can be used to
inspect a sapphire powder with default settings for temperature, packing-factor,
$d$-spacing cut-off, etc.:
\begin{lstlisting}[language={},frame=none,belowskip=0.0\baselineskip]
  ncrystal_inspectfile -a Al2O3_sg167_Corundum.ncmat
\end{lstlisting}
The resulting plots are shown in
\reffig{inspectfilestdplots}. \Reffig{inspectfilestdplots}.a shows the
components of the total interaction cross section, including absorption since the
\texttt{-a} flag was specified. \Reffig{inspectfilestdplots}.b shows a
two-dimensional scatter-plot: at each neutron wavelength, scatterings
are sampled with statistics proportional to the scattering cross section at that
wavelength and the resulting scatter angles shown.
\begin{figure}
  \centering
  \subfloat[]{\includegraphics[width=0.85\textwidth]{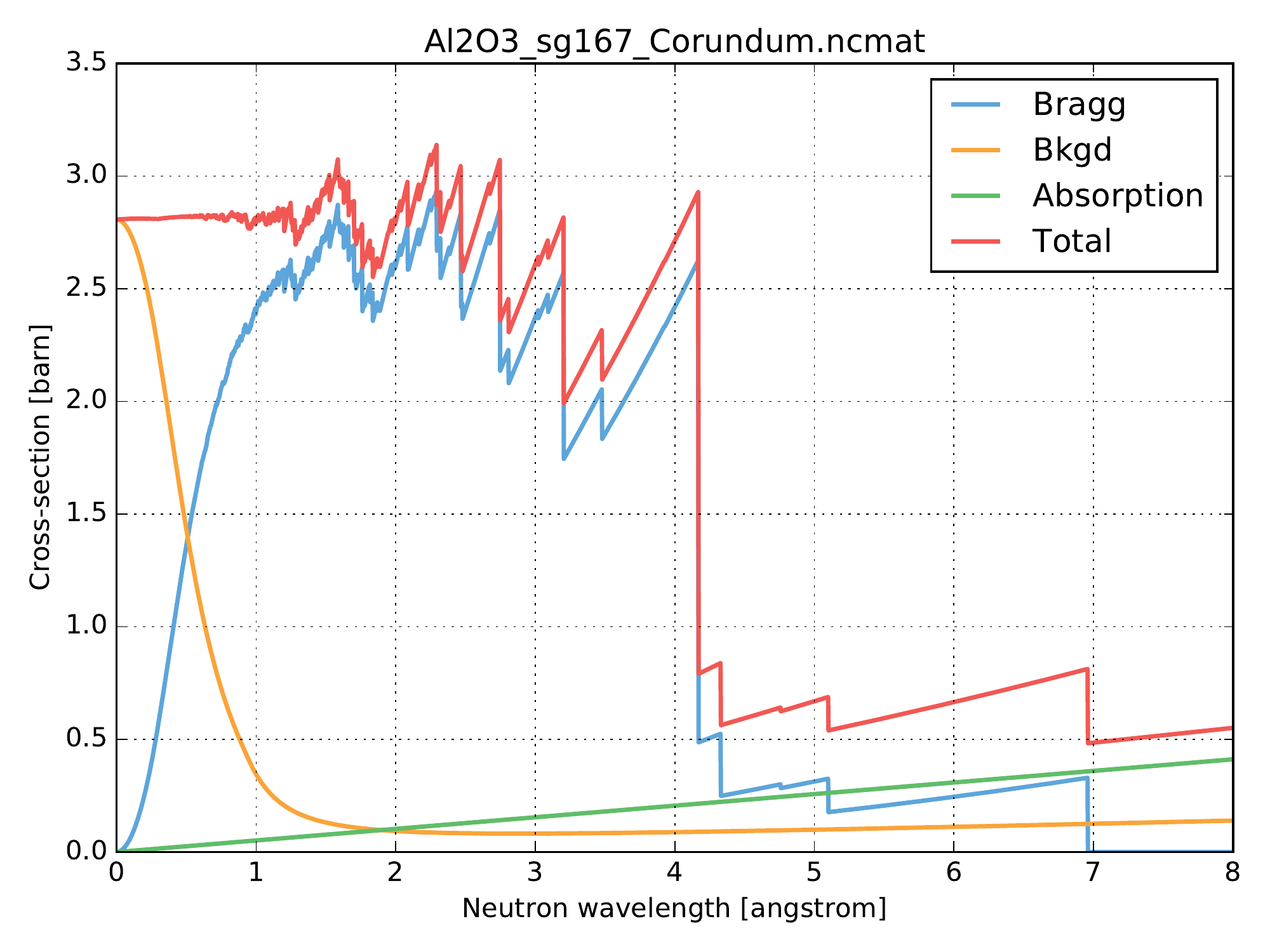}}\\
  \subfloat[]{\includegraphics[width=0.85\textwidth]{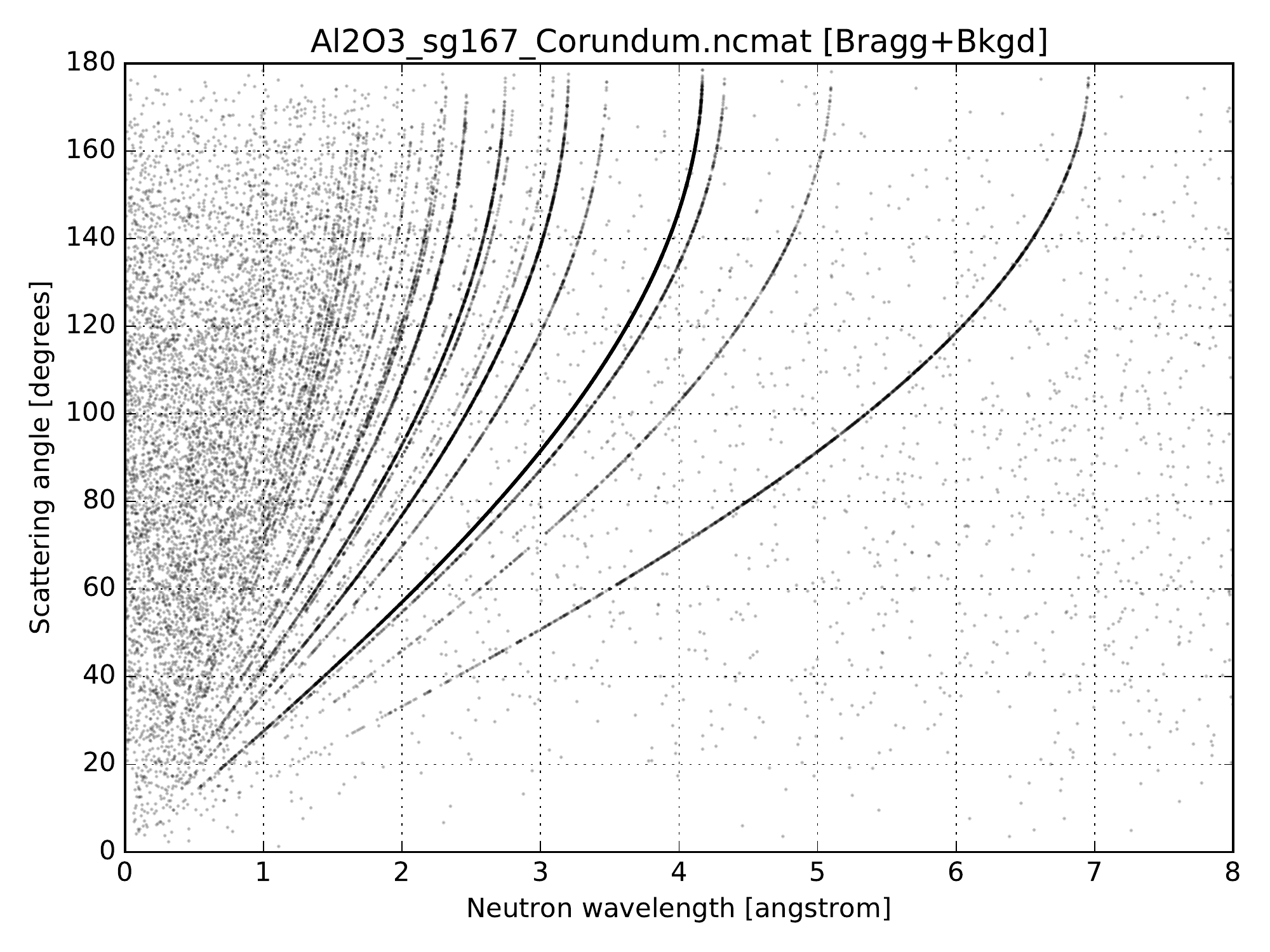}}
  \caption{Plots produced by invoking \texttt{ncrystal\_inspectfile -a} on a single
    configuration string (here
    \texttt{"Al2O3\_sg167\_Corundum.ncmat"}).}
  \labfig{inspectfilestdplots}
\end{figure}
By supplying the \texttt{-{}-dump} flag, the graphical plotting will be replaced
with a printout of loaded crystal information. For instance, the command:
\begin{lstlisting}[language={},frame=none,belowskip=0.0\baselineskip]
  ncrystal_inspectfile --dump "Cu2O_sg224_Cuprite.ncmat;dcutoff=1Aa"
\end{lstlisting}
results in the printout shown in \reflisting{inspectfiledump}.
\begin{lstlisting}[float,language={},
  label={lst:inspectfiledump},
  caption={Printout produced by invoking \texttt{ncrystal\_inspectfile -{}-dump} on a single
    configuration string (here
    \texttt{"Cu2O\_sg224\_Cuprite.ncmat;dcutoff=1Aa"}). Refer to \reftab{ncinfo}
  for an explanation of the listed parameters.}
]
---------------------------------------------------------
Space group number      : 224
Lattice spacings   [Aa] : 4.2685 4.2685 4.2685
Lattice angles    [deg] : 90 90 90
Unit cell volume [Aa^3] : 77.7725
Atoms / unit cell       : 6
---------------------------------------------------------
Atoms per unit cell (total 6):
     2 O atoms [T_Debye=385.668K, MSD=0.0187741Aa^2]
     4 Cu atoms [T_Debye=189.192K, MSD=0.0189719Aa^2]
---------------------------------------------------------
Atomic coordinates:
       O            0            0            0
       O          0.5          0.5          0.5
      Cu         0.25         0.25         0.25
      Cu         0.25         0.75         0.75
      Cu         0.75         0.25         0.75
      Cu         0.75         0.75         0.25
---------------------------------------------------------
   Density : 6.11036 g/cm3
---------------------------------------------------------
   Temperature : 293.15 kelvin
---------------------------------------------------------
Neutron cross-sections:
   Absorption at 2200m/s : 2.52006 barn
   Free scattering       : 6.43997 barn
---------------------------------------------------------
HKL planes (d_lower = 1 Aa, d_upper = inf Aa):
  H   K   L  d_hkl[Aa] Multiplicity FSquared[barn]
  0   1  -1    3.01829           12         1.2426
  1  -1  -1    2.46442            8        8.42503
  0   0   2    2.13425            6        3.14444
  1  -2  -1    1.74261           24          1.056
  0   2  -2    1.50914           12        13.0016
  0   1  -3    1.34982           24       0.897427
  1  -3  -1      1.287           24        6.06387
  2  -2  -2    1.23221            8        2.25851
  1  -3  -2     1.1408           48       0.762663
  0   0   4    1.06713            6        9.36659
  0   3  -3     1.0061           36       0.648137
---------------------------------------------------------
\end{lstlisting}
\Reffig{inspectfilecmpplots} shows how specifying more than one configuration
string at a time leads results in a plot comparing the resulting
cross sections. \Reffig{inspectfilecmpplots}.a shows the total interaction cross section of thermal neutrons in a range of
polycrystalline metals -- which might for instance be considered for a support structure in some
parts of a neutron instrument -- and is the result of the command (all on one line):
\begin{lstlisting}[language={},frame=none,belowskip=0.0\baselineskip]
  ncrystal_inspectfile -a Al_sg225.ncmat Fe_sg229_Iron-alpha.ncmat
                          Cu_sg225.ncmat Ti_sg194.ncmat
\end{lstlisting}
\Reffig{inspectfilecmpplots}.b shows a similar plot, this time for a Beryllium powder
at various temperatures, indicating the significance of beam filter cooling. It is the result of the command (all on one line):
\begin{lstlisting}[language={},frame=none,belowskip=0.0\baselineskip]
  ncrystal_inspectfile -a  "Be_sg194.ncmat;temp=100K"
                           "Be_sg194.ncmat;temp=200K"
                           "Be_sg194.ncmat;temp=300K"
\end{lstlisting}
\begin{figure}
  \centering
  \subfloat[]{\includegraphics[width=0.85\textwidth]{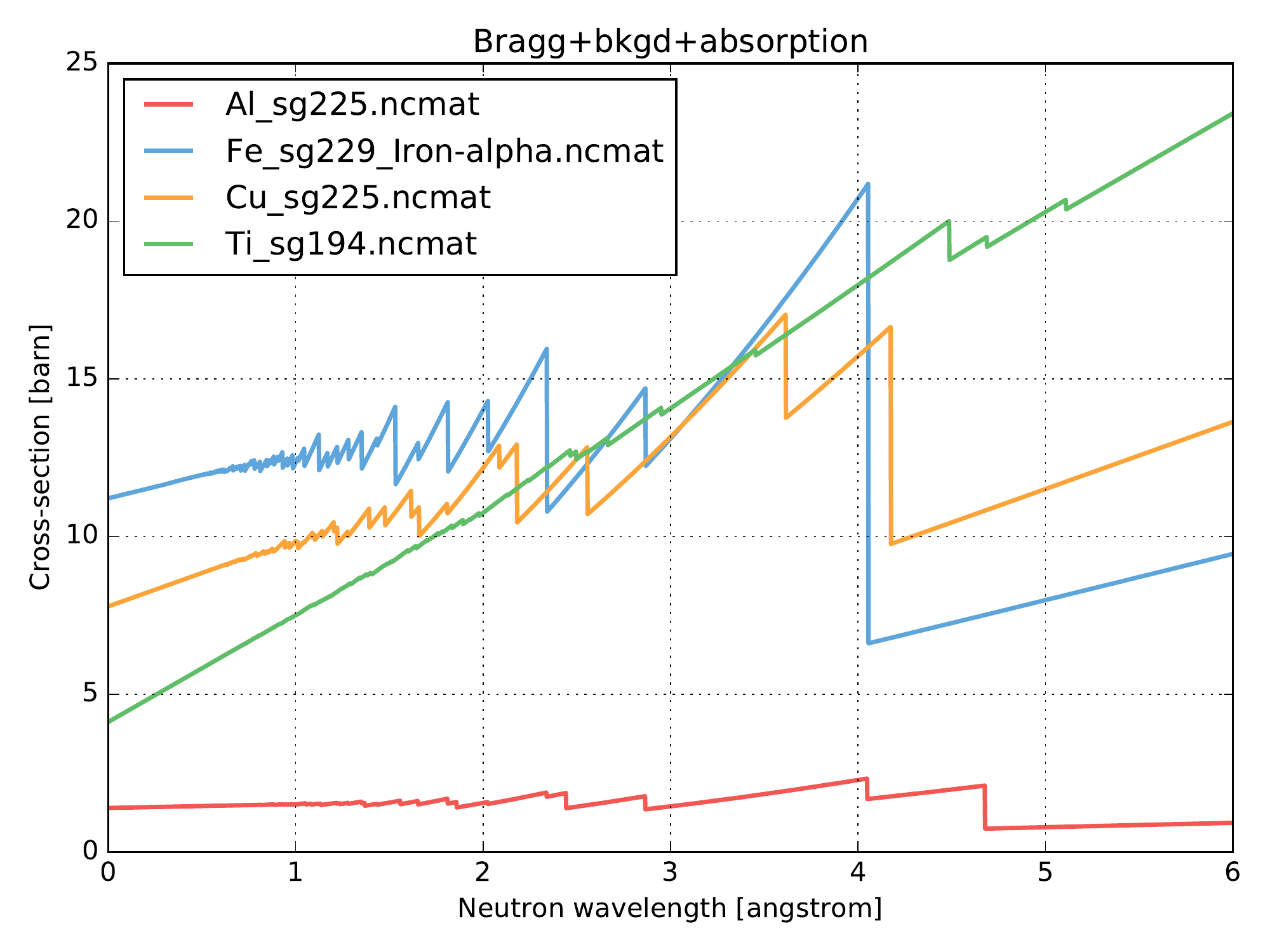}}\\
  \subfloat[]{\includegraphics[width=0.85\textwidth]{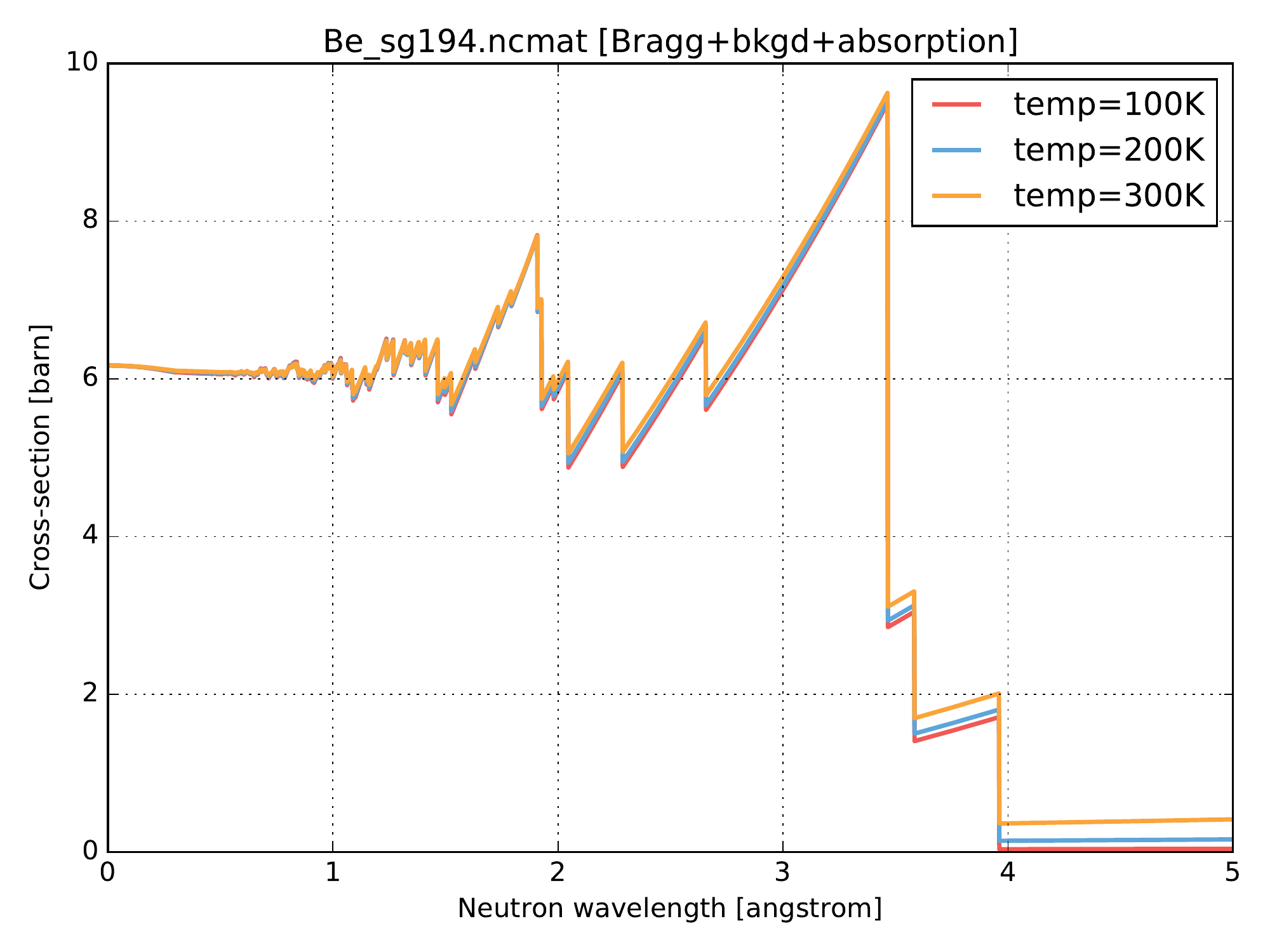}}
  \caption{Plots produced by invoking \texttt{ncrystal\_inspectfile -a} on
    multiple configuration strings. As a result, total interaction
    cross sections are shown for four different metals in (a), and for three
    beryllium powders at different temperatures in (b).}
  \labfig{inspectfilecmpplots}
\end{figure}

Finally, the command \texttt{ncrystal\_inspectfile -{}-test} can be used to
validate a given installation of \texttt{NCrystal}. Although
\texttt{ncrystal\_inspectfile} already provides a useful set of functionality,
it is expected that additional standard tools will be added in the future. In
particular it would be useful with more options for plot creation and with
utilities for assisting with the creation of configuration strings involving the
alignment of single crystals, for instance to assist in the configuration of
simulations involving neutron monochromators.

\subsection{\texttt{C++}, \texttt{C} and \texttt{Python} bindings}\labsec{interfaces::bindings}

Using the configuration strings presented in \refsec{factoriesandunifiedcfg},
the somewhat complicated \texttt{C++} code in \reflisting{ncrystalrawcpp} can be
significantly simplified as shown in \reflisting{ncrystalcfgcpp}: not only is it
fewer lines, but the only material-specific code is the one defining the
configuration string variable named \texttt{cfg}. It is thus straight-forward to
create \texttt{C++} applications or plugins which use \texttt{NCrystal} as a
backend, but lets users provide the configuration strings in whichever frontend
is relevant to the task at hand. The \emph{lingua franca} of software is,
however, the \texttt{C} programming language, and most modern programming
languages contain some facility for interfacing with such code. In order to make
\texttt{NCrystal} as widely useful as possible and able to support applications
like \texttt{McStas} (cf.\ \refsec{interfaces::mcstas}), a \texttt{C} interface
is thus provided in the header file \texttt{ncrystal.h}, and the \texttt{C}
equivalent of \reflisting{ncrystalcfgcpp} is shown in
\reflisting{ncrystalcfgc}. Although \texttt{C} does not support classes, an
object-oriented paradigm is still achieved by providing the \texttt{C} code with
\emph{handles} instead of class instance pointers. Most \texttt{C} programmers
should be familiar with such a scheme, as it is also encountered elsewhere, for
instance in the form of file handles for file I/O. Although not quite as
convenient as the \texttt{C++} interface, \texttt{C} code using the created
\texttt{NCrystal} handles essentially follow the same pattern as the equivalent
\texttt{C++} code. For instance, after retrieving the \texttt{absn} handle in
\reflisting{ncrystalcfgc}, it can be used in the following manner to provide a
cross section in the variable \texttt{xs}, here for a \SI{1.8}{\angstrom} neutron incident along the
$z$-axis:
\begin{lstlisting}[language={[ncrystal]C},frame=none,belowskip=0.0\baselineskip]
  double ekin = ncrystal_wl2ekin(1.8);
  double dir[3] = {0.0, 0.0, 1.0};
  double xs;
  ncrystal_crosssection(absn, ekin, dir, &xs);
\end{lstlisting}
\texttt{C} does not support exceptions, so when using the \texttt{C} interface,
any exceptions thrown internally in the \texttt{NCrystal} library due to errors
will by default result in an appropriate error message being printed and the
program terminated. For further details and a list of available functions,
please refer to the header file \texttt{ncrystal.h}.

\begin{lstlisting}[float,language={[ncrystal]C++},
  label={lst:ncrystalcfgcpp},
  caption={\texttt{C++} code using a configuration string to create related \texttt{NCrystal} objects.}
]
#include "NCrystal/NCrystal.hh"
namespace NC = NCrystal;//alias for convenience

int main() {

  //Create objects based on a universal configuration string:
  std::string cfg = "somefile.ncmat;temp=400K;dcutoff=0.5Aa;"
                    "bkgd=phonondebye:elastic";
  const NC::Info * info = NC::createInfo(cfg);
  const NC::Scatter * scat = NC::createScatter(cfg);
  const NC::Absorption * absn = NC::createAbsorption(cfg);

  // ...
  // code here using info, scat and absn pointers
  // ...

  return 0;
}
\end{lstlisting}

\begin{lstlisting}[float,language={[ncrystal]C},
  label={lst:ncrystalcfgc},
  caption={\texttt{C} code using a configuration string to create related \texttt{NCrystal} object handles.}
]
#include "NCrystal/ncrystal.h"

int main() {

  /* Create objects based on a universal configuration string: */
  const char * cfg = "somefile.ncmat;temp=400K;dcutoff=0.5Aa;"
                     "bkgd=phonondebye:elastic";
  ncrystal_info_t info = ncrystal_create_info(cfg);
  ncrystal_scatter_t scat = ncrystal_create_scatter(cfg);
  ncrystal_absorption_t absn = ncrystal_create_absorption(cfg);

  /* ...                                         */
  /* code here using info, scat and absn handles */
  /* ...                                         */

  return 0;
}
\end{lstlisting}

Finally, it is possible to use \texttt{NCrystal} directly from
\texttt{Python}~\cite{vanRossum:2011}, which mostly exists as a feature to support the
usage of \texttt{NCrystal} for advanced scripting, analysis and plotting work.
\texttt{Python} and \texttt{C++} share concepts like classes and exceptions, and
\texttt{Python} code using \texttt{NCrystal} thus ends up being similar and
possibly even simpler than the corresponding \texttt{C++} code, as can be seen
in the \texttt{Python} equivalent of \reflisting{ncrystalcfgcpp}, shown in
\reflisting{ncrystalcfgpy}. Additionally, for efficiency and convenience, the
\texttt{Python} interface supports vectorised access through
\texttt{Numpy}~\cite{numpy} arrays. For a simple example using this,
\reflisting{ncrystalcfgpyplot} illustrates how to create a
\texttt{Matplotlib}~\cite{matplotlib} plot of the scattering cross section as a function of
neutron wavelengths between \SI{0}{\angstrom} and \SI{10}{\angstrom}.

\begin{lstlisting}[float,language={[ncrystal]Python},
  label={lst:ncrystalcfgpy},
  caption={\texttt{Python} code using a configuration string to create related \texttt{NCrystal} objects.}
]
import NCrystal as NC

cfg = """somefile.ncmat;temp=400K;dcutoff=0.5Aa;
         bkgd=phonondebye:elastic"""
info = NC.createInfo(cfg)
scat = NC.createScatter(cfg)
absn = NC.createAbsorption(cfg)

# ...
# code here using info, scat and absn objects
# ...
\end{lstlisting}

\begin{lstlisting}[float,language={[ncrystal]Python},
  label={lst:ncrystalcfgpyplot},
  caption={\texttt{Python} code using the support in \texttt{NCrystal} for vectorised access through
    \texttt{Numpy} arrays, in order to plot scattering cross sections with \texttt{Matplotlib}.}
]
#Create NCScatter object:
import NCrystal as NC
scat = NC.createScatter("<some-cfg-here>")

#Then plot via Numpy and Matplotlib:
import numpy
import matplotlib.pyplot as plt
wl = numpy.linspace( 0.0, 10.0, num=10000 )
xs = scat.crossSectionNonOriented( NC.wl2ekin( wl ) )
plt.plot(wl,xs)
plt.show()
\end{lstlisting}

As for any other \texttt{Python} module, documentation of the \texttt{NCrystal}
module is built in and accessible via the \texttt{help()} function.  Behind the
scenes, the \texttt{Python} interface is implemented via the \texttt{C}
interface using the \texttt{ctypes} \texttt{Python} module to call directly into
the binary \texttt{NCrystal} library. Thus, no additional software dependencies
are introduced by this interface -- beyond naturally the \texttt{Python}
interpreter itself.

\subsection{\texttt{Geant4} interface}\labsec{interfaces::geant4}

Assuming \texttt{Geant4} is configured to use a physics list which include the
so-called HP (high precision) models for neutron physics, almost all comparisons
between cross sections in \texttt{Geant4} and \texttt{NCrystal} are qualitatively
equivalent to the ones shown in \Reffig{xsgeant4vsncrystal} for a magnesium
powder. Starting around the \SI{}{\kilo\electronvolt} scale, \texttt{Geant4}
provides detailed modelling of higher energy effects such as those related to
nuclear resonances. At lower energies, the cross section curves are, however,
completely smooth and featureless due to the free-gas approximation
used. \texttt{NCrystal}, on the other hand, provides detailed
structure-dependent scattering physics at the sub-\SI{}{\electronvolt} scale --
but has no capacity for modelling physics at the \SI{}{\kilo\electronvolt}
scale. At intermediate energy scales, neither nuclear resonance physics or
material structure-dependent physics introduce significant features, and there is an overlap in
predictions between \texttt{NCrystal} and \texttt{Geant4}. For absorption
processes, material structure is unimportant, and the predictions from
\texttt{NCrystal} and \texttt{Geant4} are in perfect agreement over the range
covered by \texttt{NCrystal}, due to the validity of the simple $1/v$ scaling of
such cross sections (cf.~\refsec{theory::neutronscattering}). However, unlike
\texttt{NCrystal}, \texttt{Geant4} provides detailed modelling of the secondary
particles produced in absorption reactions. Consequently, the \texttt{NCrystal}
plugin for \texttt{Geant4} does not touch the absorption physics at all, instead
focusing on replacing just the scattering physics for low energy
neutrons. Presently this is done with a global hard-coded cross over point of
\SI{5}{\electronvolt}, but the exact transition model might be revisited in the
future, since a few rare isotopes have nuclear resonances lower than this. As an
example, the experimental data in \reffig{valtotxsSn} indicates a
resonance around \SI{1.3}{\electronvolt}.

\begin{figure}
  \centering
  \includegraphics[width=1.0\textwidth]{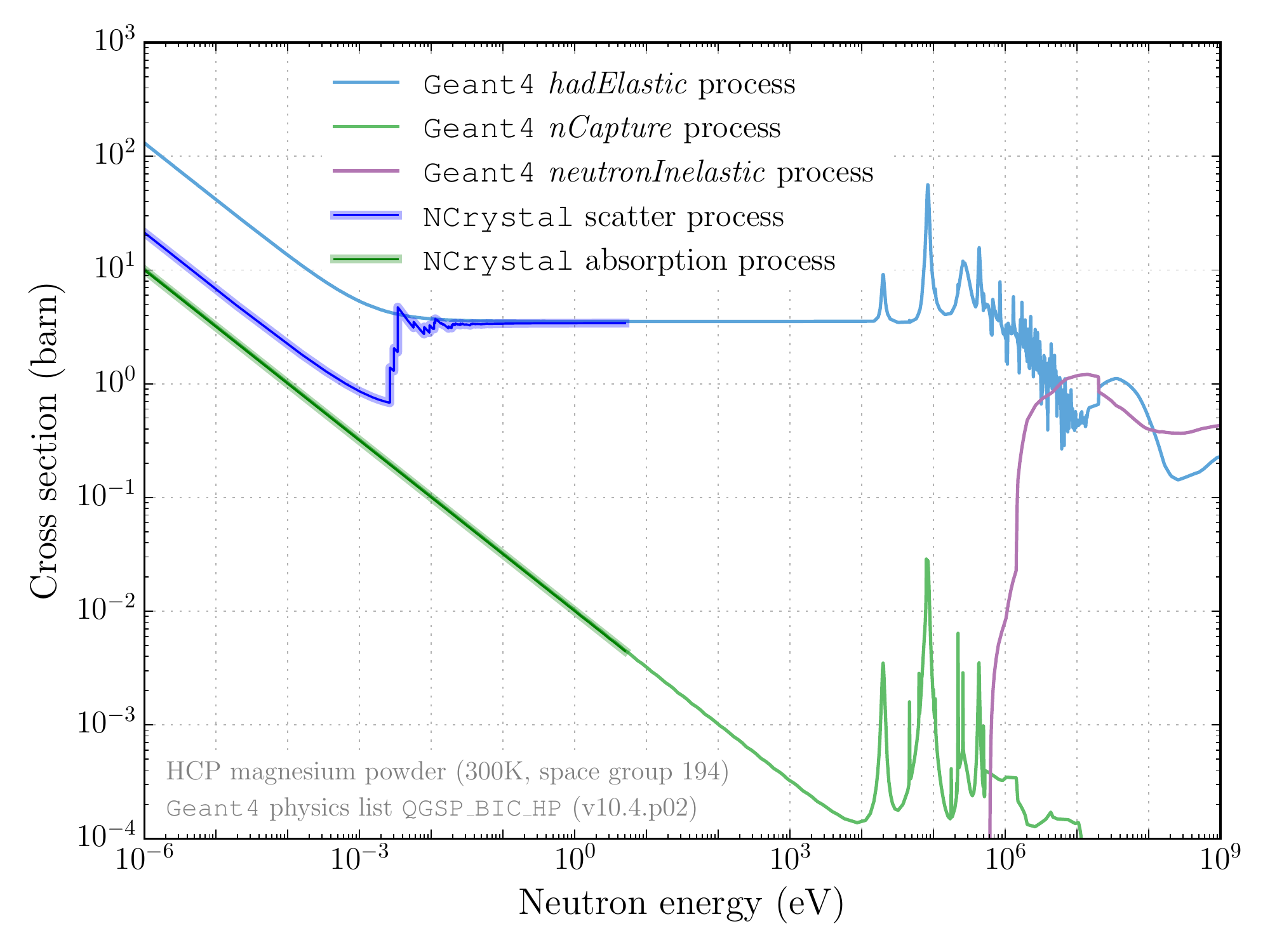}
  \caption{Neutron interaction cross sections in a magnesium powder, as
    predicted by \texttt{Geant4} and \texttt{NCrystal}. At thermal neutron
    energies, absorption in \texttt{Geant4} is handled by the \texttt{Geant4}
    process named ``nCapture'', while scattering (both elastic and inelastic) is
    handled by the process named ``hadElastic''.}
  \labfig{xsgeant4vsncrystal}
\end{figure}

\texttt{Geant4} user code requires very few changes in order to enable
\texttt{NCrystal} modelling of scattering for low energy neutrons. First of all,
the \texttt{G4NCrystal.hh} header file must be included. Next, materials for
which it is desired to use \texttt{NCrystal} to provide scattering physics for
thermal neutrons should be identified. Instances of \texttt{G4Material} for
these materials must then be created by providing appropriate \texttt{NCrystal}
configuration strings:

\begin{minipage}{\linewidth}
\begin{lstlisting}[language={[ncrystal]C++},frame=none]
  G4Material * mat =
    G4NCrystal::createMaterial("Al_sg225.ncmat;temp=200K");
\end{lstlisting}
\end{minipage}

This creates a new \texttt{G4Material}, with relevant settings for standard parameters
including atomic compositions, density and temperature, but additionally an
\texttt{NCrystal} \texttt{Scatter} class instance is attached as a
property. Created instances of \texttt{G4Material} must of course be
subsequently inserted into the simulation geometry in the usual fashion for
\texttt{Geant4}. For oriented materials, i.e.\ single crystals, it is the
\emph{local} orientation of the neutron with respect to the \texttt{Geant4}
volume which is passed to \texttt{NCrystal}. Thus, rotating volumes at the
\texttt{Geant4} level will also rotate the contained material structures, in
line with what one might intuitively expect.

Finally, the \texttt{Geant4} class instances representing physics processes must
be modified, in order to ensure that the embedded \texttt{Scatter} instances are
queried at the correct points during the simulations. This is presently done
via a dynamic modification of the already loaded physics list, with code similar
to:
\begin{lstlisting}[language={[ncrystal]C++},frame=none,belowskip=0.0\baselineskip]
  runManager->Initialize();//Initialise G4RunManager
  G4NCrystal::install();//Install NCrystal into G4 physics list
  runManager->BeamOn(1000);//simulate 1000 events
\end{lstlisting}
This run-time modification of the physics list allows the usage of
\texttt{NCrystal} with any existing physics list in which HP models have been
activated. Although highly flexible, it is planned to also allow
\texttt{NCrystal} to be used in a manner more customary in \texttt{Geant4},
hard-coding it into physics lists at compilation time. Additionally, it is
intended that \texttt{NCrystal} should eventually be integrated into
\texttt{Geant4} releases, making the combined functionality available out of the
box.

As an illustration, \reffig{geantncrystal3dexample} shows a visualisation of a
\texttt{Geant4} simulation in which the \texttt{NCrystal} plugin has been used
to set up a single crystal monochromator. Although the simplistic example does
not do justice to the capabilities of \texttt{Geant4} to support arbitrarily
complicated geometries, the novel potential for physics modelling with the setup
is clear: the orientation of the crystal ensures that those neutrons in the
incoming white beam possessing a compatible wavelength, will be reflected at
exactly \SI{135}{\degree} in the $xy$-plane, which is the defining feature of a
single crystal neutron monochromator. Multiple scattering and geometrical
boundaries are naturally accounted for, and the reflected neutrons exhibit a
characteristic ``zig-zag'' walk,\footnote{This follow from the fact that the
  planes $(h,k,l)$ and $(-h,-k,-l)$ will have identical squared form factors and
  opposite plane normals. A neutron scattering on the $(h,k,l)$ plane will
  therefore always subsequently satisfy the Bragg condition for scattering on the $(-h,-k,-l)$
  plane, and this second interaction will scatter it back in its original
  direction.} and corresponding shifts in positions before leaving the
crystal. Additionally featured are realistic processes such as both inelastic
scatterings and absorption processes which results in the emission of energetic
gamma particles, with a few subsequent Compton scatterings and pair conversions adding
electrons and positrons into the mix.

\begin{figure}
  \centering
  \includegraphics[width=1.0\textwidth]{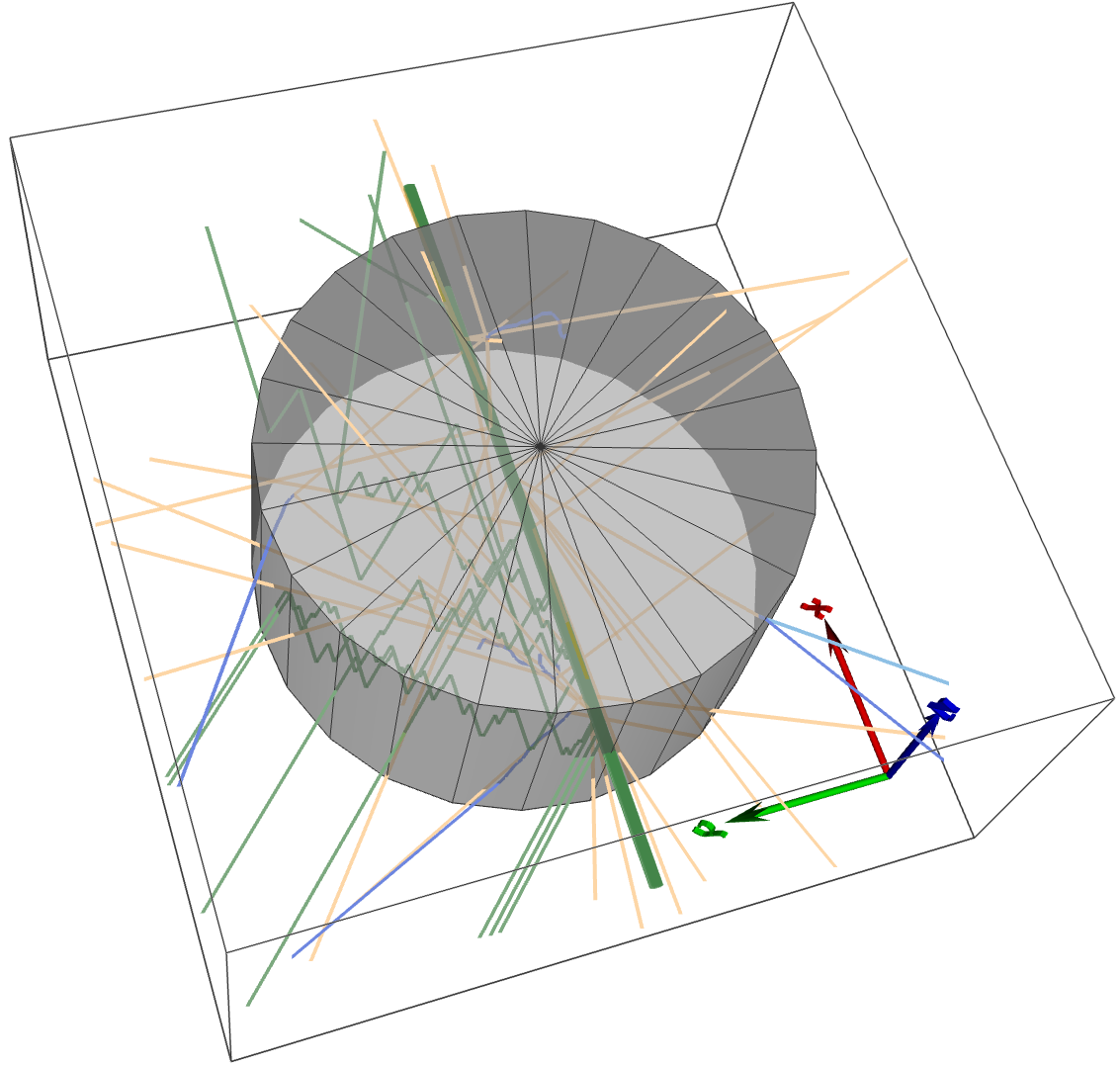}
  \caption{Simulation with \texttt{Geant4} and \texttt{NCrystal} in which a polychromatic beam
    of neutrons (green), travelling along the $x$-axis, enters a cylindrical
    silicon crystal of radius \SI{1}{\milli\metre}, oriented such that neutrons
    with $\lambda\approx\SI{5.79}{\angstrom}$ will be scattered by
    \SI{135}{\degree} in the $xy$-plane by the lattice plane with Miller
    index $111$. Other particles appearing are gammas (yellow), electrons
    (blue), and positrons (light blue). Visualisation created with viewer
    from~\cite{simtoolskelly2018,dgcodechep2013} and used \texttt{Geant4}
    physics list \texttt{QGSP\_BIC\_HP} (v10.0.p03).}
  \labfig{geantncrystal3dexample}
\end{figure}

Another example is shown in \reffig{geant4ncrystal3ddebyescherrer}, where it is
illustrated how scattering of a monochromatic pencil beam of neutrons in an
(untextured) polycrystalline sample changes qualitatively when \texttt{NCrystal}
is enabled: instead of diffuse scattering due to the free-gas approximation,
proper scattering into Debye-Scherrer cones by crystal planes is observed.

\begin{figure}
  \centering
  \subfloat[]{\includegraphics[width=0.49\textwidth]{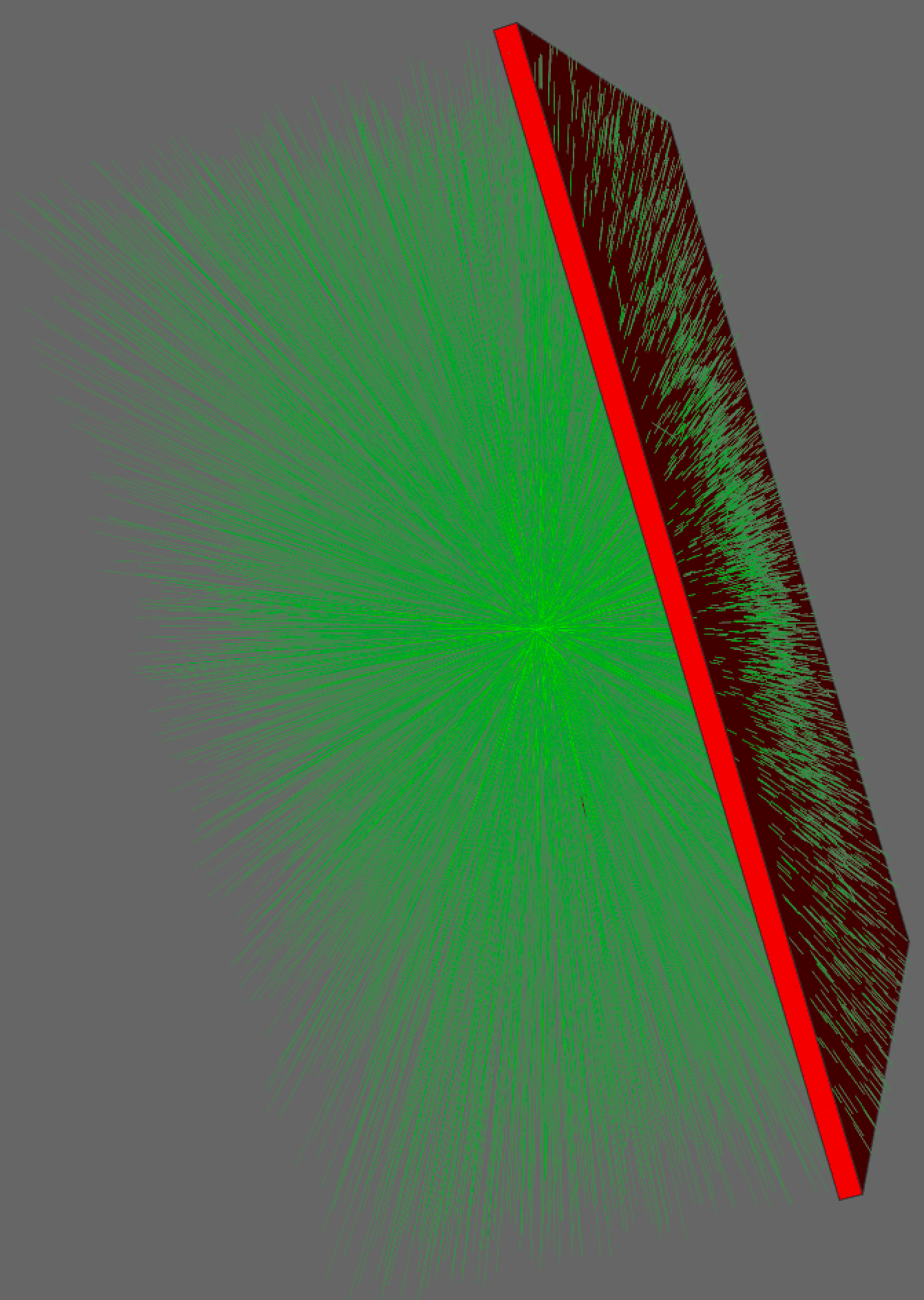}}\hspace*{0.0199\textwidth}
  \subfloat[]{\includegraphics[width=0.49\textwidth]{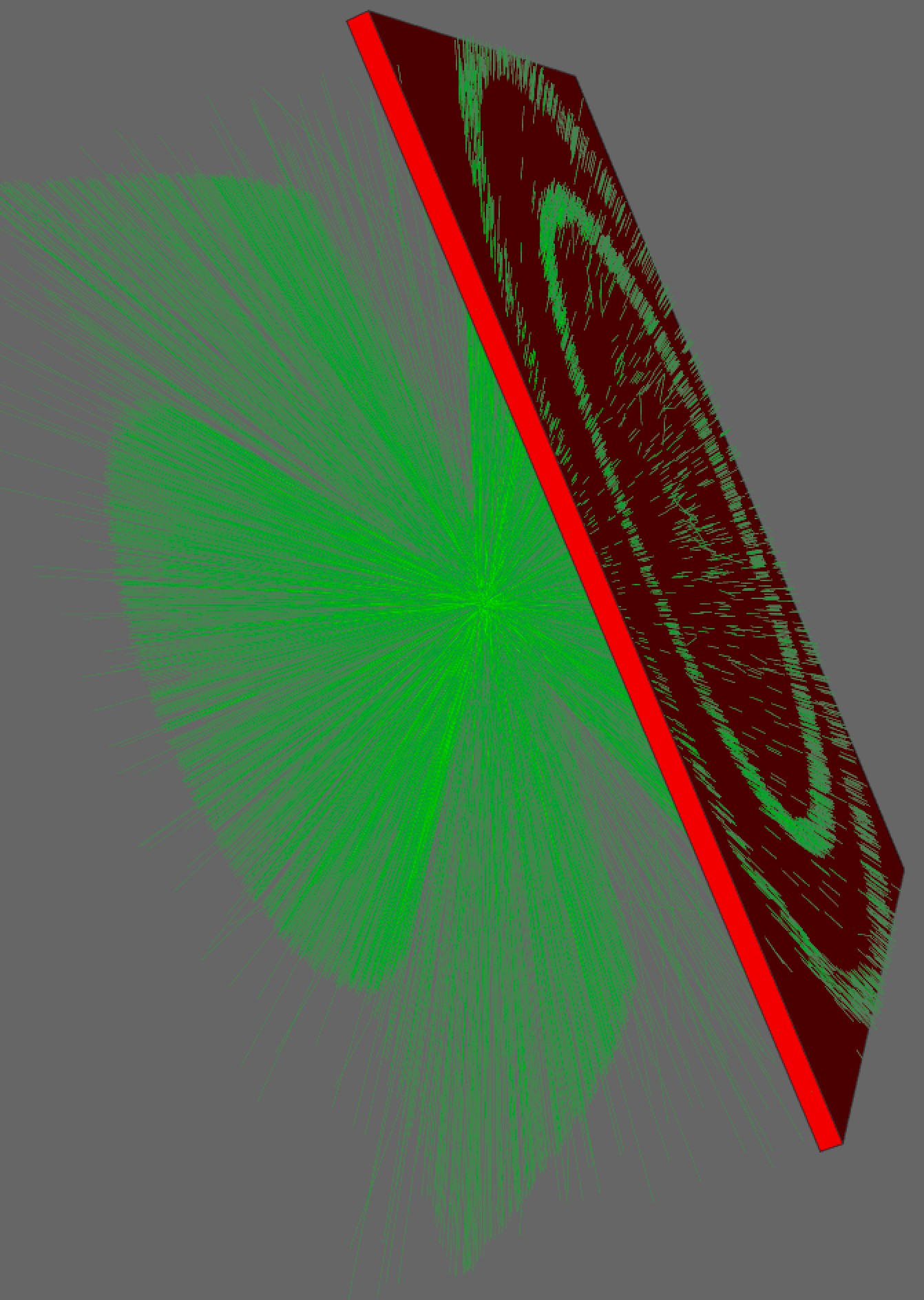}}
  \caption{Simulation with \texttt{Geant4} in which a monochromatic beam of
    neutrons (green), with wavelength $\SI{2.2}{\angstrom}$, enters from the
    left and interacts with a small sample of polycrystalline aluminium. For
    clarity, other particle types such as gamma particles created as by-product
    of absorption events, are not shown. In (a) is shown the diffuse scattering
    resulting from the free-gas approximation when using \texttt{Geant4} out of
    the box with physics list \texttt{QGSP\_BIC\_HP} (v10.0.p03). In (b) is
    shown the more realistic scattering into Debye-Scherrer cones, resulting
    from using \texttt{NCrystal} to provide polycrystalline structure to the
    sample.  Visualisation created with viewer
    from~\cite{simtoolskelly2018,dgcodechep2013}.}
  \labfig{geant4ncrystal3ddebyescherrer}
\end{figure}

\subsection{\texttt{McStas} interface}\labsec{interfaces::mcstas}

In \texttt{McStas} simulations, thermal neutrons are passed through an ordered
list of \emph{components} configured by users in a so-called \texttt{instrument
  file}. Components typically represent actual in-beam elements found at the
modelled neutron instrument such as: source, optical guides, choppers,
filters, monochromators, analysers, samples, or detectors. Each component is
responsible for modelling both geometrical and physics effects, and in addition
to scattering, absorption physics can be implemented either by
trajectory termination or intensity reduction.

The \texttt{NCrystal} plugin for \texttt{McStas} is provided as a component
with the name \texttt{NCrystal\_sample}, but it can be used to model a variety of elements
in addition to samples, including filters, monochromators, and analysers. For
now, it accepts a single configuration string for \texttt{NCrystal} and
implements a material in either a spherical, cylindrical, or box-shaped
geometry. By default it implements absorption via intensity reduction and allows
for multiple scattering interactions, but both of these aspects are
configurable. Full usage instructions are available via the usual \texttt{mcdoc}
documentation system of \texttt{McStas}. The component is implemented by using
the \texttt{C} bindings for \texttt{NCrystal}.

\Reflisting{examplemcstasinstr} shows an example of a \texttt{McStas} instrument
file in which neutrons from a simplistic source model are reflected onto a
cylindrical \texttt{NCrystal} sample of yttrium-oxide powder by a box-shaped
\texttt{NCrystal} copper monochromator aligned to reflect on the $hkl=002$
plane. Finally, neutrons reflected by the powder sample are recorded by a
``banana-shaped'' detector array. The beam-monochromator setup is tuned to provide a
\SI{90}{\degree} scatter angle ($\thetabragg=\SI{45}{\degree}$), when the
incident neutron wavelength fulfils the \texttt{Bragg} condition
$\lambda=2d_{002}\sin\thetabragg$, evaluating to \SI{2.55616}{\angstrom} when
using a value $d_{002}=\SI{1.80748}{\angstrom}$ -- which could have been
extracted programmatically but for simplicity it was in this case determined by
the user via the interactive tool described in \refsec{interfaces::cmdline}.
\Reffig{mcstastracking} shows a 3D visualisation of the resulting simulation
using \texttt{McStas} 2.4.1, while \reffig{mcstaspowderplot} shows the
corresponding diffraction pattern observed in the modelled detector array.

\begin{lstlisting}[float,language={[mccode]C},
  label={lst:examplemcstasinstr},
  caption={Simple \texttt{McStas} instrument file using \texttt{NCrystal} for
    monochromator and sample.}
]
DEFINE INSTRUMENT example()

TRACE

COMPONENT origin = Progress_bar()
  AT (0,0,0) RELATIVE ABSOLUTE

COMPONENT source = Source_div
  ( lambda0 = 2.55616, dlambda = 0.01, xwidth = 0.001,
    yheight = 0.001, focus_aw = 0.4, focus_ah = 0.4 )
  AT (0,0,0.3) RELATIVE origin

COMPONENT arm1 = Arm()
  AT (0,0,0.5) RELATIVE source ROTATED (0,45,0) RELATIVE source

COMPONENT monochromator = NCrystal_sample
  ( xwidth = 0.05, yheight = 0.05, zdepth = 0.003,
    cfg = "Cu_sg225.ncmat;mos=0.3deg;bkgd=0"
          ";dir1=@crys_hkl:0,0,2@lab:0,0,1"
          ";dir2=@crys_hkl:1,0,0@lab:0,1,0" )
  AT (0,0,0) RELATIVE arm1

COMPONENT arm2 = Arm()
  AT (0,0,0) RELATIVE arm1 ROTATED (0,-90,0) RELATIVE source

COMPONENT powder_sample = NCrystal_sample
  ( yheight = 0.03, radius = 0.01,
    cfg = "Y2O3_sg206_Yttrium_Oxide.ncmat;packfact=0.8" )
  AT (0,0,0.4) RELATIVE arm2

COMPONENT detectors = Monitor_nD
  ( options = "banana, angle limits=[10 170], bins=500",
    radius = 0.2, yheight = 0.2 )
  AT (0,0,0) RELATIVE powder_sample

END
\end{lstlisting}

\begin{figure}
  \centering
  \includegraphics[width=1.0\textwidth]{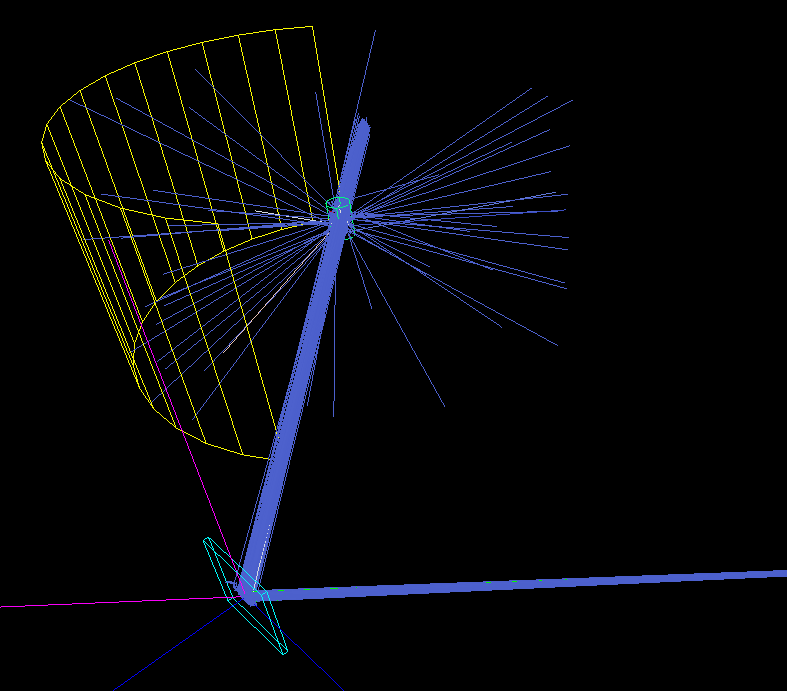}
  \caption{\texttt{McStas} visualisation of the simulation resulting from the
    instrument file shown in \reflisting{examplemcstasinstr}. Neutrons impinge
    on the monochromator box (light-blue) from the right, and are reflected
    in the horizontal plane towards the cylindrical powder sample (also light-blue). The
    detector array (yellow) captures parts of the Debye-Scherrer cones created
    by reflections in the sample.}
  \labfig{mcstastracking}
\end{figure}

\begin{figure}
  \centering
  \includegraphics[width=0.9\textwidth]{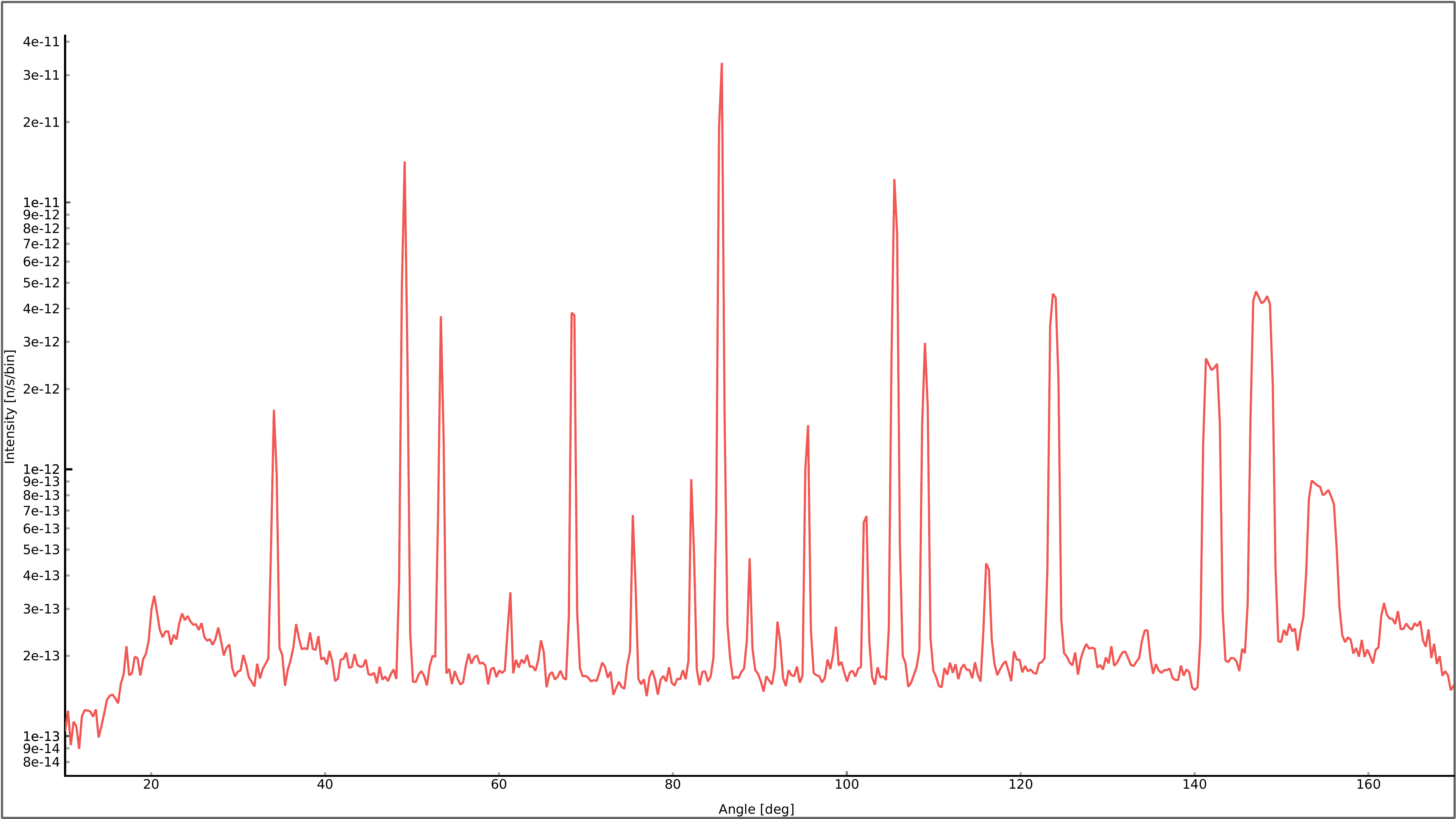}
  \caption{\texttt{McStas} diffraction pattern resulting from a simulation of
    \num{e7} source neutrons using the instrument file shown in
    \reflisting{examplemcstasinstr}. The shown intensity in each of the 500 bins
    corresponds to a source flux of
    \SI{1}{neutron/\second/\square\cm/\steradian/\angstrom}.}
  \labfig{mcstaspowderplot}
\end{figure}

In addition to benefiting from future improvements to the \texttt{NCrystal}
library, it is foreseen that the \texttt{NCrystal\_sample} \texttt{McStas}
component itself might also be further enhanced at a later state. In particular,
it would be desirable to implement variance reduction techniques, at least in
terms of making it possible to focus outgoing neutrons towards the next
down-stream component. It is also likely that use-cases for more advanced
geometrical layouts will arise, and the code has consequently been structured in
a way which makes it straight-forward to add such features.

\section{Outlook}\labsec{outlook}

The presented toolkit for thermal neutron transport is arguably unique in its
attention to interfaces and capability for integration into various technical
contexts and is already in version 1.0.0 very capable in terms of modelling of interactions in
single crystals and crystal powders, and has already been used to enable a
range of interesting studies
(e.g.~\cite{simtoolskelly2018,xxcaisampling2018,santoro2015,eszternssmic2017,taggingmessi2017preprint,kellyrates2018,mauri2018,galgoczi2018,dian2019}).

Nonetheless, work has already begun on several improvements to both physics
models and the framework itself. Firstly, as mentioned in
\refsec{theory::inelastic}, the capabilities for modelling of inelastic and
incoherent scattering should see significant enhancements -- with the possibility
of supporting liquids or polymers to some extent. These models, and those
implementing Bragg diffraction, will be described in detail in future dedicated
publications.

Next, it is the plan to carry out framework extensions and refactorisations
which will allow \texttt{NCrystal} to support enhanced material realism, by
making the exact composition of materials and crystals more customisable. Once
implemented, it should on one hand become possible to support multi-phase
materials -- needed for realistic multi-phase metal alloys or crystal powders
suspended in liquids -- and on the other hand the composition of each phase
should become more flexible as well, allowing for enriched materials or chemical
disorder. That would enable modelling of crystals in which some sites are not
fully occupied in all cells -- or occasionally occupied with elements playing
the role of contaminants or dopants.

Several more technical developments are envisioned as well: planned interface
extensions will enable better support of multi-threaded applications (such as
\texttt{ANTS2} or multi-threaded builds of \texttt{Geant4}), and several
use-cases have been identified where it would be advantageous to be able to
initialise crystal data directly from process memory rather than needing on-disk
files. In the longer run, once all relevant platforms and applications support
it, it is also planned to drop the support for the \texttt{C++98} standard, in
order to better benefit from modern \texttt{C++} features and better
cross platform support introduced in \texttt{C++11} and beyond.

Beyond that, the future directions will depend on resources and community
interest. At the very least the library of data files and the list of Monte
Carlo applications with \texttt{NCrystal} support are both expected to
expand. But given sufficient interest and contributions new ambitious physics
models could be added, ranging from treatment of texture, bent crystals and new
anisotropic mosaicity models to better facilities for dealing with nuclear
resonances or branching into new areas like magnetic spin-dependent interactions
or support of X-ray physics. Input, feedback, ideas or contributions are
gratefully received via the \texttt{NCrystal} website~\cite{ncrystalwww}.

\section*{Acknowledgements}

This work was supported in part by the European Union's Horizon 2020 research
and innovation programme under grant agreement No 676548 (the BrightnESS
project). The authors would like to thank the following colleagues for valuable
contributions, testing, feedback, ideas or other support: E. Dian,
G. Galg\'{o}czi, R. Hall-Wilton, K. Kanaki, M. Klausz, E. Klinkby,
E. B. Knudsen, J.I. M\'{a}rquez Dami\'{a}n, V. Maulerova, A. Morozov,
V. Santoro, and P. Willendrup.

\section*{References}

\bibliography{refs}

\end{document}